\documentclass[aps,prc,showpacs,superscriptaddress,reprint]{revtex4-2}
\usepackage{rotating}
\usepackage{epsfig} 
\usepackage{color}
\usepackage{xcolor}

\usepackage[normalem]{ulem}

\usepackage{url}

\usepackage{multirow}

\usepackage{amsmath}

\usepackage{lineno}
\usepackage{graphicx}

\usepackage{hyperref}

\graphicspath{ {./plot/}, {./image/} }
\DeclareGraphicsExtensions{ .pdf, .png}

\makeatletter
\renewcommand*{\p@subsection}{}
\renewcommand*{\p@subsubsection}{}
\makeatother

\begin{document}

%------------------------------------------------------------------------------------------------%

\newcommand{\der}{\text{d}}
\newcommand{\rp}{{\textsc{rp}}}
\newcommand{\pt}{p_{T}}
\newcommand{\gevc}{GeV/$c$}
\newcommand{\Ru}{{\rm Ru}}
\newcommand{\Zr}{{\rm Zr}}
\newcommand{\Ntrk}{N_{\rm trk}}

\newcommand{\qcd}{{\textsc{qcd}}}
\newcommand{\pos}{{\textsc{os}}}
\newcommand{\pss}{{\textsc{ss}}}
\newcommand{\twp}{{\rm 2p}}
\newcommand{\thp}{{\rm 3p}}
\newcommand{\nf}{{\text{nf}}}
\newcommand{\deta}{\Delta\eta}
\newcommand{\dphi}{\Delta\phi}
\newcommand{\cme}{{\textsc{cme}}}
\newcommand{\ampt}{{\textsc{ampt}}}
\newcommand{\hijing}{{\textsc{hijing}}}
\newcommand{\geant}{{\textsc{geant}}}
\newcommand{\poi}{{\textsc{poi}}}
\newcommand{\snn}{\sqrt{s_{_{\textsc{NN}}}}}
\newcommand{\srp}{{\textsc{rp}}}
\newcommand{\ssp}{{\textsc{sp}}}
\newcommand{\spp}{{\textsc{pp}}}
\newcommand{\sep}{{\textsc{ep}}}
\newcommand{\dg}{\Delta\gamma}
\newcommand{\enf}{\epsilon_{\rm nf}}
\newcommand{\fcme}{f_{\textsc{cme}}}
\newcommand{\minv}{m_{\rm inv}}
\newcommand{\bkgd}{\text{bkgd}}
\newcommand{\signal}{\text{signal}}
\newcommand {\mean}[1]   {\left\langle{#1}\right\rangle}

\newcommand{\fhr}{h_{\text{r}}}
\newcommand{\fhm}{h_{\text{m}}}
\newcommand{\fhc}{h_{\text{c}}}
\newcommand{\fhp}{h_{\text{p}}}
\newcommand{\ffr}{f_{\text{r}}}
\newcommand{\ffc}{f_{\text{c}}}
\newcommand{\ffnf}{f_{\text{nf}}}
\newcommand{\fftf}{f_{\text{tf}}}
\newcommand{\fgr}{g_{\text{r}}}
\newcommand{\fgm}{g_{\text{m}}}
\newcommand{\fgc}{g_{\text{c}}}

\newcommand {\red}[1]   {\textcolor{red}{#1}}
\newcommand {\blue}[1]  {\textcolor{blue}{#1}}
\newcommand {\green}[1] {\textcolor{green}{#1}}
\newcommand {\gray}[1] {\textcolor{gray}{#1}}
\newcommand {\orange}[1] {\textcolor{orange}{#1}}

%------------------------------------------------------------------------------------------------%

%\linespread{1.6}
\title{
%Estimate of the nonflow background in the isobar measurement Ru+Ru/Zr+Zr ratio of $N\Delta\gamma/v_{2}$ at STAR
Estimate of Background Baseline and Upper Limit on the Chiral Magnetic Effect in Isobar Collisions at $\snn=200$~GeV at the Relativistic Heavy-Ion Collider
}

\affiliation{Abilene Christian University, Abilene, Texas   79699}
\affiliation{Alikhanov Institute for Theoretical and Experimental Physics NRC "Kurchatov Institute", Moscow 117218}
\affiliation{Argonne National Laboratory, Argonne, Illinois 60439}
\affiliation{American University in Cairo, New Cairo 11835, Egypt}
\affiliation{Ball State University, Muncie, Indiana, 47306}
\affiliation{Brookhaven National Laboratory, Upton, New York 11973}
\affiliation{University of Calabria \& INFN-Cosenza, Rende 87036, Italy}
\affiliation{University of California, Berkeley, California 94720}
\affiliation{University of California, Davis, California 95616}
\affiliation{University of California, Los Angeles, California 90095}
\affiliation{University of California, Riverside, California 92521}
\affiliation{Central China Normal University, Wuhan, Hubei 430079 }
\affiliation{University of Illinois at Chicago, Chicago, Illinois 60607}
\affiliation{Chongqing University, Chongqing, 401331}
\affiliation{Creighton University, Omaha, Nebraska 68178}
\affiliation{Czech Technical University in Prague, FNSPE, Prague 115 19, Czech Republic}
\affiliation{National Institute of Technology Durgapur, Durgapur - 713209, India}
\affiliation{ELTE E\"otv\"os Lor\'and University, Budapest, Hungary H-1117}
\affiliation{Frankfurt Institute for Advanced Studies FIAS, Frankfurt 60438, Germany}
\affiliation{Fudan University, Shanghai, 200433 }
\affiliation{Guangxi Normal University, Guilin, 541004}
\affiliation{University of Heidelberg, Heidelberg 69120, Germany }
\affiliation{University of Houston, Houston, Texas 77204}
\affiliation{Huzhou University, Huzhou, Zhejiang  313000}
\affiliation{Indian Institute of Science Education and Research (IISER), Berhampur 760010 , India}
\affiliation{Indian Institute of Science Education and Research (IISER) Tirupati, Tirupati 517507, India}
\affiliation{Indian Institute Technology, Patna, Bihar 801106, India}
\affiliation{Indiana University, Bloomington, Indiana 47408}
\affiliation{Institute of Modern Physics, Chinese Academy of Sciences, Lanzhou, Gansu 730000 }
\affiliation{University of Jammu, Jammu 180001, India}
\affiliation{Joint Institute for Nuclear Research, Dubna 141 980}
\affiliation{Kent State University, Kent, Ohio 44242}
\affiliation{University of Kentucky, Lexington, Kentucky 40506-0055}
\affiliation{Lawrence Berkeley National Laboratory, Berkeley, California 94720}
\affiliation{Lehigh University, Bethlehem, Pennsylvania 18015}
\affiliation{Max-Planck-Institut f\"ur Physik, Munich 80805, Germany}
\affiliation{Michigan State University, East Lansing, Michigan 48824}
\affiliation{National Research Nuclear University MEPhI, Moscow 115409}
\affiliation{National Institute of Science Education and Research, HBNI, Jatni 752050, India}
\affiliation{National Cheng Kung University, Tainan 70101 }
\affiliation{The Ohio State University, Columbus, Ohio 43210}
\affiliation{Panjab University, Chandigarh 160014, India}
\affiliation{NRC "Kurchatov Institute", Institute of High Energy Physics, Protvino 142281}
\affiliation{Purdue University, West Lafayette, Indiana 47907}
\affiliation{Rice University, Houston, Texas 77251}
\affiliation{Rutgers University, Piscataway, New Jersey 08854}
\affiliation{University of Science and Technology of China, Hefei, Anhui 230026}
\affiliation{South China Normal University, Guangzhou, Guangdong 510631}
\affiliation{Sejong University, Seoul, 05006, South Korea}
\affiliation{Shandong University, Qingdao, Shandong 266237}
\affiliation{Shanghai Institute of Applied Physics, Chinese Academy of Sciences, Shanghai 201800}
\affiliation{Southern Connecticut State University, New Haven, Connecticut 06515}
\affiliation{State University of New York, Stony Brook, New York 11794}
\affiliation{Instituto de Alta Investigaci\'on, Universidad de Tarapac\'a, Arica 1000000, Chile}
\affiliation{Temple University, Philadelphia, Pennsylvania 19122}
\affiliation{Texas A\&M University, College Station, Texas 77843}
\affiliation{University of Texas, Austin, Texas 78712}
\affiliation{Tsinghua University, Beijing 100084}
\affiliation{University of Tsukuba, Tsukuba, Ibaraki 305-8571, Japan}
\affiliation{University of Chinese Academy of Sciences, Beijing, 101408}
\affiliation{Valparaiso University, Valparaiso, Indiana 46383}
\affiliation{Variable Energy Cyclotron Centre, Kolkata 700064, India}
\affiliation{Wayne State University, Detroit, Michigan 48201}
\affiliation{Wuhan University of Science and Technology, Wuhan, Hubei 430065}
\affiliation{Yale University, New Haven, Connecticut 06520}

\author{M.~I.~Abdulhamid}\affiliation{American University in Cairo, New Cairo 11835, Egypt}
\author{B.~E.~Aboona}\affiliation{Texas A\&M University, College Station, Texas 77843}
\author{J.~Adam}\affiliation{Czech Technical University in Prague, FNSPE, Prague 115 19, Czech Republic}
\author{J.~R.~Adams}\affiliation{The Ohio State University, Columbus, Ohio 43210}
\author{G.~Agakishiev}\affiliation{Joint Institute for Nuclear Research, Dubna 141 980}
\author{I.~Aggarwal}\affiliation{Panjab University, Chandigarh 160014, India}
\author{M.~M.~Aggarwal}\affiliation{Panjab University, Chandigarh 160014, India}
\author{Z.~Ahammed}\affiliation{Variable Energy Cyclotron Centre, Kolkata 700064, India}
\author{A.~Aitbaev}\affiliation{Joint Institute for Nuclear Research, Dubna 141 980}
\author{I.~Alekseev}\affiliation{Alikhanov Institute for Theoretical and Experimental Physics NRC "Kurchatov Institute", Moscow 117218}\affiliation{National Research Nuclear University MEPhI, Moscow 115409}
% ...

\collaboration{STAR Collaboration}\noaffiliation

%\author{The STAR Collaboration}
%\author{Yicheng Feng}
%\email{feng216@purdue.edu}
%\address{Department of Physics and Astronomy, Purdue University, West Lafayette, IN 47907, USA}
%\author{Fuqiang Wang}
%\email{fqwang@purdue.edu}
%\address{Department of Physics and Astronomy, Purdue University, West Lafayette, IN 47907, USA}
%\author{CME Focus Group}
\date{\today} %this is useful in drafting stage
%\draftversion{12}

%------------------------------------------------------------------------------------------------%

\begin{abstract}
For the search of the chiral magnetic effect (CME), STAR previously presented the results from isobar collisions (${^{96}_{44}\text{Ru}}+{^{96}_{44}\text{Ru}}$, ${^{96}_{40}\text{Zr}}+{^{96}_{40}\text{Zr}}$) obtained through a blind analysis. The ratio of results in Ru+Ru to Zr+Zr collisions for the CME-sensitive charge-dependent azimuthal correlator ($\Delta\gamma$), normalized by elliptic anisotropy ($v_{2}$), was observed to be close to but systematically larger than the inverse multiplicity ratio.
The background baseline for the isobar ratio, $Y = \frac{(\Delta\gamma/v_{2})^{\text{Ru}}}{(\Delta\gamma/v_{2})^{\text{Zr}}}$, is naively expected to be $\frac{(1/N)^{\Ru}}{(1/N)^{\Zr}}$; however, genuine two- and three-particle correlations are expected to alter it.
We estimate the contributions to $Y$ from those correlations, utilizing both the isobar data and  \hijing\ simulations.
After including those contributions, we arrive at a final background baseline for $Y$, which is consistent with the isobar data.
We extract an upper limit for the CME fraction in the $\Delta\gamma$ measurement of approximately $10\%$ at a $95\%$ confidence level on in isobar collisions at $\sqrt{s_{\textsc{nn}}} =$ 200 GeV, with an expected 15\% difference in their squared magnetic fields. %with a realistic assumption of 15\% isobar difference between their magnetic fields ($\delta B^{2}/B^{2} = 15\%$). %stronger magnetic field in Ru+Ru than Zr+Zr ($\delta B^{2}/B^{2} = 15\%$). 

\end{abstract}

%\pacs{25.75.-q, 25.75.-Gz, 25.75.-Ld} 

%------------------------------------------------------------------------------------------------%

\maketitle

%------------------------------------------------------------------------------------------------%

\section{Introduction} \label{sec:introduction}

%The chiral magnetic effect (CME) refers to an electric current (charge separation) along the strong magnetic field produced in relativistic heavy-ion collisions, arising from chirality-imbalanced, $\mathcal{P}$- and $\mathcal{CP}$-odd metastable domains~\cite{Fukushima:2008xe}.The formation of such domains has been predicted by quantum chromodynamics (QCD) to occur at high temperatures in those collisions because of vacuum fluctuations~\cite{Morley:1983wr,Kharzeev:1998kz,Kharzeev:2004ey,Kharzeev:2007jp} and may be pertinent to the matter-antimatter asymmetry of our universe~\cite{RevModPhys.76.1}. 
The quantum chromodynamics (QCD) predicts vacuum fluctuations, rendering nonzero topological charges in a local domain with odd $\mathcal{P}$ and $\mathcal{CP}$ symmetries~\cite{Morley:1983wr,Kharzeev:1998kz,Kharzeev:2004ey,Kharzeev:2007jp}, which may be pertinent to the matter-antimatter asymmetry of our universe~\cite{RevModPhys.76.1}. As a result, there would be more particles with one certain chirality than the other, which is called chirality imbalance. If there is a strong magnetic field, the spins of particles would be locked either parallel or anti-parallel to the magnetic field direction, depending on their charges. Then, with the same chirality imbalance and opposite spin directions, the positive and negative charged particles would have opposite momentum directions. This charge separation phenomenon is called the chiral magnetic effect (CME)~\cite{Fukushima:2008xe}. 

In heavy-ion collisions, we refer to the collided nucleons as participants and the others as spectators. From the collision geometry, the magnetic field created by the spectator protons is generally perpendicular to the reaction plane (\srp, spanned by the impact parameter and beam direction), so the CME searches usually take \srp\ as the reference direction. The particle distribution can be expanded w.r.t.~\srp\ in azimuth ($\psi_{\rp}$) into a Fourier series
\begin{equation} \label{eq:dist}
\frac{2\pi}{N} \frac{\der N}{\der \phi} = 1 + 2 v_{1}\cos(\phi-\psi_{\srp}) + 2 v_{2} \cos2(\phi-\psi_{\srp}) + \ldots\,,
\end{equation}
where $N$ is the number of particles and $\phi$ is the azimuthal angle in the plane perpendicular to the beam axis.

The charge-dependent azimuthal correlator $\Delta\gamma$ is used to measure the charge separation of CME~\cite{Voloshin:2004vk}. Its definition is as follows: 
% \begin{equation} \label{eq:dg:rp}
% \begin{split}
%     \gamma_{\alpha\beta} =& \langle \cos(\phi_{\alpha} + \phi_{\beta} - 2\psi_{\srp}) \rangle\,, \\
%     \Delta\gamma =& \gamma_{\pos} - \gamma_{\pss}\,,
% \end{split}
% \end{equation}
\begin{equation} \label{eq:dg:rp}
    \gamma_{\alpha\beta} \equiv \langle \cos(\phi_{\alpha} + \phi_{\beta} - 2\psi_{\srp}) \rangle\,.
\end{equation}
%$\psi_{\srp}$ is the azimuth of \srp (defined by the beam and impact parameter directions).  
Here, the subscripts $\alpha$, $\beta$ represent two different particles of interest (\poi) in the same event, and their electric charge signs determine whether the pair ($\alpha\beta$) is opposite-sign (\pos) or same-sign (\pss). For a given pair sign (\pos\ or \pss), the angle brackets $\langle \cdots \rangle$ represent the average over those pairs within one event and then further averaged over multiple events in each centrality. 
The signature of CME is that $\gamma_\pos>0$ and $\gamma_\pss<0$ with the same magnitude~\cite{Voloshin:2004vk}. 
However, there are correlation backgrounds that can also cause the $\gamma$ correlators to deviate from zero~\cite{Voloshin:2004vk}. 
Many of these backgrounds are charge independent, such as those arising from momentum conservation. 
In order to remove those charge-independent backgrounds, the difference between \pos\ and \pss\ is computed as follows: 
\begin{equation}
    \Delta\gamma \equiv \gamma_{\pos} - \gamma_{\pss}\,.
\end{equation}
Experiments have focused on measuring the $\dg$ observable. Large signals of $\dg$ have been measured at the Relativistic Heavy-Ion Collider (RHIC)~\cite{Abelev:2009ac,Abelev:2009ad,Abelev:2012pa,Adamczyk:2013hsi,Adamczyk:2013kcb,Adamczyk:2014mzf,STAR:2019xzd,STAR:2020gky,STAR:2021pwb} and the Large Hadron Collider (LHC)~\cite{Khachatryan:2016got,Sirunyan:2017quh,Acharya:2017fau,ALICE:2020siw}.

Although charge-independent backgrounds are canceled in $\dg$, backgrounds remain from charge-dependent two-particle (2p) correlations coupled with elliptic flow of those correlation sources such as resonance decays and jets~\cite{Voloshin:2004vk,Wang:2009kd,Bzdak:2009fc,Schlichting:2010qia,STAR:2022ahj}.
These backgrounds can be expressed as
\begin{equation}
    \gamma_\pos^{\rm bkgd}=\frac{N_\twp}{N_\pos}\mean{\cos(\phi_{\alpha}+\phi_{\beta}-2\phi_{\twp})}_{\twp} v_{2,\twp}\,.
    \label{eq:bkgd}
\end{equation}
Here, $N_\twp$, $\phi_{\twp}$, and $v_{2,\twp}$ represent the number, azimuthal angle, and elliptic flow parameter of those \pos\ $\twp$ correlation sources (e.g., $\rho$ resonances), respectively.
Elliptic flow $v_2$ is defined as
\begin{equation} \label{eq:v2:rp}
    v_{2} = \mean{\cos(2\phi - 2\psi_{\srp})}\,,
\end{equation}
which is also a coefficient in Eq.~(\ref{eq:dist}).
It has been found that those backgrounds dominate charge-separation measurements~\cite{STAR:2021pwb,STAR:2022ahj}.

To mitigate these backgrounds, 
STAR conducted experiments of isobar collisions ${^{96}_{44}\text{Ru}}+{^{96}_{44}\text{Ru}}$ and ${^{96}_{40}\text{Zr}}+{^{96}_{40}\text{Zr}}$ at $\snn = 200$ GeV in 2018~\cite{STAR:2021mii}. 
%The two isobar species were initially expected to have similar background in CME measurements due to the same nucleon number, while Ru+Ru was expected to have larger CME signal due to more protons.
%The choice of these two isobaric species was based on the initial expectation that they would exhibit similar backgrounds in CME measurements due to having the same nucleon number. However, it was anticipated that Ru+Ru collisions would yield a larger CME signal due to having more protons.
%If so, the charge-independent quantities like multiplicity and elliptic flow $v_{2}$ would be ideally the same between the two isobars.
The choice of these two isobaric species was based on the strategy of keeping the background constant while changing the signal. 
It was anticipated that the CME-related signal would be larger in Ru+Ru due to the larger number of protons creating a stronger magnetic field.  
It was also initially expected that with the same number of nucleons in Zr and Ru, flow-related backgrounds would be the same in the two species' collisions, ideally with charge-independent quantities like multiplicity and elliptic flow ($v_{2}$) the same between the two isobars. 
However, the STAR data shows a few percent difference between the two isobaric species for both quantities. 
This discrepancy arises because Ru and Zr have different nuclear sizes and structures, which were  predicted by energy density functional theory (DFT) calculations~\cite{Xu:2017zcn,Li:2018oec,Xu:2021vpn}. 
For the charge-dependent and CME-sensitive quantity, the STAR data show the isobar ratio (Ru+Ru/Zr+Zr) of $\Delta\gamma / v_{2}$ below unity~\cite{STAR:2021mii}. This seems opposite to the initial expectation of larger CME in Ru+Ru but, in fact, results from the different multiplicities. 
If the number of background correlation sources is proportional to multiplicity ($N$), then background $\Delta\gamma$ would be diluted by $N$ because $\Delta\gamma$ is a pair-wide average and the pair multiplicity $N_\pos,N_\pss\propto N^2$. 
Then, the naive background baseline of $\frac{(\Delta\gamma / v_{2})^{\Ru}}{(\Delta\gamma / v_{2})^{\Zr}}$ would be $\frac{(1/N)^{\Ru}}{(1/N)^{\Zr}}$, and Ref.~\cite{STAR:2021mii} finds that the former is larger than the latter, though with large uncertainties, suggesting a small CME signal~\cite{Kharzeev:2022hqz}.

%The inverse multiplicity scaling of $\dg/v_{2}$ is only approximate and precise only when the number of background sources is proportional to multiplicity, $N_\twp\propto N$. This assumption may not hold for the two isobar collision systems due to their slightly  different energy densities. 
However, this $1/N$ scaling for $\Delta\gamma / v_{2}$ is approximate. 
It assumes that the number of background sources is proportional to multiplicity, $N_\twp\propto N$. 
This assumption may not hold for the two isobar collision systems and may be violated differently in the two due to their slightly different energy densities. 
A more precise way of scaling may be to divide $\Delta\gamma / v_{2}$ by $r = N_\twp/N_\pos$ (Sec.~\ref{sec:r}).
Then, the STAR data show the isobar ratio of $\frac{1}{r} \Delta\gamma / v_{2}$ below unity~\cite{STAR:2021mii}, indicating more complications to be considered~\cite{Feng:2022yus}.
The fact that scaling by $r$ does not give the same result as scaling by multiplicity indicates that the assumption of proportionality is not necessarily valid--there are residual backgrounds in addition to the naive baseline of unity.

Consequently, it is critical to pin down the exact background baseline in order to access information on the CME from the $\dg$ measurements in isobar collisions. It is the goal of this paper and Ref.~\cite{STAR:2023gzg} to arrive at a rigorous estimate of the background baseline and examine what the isobar data entail with respect to the possible CME signal. 
Reference~\cite{STAR:2023gzg} summarizes the essential findings, and this paper provides necessary details of the analysis work.

%------------------------------------------------------------------------------------------------%

\section{Background baseline} \label{sec:methodology}

In heavy-ion experiments, \srp\ is unknown. One often reconstructs an event plane (\sep) from the particle momentum distribution, taking this reconstructed \sep\ as a proxy for \srp. More elegantly, one exploits particle cumulants; 
instead of reconstructing an \sep, one can calculate $v_{2}$ from the two-particle correlator
\begin{equation} \label{eq:v2:2p}
    v_{2}^* = \sqrt{ \mean{\cos 2(\phi_{\alpha} - \phi_{\beta})} }\,.
\end{equation}
This $v_{2}$ measurement includes nonflow backgrounds, such as two-particle correlations from  jets or resonance decays, whose contribution we quantify by  
\begin{equation} \label{eq:enf}
    \enf = \frac{v_{2,\nf}^{2}}{v_{2}^{2}} = \frac{{v_{2}^{*}}^{2}}{v_{2}^{2}} - 1 \,. 
\end{equation}
Here, $v_{2}^{*}$ stands for the measurement that includes nonflow correlations unrelated to the global collision geometry, while $v_{2}$ refers to the \textit{true} flow.
Note that the cumulant measurement of $v_2^*$ is similar to that obtained by reconstructing an \sep, but with subtle differences~\cite{Adams:2004bi,Bilandzic:2008nx}.  
It should also be noted that the elliptic flow measured by the \sep\ method is also contaminated by nonflow, as the \sep\ is reconstructed using final-state particles~\cite{Poskanzer:1998yz}. 
However, the decomposition of nonflow from flow in the \sep\ method is less straightforward than that in the cumulant method~\cite{Ollitrault:2009ie}.

Similarly, instead of calculating the $\gamma$ correlation using Eq.~\ref{eq:dg:rp}, one can measure a three-particle (3p) correlator
\begin{equation} \label{eq:dg:3p}
    C_{3, \alpha\beta} = \langle \cos(\phi_{\alpha} + \phi_{\beta} - 2\phi_{c}) \rangle\,. 
\end{equation}
Here, $c$ represents a third particle that is different from $\alpha$, $\beta$.  The $\gamma$ correlator can be calculated using  $\gamma_{\alpha\beta} = C_{3, \alpha\beta} / v_{2,c}$, where $v_{2,c}$ is the elliptic flow of particle of type $c$, given by Eq.~(\ref{eq:v2:rp}). 
Effectively, particle $c$ serves as the event plane, and its resolution is simply equal to the particle's elliptic flow $v_{2,c}$.
In practice, $\gamma$ is determined by dividing by the measured quantity $v_2^*$,
\begin{equation} \label{eq:g:c3}
    \gamma_{\alpha\beta} = C_{3, \alpha\beta} / v_2^* \,.
\end{equation}
% and the average over $c$ serves as an estimate to $\Psi_{\srp}$, whose resolution is equal to its elliptic flow $v_{2}$.

The main background in $C_3$ ($=C_{3,\pos}-C_{3,\pss}$) arises from the flow-induced background~\cite{STAR:2021pwb,STAR:2022ahj}.  In this scenario, some of the \poi's are correlated with one another via a 2p source and these 2p sources are all correlated with each particle $c$ through the global flow correlation. 
This flow-induced background is described by Eq.~(\ref{eq:bkgd}). 
In addition to this flow-induced background, there is contamination in $C_3$ from genuine 3p correlations, where the three particles $\alpha$, $\beta$ and $c$ are intrinsically correlated. Thus, the background contributions to the 3p correlators can be expressed as: 
\begin{equation} 
\begin{array}{llllll}
    C_{3,\pos}^\bkgd=&\frac{N_\pss}{N_\pos}&\gamma_\pss v_{2,c} & +\frac{N_\twp}{N_\pos}C_{\twp,\pos} v_{2,\twp}v_{2,c} &+&\frac{N_{\thp,\pos}}{N_\pos N_c}C_{\thp,\pos}\,,\\ 
    C_{3,\pss}^\bkgd=&                     &\gamma_\pss v_{2,c} &                                                     &+&\frac{N_{\thp,\pss}}{N_\pss N_c}C_{\thp,\pss}\,,
    \label{eq:c3bkgd}
\end{array}
\end{equation}
where the first terms in both lines are the charge-independent backgrounds, such as momentum conservation, which will be largely canceled out when taking $\pos-\pss$.
The shorthand notations stand for
\begin{equation} \label{eq:c-shorthand}
\begin{array}{lcll}
C_{\twp,\pos}&=&\mean{\cos(\phi_{\alpha}+\phi_{\beta}-2\phi_{\twp})}_{\twp,\pos}&,\\
C_{\thp,\pos}&=&\mean{\cos(\phi_{\alpha}+\phi_{\beta}-2\phi_{c})}_{\thp,\pos}&,\\
C_{\thp,\pss}&=&\mean{\cos(\phi_{\alpha}+\phi_{\beta}-2\phi_{c})}_{\thp,\pss}&
\end{array}
\end{equation}
where the average $\langle \cdots \rangle_{\twp,\pos}$ runs only over the correlated background pairs with parent cluster azimuth $\phi_\twp$.  The averages $\langle \cdots \rangle_{\thp,\pos}$ and $\langle \cdots \rangle_{\thp,\pss}$ run over only the correlated background triplets, whose multiplicities are $N_{\thp,\pos}$ and $N_{\thp,\pss}$, respectively. In other words, these quantities characterize the angular properties of the correlated clusters and are not diluted by combinatorial multiplicities. 
In Eq.~(\ref{eq:c3bkgd}), the set of \pos\ pairs consists of two components: the first component is the  correlated 2p \pos\ pairs,  while the second component comprises the remaining \pos\ pairs that are identical to the \pss\ pairs (in terms of the $\gamma$ quantity). The number of  correlated 2p pairs is denoted by $N_\twp=N_\pos-N_\pss$. 
%
%We note that the details of 2p correlation sources can be complicated (e.g., they can also come from multi-particle decays of clusters). However, our formulations in Eqs.~\ref{eq:c3bkgd} and \ref{eq:c-shorthand} do not rely on those details, but refer only to the overall difference of those correlations in \pos-\pss\ that matters in CME analysis. 
The correlated 3p triplets also contribute nonflow background to $\Delta\gamma$ and are treated as a separate term in Eq.~(\ref{eq:c3bkgd}). 
It is worth noting that the $v_2$'s in Eq.~(\ref{eq:c3bkgd}) are the true elliptic flows, as they arise from the correlations between the common global symmetry of the correlated 2p source ($v_{2,\twp}$) and the particle $c$ ($v_{2,c}$), giving rise to the contribution to $C_3$.
%$C_{\twp,\pos}=\mean{\cos(\phi_{\alpha}+\phi_{\beta}-2\phi_{\twp})}$
In this study, $v_{2,c} = v_{2}$ because the $c$ particles use the same cut as \poi.

The $\gamma$ correlators are calculated from $C_3$ by Eq.~(\ref{eq:g:c3}).
The backgrounds in $\Delta\gamma / v_{2}^{*}$ can be expressed using Eq.~(\ref{eq:c3bkgd}) as
\begin{equation} \label{eq:dgbkgd}
    \frac{\dg_{\rm bkgd}}{v_2^*}=\frac{C_\twp}{N}\cdot\frac{v_2^2}{v_2^{*2}}+\frac{C_\thp}{N}\cdot\frac{1}{N_cv_2^{*2}}.
    %=C_\twp\frac{1+\frac{C_\thp/C_\twp}{Nv_2^2}}{1+\epsilon_{\rm nf}}\,.
\end{equation}
The first term comes from the two-particle (2p) nonflow backgrounds (like resonance decay daughter pairs, and \pos\ pairs from intra-jet correlation)
\begin{equation} \label{eq:c2p}
    \frac{C_{\twp}}{N} = \frac{N_{\twp}}{N_\pos} \left(C_{\twp,\pos}\frac{v_{2,\twp}}{v_{2}}- \frac{\gamma_{\pss}}{v_{2}} \right) \,,
\end{equation}
where $v_{2,\twp}$ is the elliptic flow parameter of the correlated pairs (like resonance decays).
%We note that besides those major sources, there could be other resonances decaying into \pss\ pairs (e.g., $\Delta^{++} \rightarrow p^{+} \pi^{+}$), which similarly obey the multiplicity and flow scaling and effectively is just a reduce in the $C_{\twp,\pos}$ value. 
%It is the total effect of all resonance decays in the \pos-\pss\ observable that matters, and that is extracted in the data analysis. 
%Since opposite-sign pions are the dominant contribution, we use the language of ``$\rho$'' in a few places in the paper as an illustrative example (Sec.~\ref{sec:r}). 
The second term arises from the three-particle (3p) nonflow backgrounds, such as jets
\begin{equation} \label{eq:c3p}
    \frac{C_\thp}{N}= \frac{N_{\thp,\pos}}{N_\pos}C_{\thp,\pos}-\frac{N_{\thp,\pss}}{N_\pss}C_{\thp,\pss} \,,
\end{equation}
where the $N$ is the multiplicity of \poi, and $N_c$ is that of particle $c$ (in this analysis $N=N_c$). 
%Eqs.~(\ref{eq:c2p}) and (\ref{eq:c3p}) may appear asymmetric, but they are just different ways to express the same thing, which is the differences between \pos\ and \pss. 
Both the Eqs.~(\ref{eq:c2p}) and (\ref{eq:c3p}) have their corresponding \pss\ contributions subtracted from their \pos\ components, as how $\Delta\gamma$ is defined. 

With this decomposition, the isobar ratio due to these background sources can be calculated for $\Delta\gamma / v_{2}^{*}$. 
After approximation to the leading order, the expression becomes
\begin{widetext}
\begin{equation} \label{eq:Y}
    Y_{\bkgd}\equiv\frac{\left(\dg_{\rm bkgd}/v_2^*\right)^{\Ru}}{\left(\dg_{\rm bkgd}/v_2^*\right)^{\Zr}}
    \approx 1+\frac{\delta (C_\twp/N)}{C_\twp/N}-\frac{\delta\enf}{1+\enf} \\ 
    +\frac{1}{1+\frac{Nv_2^2}{C_\thp/C_\twp}}\left(\frac{\delta C_\thp}{C_\thp}-\frac{\delta C_\twp}{C_\twp}-\frac{\delta N}{N}-\frac{\delta v_2^2}{v_2^2}\right)\,,
\end{equation}
\end{widetext}
where $\delta X\equiv X^{\Ru}-X^{\Zr}$ for any $X = C_{\thp}$, $C_{\twp}$, etc., while all other quantities without ``$\delta$'' refer to those in Zr+Zr.

%We note that global spin alignment of $\rho$ mesons can present an additional background to the CME~\cite{Shen:2022gtl}. Effect of such a background on isobar measurements needs to be assessed in future works. 
We note that the details of 2p and 3p correlation sources can be complicated. For example, besides 2- and 3-body decays of resonances, they can also come from multi-particle decays of clusters or jets. The \pos\ and \pss\ background pairs, besides the extra \pos\ pairs from decays of neutral objects, may still not be strictly symmetric (e.g., decays from the $\Delta$ resonances). In addition, the decay kinematics themselves can be altered in heavy-ion collisions, such as by the possible global spin alignment of $\rho$ mesons~\cite{Shen:2022gtl}. Our analysis formalisms, however, do not rely on those details, but only on the overall \pos-\pss\ difference in the $\gamma$ correlators and the overall nonflow contribution to $v_{2}^{*}$. 

%------------------------------------------------------------------------------------------------%

\section{Analysis} \label{sec:analysis}

In order to estimate the background baseline, we need to analyze the isobar data to assess the quantities required by Eq.~(\ref{eq:Y}). 
Since the purpose is to estimate the background in the measurements of the STAR isobar blind analysis, we use the same datasets, follow the same event and track selections, and apply identical analysis cuts as described in Ref.~\cite{STAR:2021mii}. 
% In our nonflow assessments, we use the specific cuts and analysis method corresponding to each measurement.

The majority of the analysis cuts are the same for all $\dg/v_2$ measurements in the blind analysis. 
These are as follows: On the event level, the minimum-bias trigger is used--correlated hits in both Vertex Position Detectors (VPD) within a time window. 
%The first set of these cuts is the trigger selection for minimum-bias events. 
The primary vertex reconstructed by the Time Projection Chamber (TPC)~\cite{STAR:1997sav, Anderson:2003ur} is also required to have a longitudinal $z$ position between $-35$~cm and 25~cm ($-35 < V_{z} < 25$ cm) and a transverse position to be within 2 cm ($V_{\perp} < 2$ cm) with respect to the center of TPC. In addition, the interaction position measured online by the Vertex Position Detector (VPD)~\cite{Llope:2003ti} is required to be within 5~cm from the reconstructed primary vertex along the beam ($|V_{z} - V_{z}^{\textsc{vpd}}|<5$~cm).
On the track level, the reconstructed particle tracks must have more than 15 space points measured in the TPC.
As the CME is a signal in primordial particles, the track's distance of closest approach to the primary vertex (DCA) is required to be less than 3 cm ($\text{DCA}<3$ cm). 
The particle transverse momentum is restricted within $0.2<\pt<2$~\gevc.
The pseudorapidity range is limited within $|\eta|<1$ for the full-event analyses. 
For the subevent analyses, where the event is divided into 
two subevents based on their $\eta$ ranges, the measurements in the STAR blind analysis used slightly different ranges. However, for the estimates reported in this paper, all subevent measurements used the same $\eta$ ranges; specifically,  $-1<\eta<-0.1$ was considered as the East subevent, and $0.1<\eta<1$ was considered as the West subevent. 
The centrality is defined by the number of tracks in the rapidity range $-0.5<\eta<0.5$. 

In the STAR blind analysis of the isobar data ~\cite{STAR:2021mii}, a total of seven $\Delta\gamma / v_{2}$ measurements were performed by four research Groups. Among these measurements, four utilized the two-particle cumulant method for the $v_2$ measurement and the three-particle cumulant method for the $\dg$ measurement. The remaining three measurements used the event-plane method. While both the cumulant and event-plane methods produced similar results, the event-plane method posed significantly more challenges in distinguishing and accounting for nonflow contributions compared to the cumulant method, as aforementioned. 
%but the nonflow effects are harder to assess. 
For this reason, we concentrate on the former four measurements, namely, the full-event measurements from Group-2 and Group-3, and the subevent measurements from Group-2 and Group-4.
The analysis details of the four measurements differ slightly, and those details are tabulated in the first block of Table~\ref{tab} for the corresponding measurements.
In our background baseline estimates for the four measurements, we maintain consistency by using the identical analysis cuts for each corresponding measurement.
The block in Table~\ref{tab} also includes the results for the average over the 20-50\% centrality range for the isobar ratios of $\dg/v_2$, denoted as $Y \equiv \mean{\frac{(\dg/v_2)^\Ru}{(\dg/v_2)^\Zr}}$, which were obtained from the STAR blind analysis~\cite{STAR:2021mii}. 
%To calculate the  average, each term was individually weighted using the inverse statistical uncertainty squared~\cite{STAR:2021mii}. 
The average of each term in Eq.~\ref{eq:Y} over centrality bins is calculated separately using the inverse statistical uncertainty squared as weight, and then summed together to get the background baseline. 
Equation~(\ref{eq:Y}) suggests categorizing the nonflow contributions to the background into three ingredients: (1) $\delta(C_{\twp}/N)/(C_{\twp}/N)$ which characterizes the relative difference of flowing clusters between the two isobars, (2) Differences that arise from using $v_{2}^{*}$ rather than true flow in the calculation of $\Delta\gamma$, characterized by $\enf$, and (3) differences in the relative amounts (or character) of three particle clusters between the isobars.  In the next section we will discuss each of these three in turn.

%------------------------------------------------------------------------------------------------%

\subsection{Pair versus single multiplicity difference} \label{sec:r}

\begin{figure*}
	\includegraphics[width=0.45\linewidth]{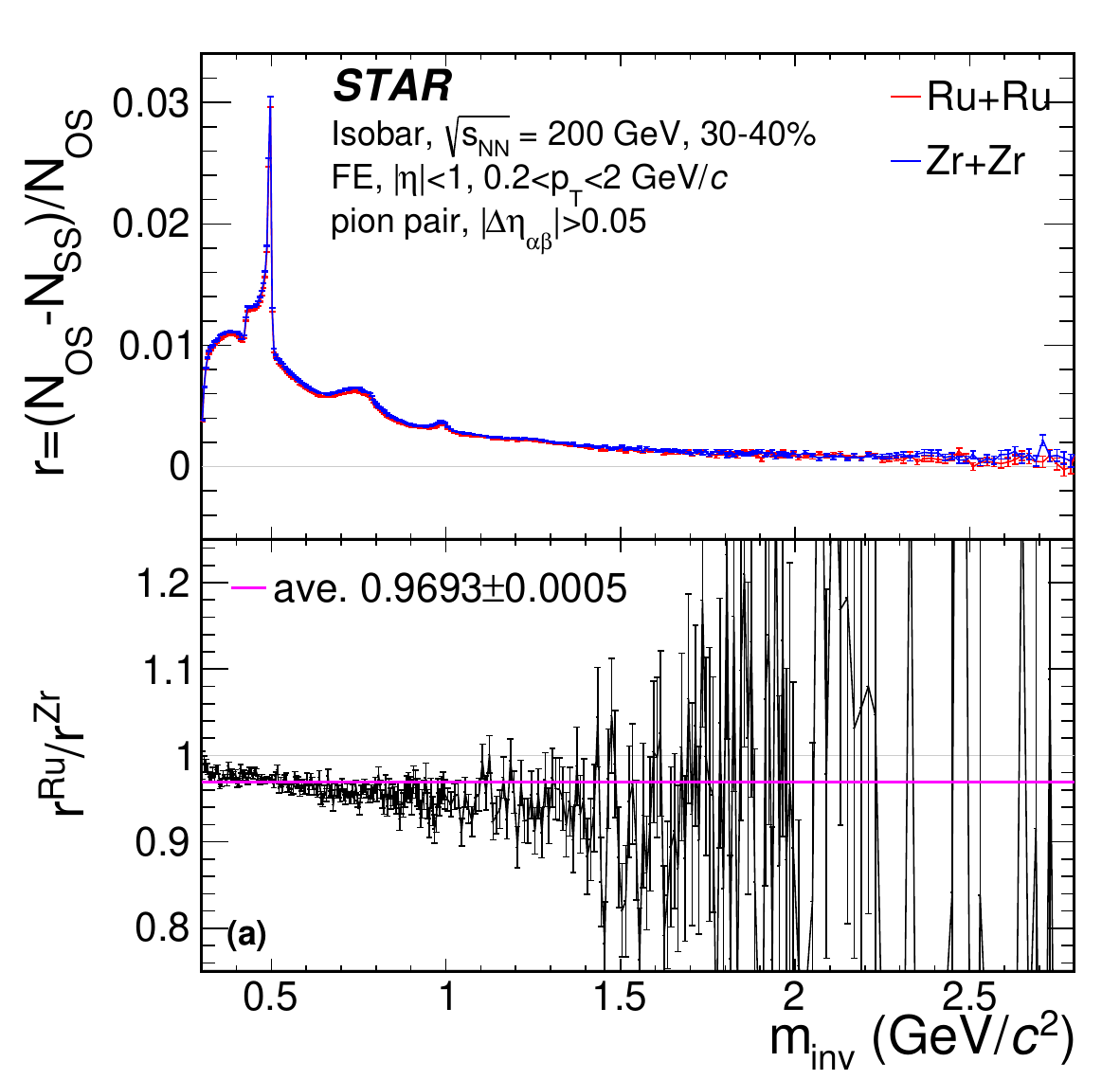}
	\includegraphics[width=0.45\linewidth]{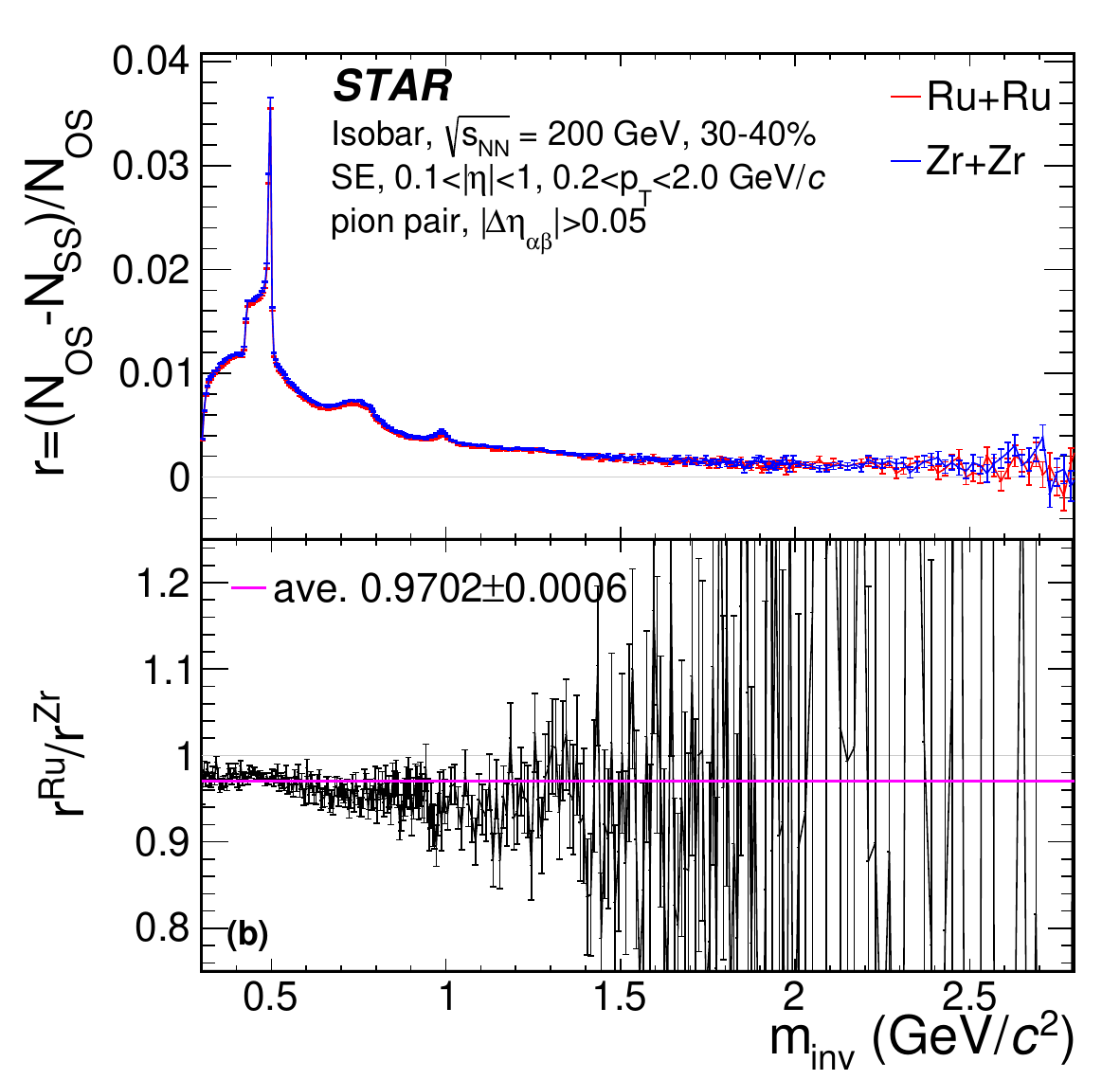}
	\caption{The relative excess of \pos\ over \pss\ pion pair multiplicity, $r=(N_{\pos}-N_{\pss})/N_{\pos}$, as a function of the pair invariant mass ($\minv$) for the 30-40\% centrality in Ru+Ru and Zr+Zr collisions (upper panels). The lower panels show the ratio of $r$ in Ru+Ru to that in Zr+Zr collisions. Full-event results are shown on the left, corresponding to the Group-3 analysis cuts. Subevent results are shown on the right, corresponding to the Group-2 analysis cuts. Only statistical uncertainties are shown. 
	%To identify a track to be pion, TPC and TOF hits are required to match for this track, specific energy loss in the TPC differs from pion expectation within 3 standard deviations, TOF-measured mass $m^{2}<0.1 \text{ GeV}^{2}/c^{4}$, and transverse momentum $0.2 < \pt < 1.8 \text{ GeV}/c$. 
	The pion pair is required to have an $\eta$ gap $\Delta\eta_{\alpha\beta}>0.05$. For subevent method, the pion pair needs to come from the same subevent.}
	\label{fig:rminv}
\end{figure*}

\begin{figure*}
	\includegraphics[width=0.45\linewidth]{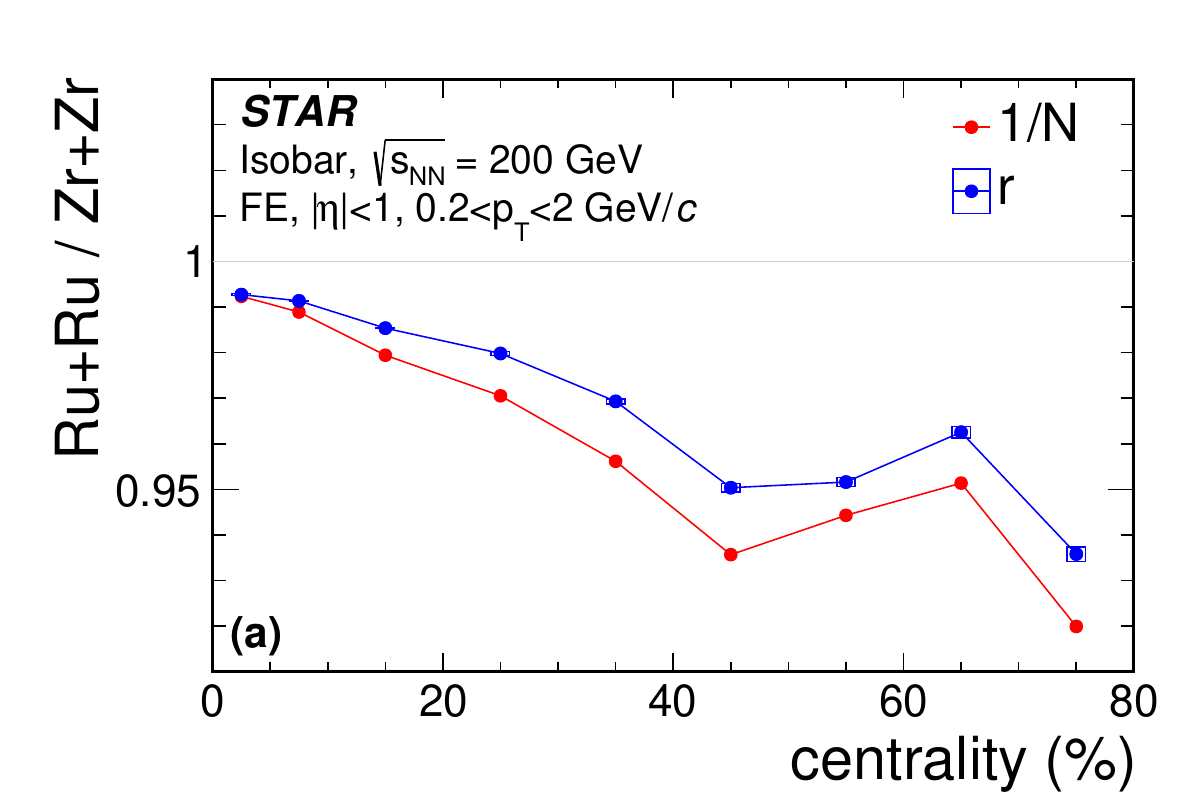}
	\includegraphics[width=0.45\linewidth]{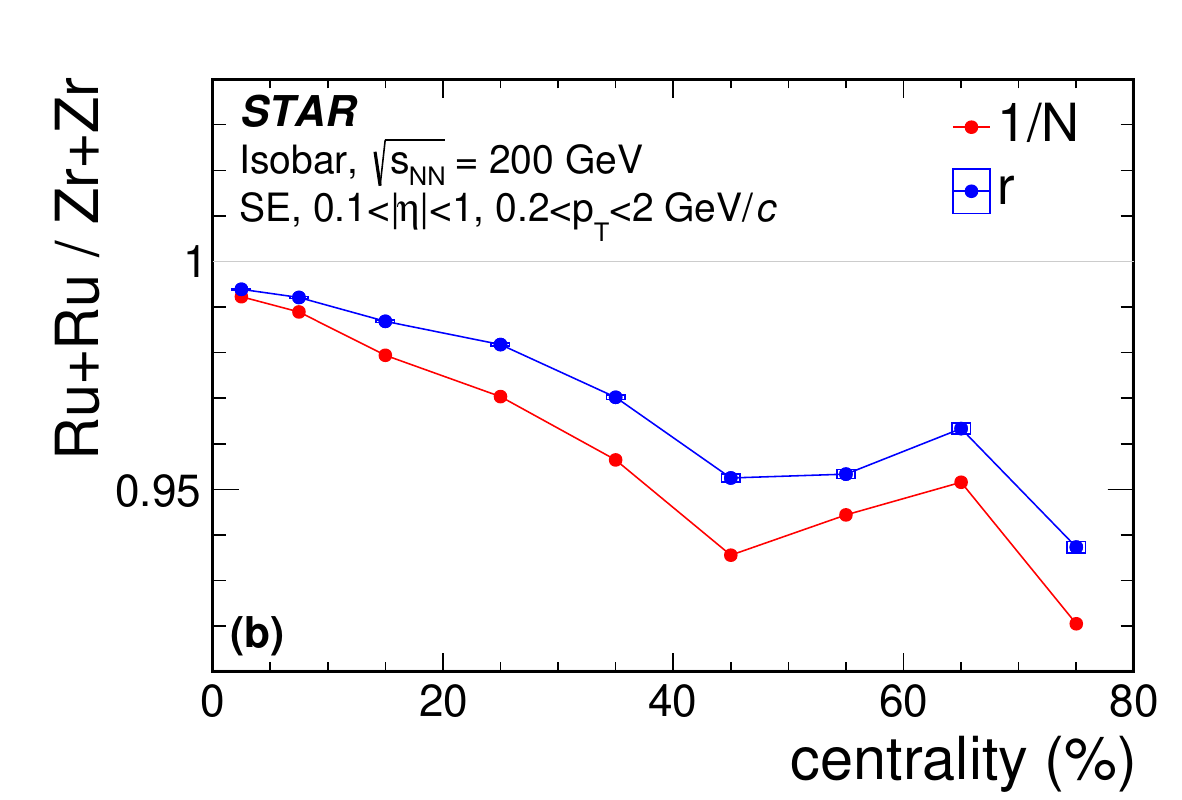}
	\caption{The Ru+Ru to Zr+Zr ratio of $r=(N_{\pos}-N_{\pss})/N_{\pos}$ and that of the efficiency-corrected inverse multiplicity $1/N$ as functions of centrality for full-event (left panel, with Group-3 cuts) and subevent (right panel, with Group-2 cuts) methods. %The lower panels show the difference between the $r$ ratio and the $1/N$ ratio, which is one of the contributions to the $Y$ baseline deviation from unity. 
    Vertical bars are statistical uncertainties; hollow boxes on $r$ data points are  systematic uncertainties. 
	%Here, only $r$ curves have systematic uncertainty, but too small to be visible.
    }
	\label{fig:rn}
\end{figure*}

The main background in a $\Delta\gamma$ measurement comes from the charge-independent correlated pairs, such as resonance decays and intra-jet correlations.
The number of those pairs is denoted as $N_{\twp} = N_{\pos} - N_{\pss}$, the excess of \pos\ pairs over \pss, so the charge-independent part is removed (similar to $\Delta\gamma = \gamma_{\pos} - \gamma_{\pss}$). 
The relative excess is 
\begin{equation} \label{eq:r}
    r = \frac{N_{\twp}}{N_{\pos}} = \frac{N_{\pos}-N_{\pss}}{N_{\pos}} \,.
\end{equation}
This is reflected in the background contribution to $\dg$ of Eq.~(\ref{eq:Y}).

The $C_\twp$ in Eq.~(\ref{eq:Y}) refers to the background contribution, and the most relevant quantity is $\delta C_\twp$. 
The ZDC (Zero Degree Calorimeter)~\cite{Adler:2003sp} measurement of $N\dg/v_2$ is close to the background $C_\twp$ because the possible CME signal is small and ZDC has a large $\eta$-gap with TPC (Sec.~\ref{sec:3p}). However, the difference in CME signal contributions in $\delta C_\twp$ between the isobar systems may not be negligible compared to that in the background contributions, which is what we need for the baseline estimation. Therefore, we cannot simply take the difference between the ZDC measurements in Ru+Ru and Zr+Zr collisions. 
Even without this complication, %the statistical uncertainties are too large in those ZDC $C_{\twp}$ measurements, because of the low resolution of ZDC event plane, for the precision we need to achieve in the background baseline estimate. 
due to the low resolution of the ZDC event plane, the statistical uncertainties of the ZDC $C_{\twp}$ measurements are too large for the precision needed to achieve the background baseline estimate. 
%Then their isobar ratio could be even worse in statistical fluctuations, so we need to find an alternative way to estimate it. 
Therefore, we go to the $C_\twp/N$ term itself as defined in Eq.~(\ref{eq:c2p}). 
%The general idea is to assume the quantity in the parentheses of Eq.~(\ref{eq:c2p}), i.e., $\left(C_{\twp,\pos}\frac{v_{2,\twp}}{v_2}-\frac{\gamma_\pss}{v_2}\right)$, is the same for Ru+Ru and Zr+Zr, then 
The $C_{\twp,\pos}$ represents average angular correlations per 2p cluster (determined by decay kinematics in the case of a resonance decay), and is therefore insensitive to the collision species. The $v_{2}$'s should well scale between various particle/cluster types. The $\gamma_{\pos}/v_{2}$ is relatively insignificant compared to $C_{\twp,\pos}$. It is therefore reasonably safe to assume that the quantity in the parentheses of Eq.~(\ref{eq:c2p}), i.e., $\left(C_{\twp,\pos}\frac{v_{2,\twp}}{v_2}-\frac{\gamma_\pss}{v_2}\right)$, is the same for Ru+Ru and Zr+Zr. Thus, we have
% \begin{equation} \label{eq:dc2p}
% \begin{split}
%     C_{\twp} \propto& N N_{\twp} / N_{\pos} = N r\,, \\
%     \frac{\delta C_{\twp}}{C_{\twp}} \equiv& \frac{ C_{\twp}^{\text{Ru+Ru}} - C_{\twp}^{\text{Zr+Zr}} }{C_{\twp}^{{\text{Zr+Zr}}}}
%     = \frac{(Nr)^{\text{Ru+Ru}} }{(Nr)^{\text{Zr+Zr}}} - 1 \\
%     \approx& \frac{\delta N}{N} + \frac{\delta r}{r}\,,
% \end{split}
% \end{equation}
\begin{equation} \label{eq:dc2p}
    \frac{\delta(C_{\twp}/N)}{C_{\twp}/N} \approx \frac{\delta r}{r}\,.
\end{equation}

In this analysis, we only use pions to calculate $r$, because the main source of this background is the $\rho$ meson decay and the baryon stopping effect does not affect the produced pions.
To identify a track as a pion, the TPC reconstructed track is required to match with a TOF (Time Of Flight detector)~\cite{Llope:2003ti} hit, and additional selections are the TPC energy loss deviation $n \sigma_{\pi} < 3$, TOF mass $m^{2}<0.1 \text{ GeV}^{2}/c^{4}$, and transverse momentum $0.2 < \pt < 1.8 \text{ GeV}/c$.
To study the invariant pair mass ($\minv$) dependence, $r$ is calculated as a function of $\minv$ (Fig.~\ref{fig:rminv}) for Ru+Ru and Zr+Zr separately, and then the isobar ratio is taken between the two. A constant fit is used to extract the average of the ratio. 
By applying the same procedure to each centrality bin, we can get their centrality dependence as shown in Fig.~\ref{fig:rn}. 

The quantity in the parentheses of Eq.~(\ref{eq:c2p}), assumed to be equal between the two isobar systems, may have a $\minv$ dependence. $Y_{\bkgd}$ therefore can be described by the average $\delta r /r$ weighted by a $\minv$-dependent function. We estimate the systematic uncertainty by the variation in $r$ obtained by changing the fit range to $\minv < 1 \text{ GeV}/c^{2}$.
%Considering the invariant mass dependence, we take a systematic variation by requiring $\minv < 1 \text{ GeV}/c^{2}$ for this fit. 
%This variation eventually contributes to the systematics of the final result $Y_{\bkgd}$, which is listed with other main sources of systematic uncertainties in Table~\ref{tab:sys}. 
%The difference between this variation and the default is one-sided but expanded to be symmetric in the calculation of systematics, which is listed with other main sources of systematic uncertainties in Table~\ref{tab:sys}. 
The resultant difference from the default is expanded to be symmetric and assigned as part of the systematic uncertainty. This is listed in Table~\ref{tab:sys} together with other main sources of systematic uncertainties.

The isobar ratio of $1/N$ is also plotted in Fig.~\ref{fig:rn}, which is different from the isobar ratio of $r$ as mentioned in Sec.~\ref{sec:introduction} and observed in Ref.~\cite{STAR:2021mii}. 
This difference arises from the fact that the background source multiplicity (such as the $\rho$ mesons) does not scale identically with multiplicity for Ru+Ru and Zr+Zr collisions. As a result, the background does not strictly scale as the inverse multiplicity, and thus $Y_{\bkgd}$ deviates from unity. The amount of deviation is the difference between the two curves in each panel of Fig.~\ref{fig:rn}.

\begin{table*}[]
\caption{The main sources of systematic uncertainties in $Y_{\bkgd}$. Those contributions from one-sided variations are expanded to be symmetric in the calculation of systematic uncertainties. %The comparison of Group-3 $v_{2}^{*}$ with Ref.~\cite{STAR:2021mii} is included in calculation but not listed below, since it is a minor effect.
} %The total systematic uncertainty is the quadrature sum of the individual ones, e.g., $s = \sqrt{\sum_{i}s_{i}^{2}}$.}
\label{tab:sys}
\begin{tabular}{c|c|cc|cc}
\hline
 $Y_{\bkgd}$ syst.~source/variation & affected quantity & Group-2 FE & Group-3 FE & Group-2 SE & Group-4 SE \\ \hline
 $m_{\text{inv}}<1$ GeV/$c^{2}$ fit range & $r$ & $\pm0.00061$ & $\pm0.00061$ & $\pm0.00055$ & $\pm0.00052$ \\
 flow decorrelation $\pm3\%$ & $v_{2}$, $\enf$ & $\pm0.00040$ & $\pm0.00040$ & $\pm0.00030$ & $\pm0.00035$ \\
 %flow decorr. $+3\%$ & $v_{2}$, $\enf$ & 0.00028 & 0.00028 & 0.00021 & 0.00024 \\
 %flow decorr. $-3\%$ & $v_{2}$, $\enf$ & 0.00029 & 0.00029  & 0.00022 & 0.00025 \\
 alternative 2D fits & $v_{2}$, $\enf$ & $\pm0.00016$ & $\pm0.00107$ & $\pm0.00058$ & $\pm0.00065$ \\
% check with Ref.~\cite{STAR:2021mii} & $v_{2}^{*}$, $\enf$ & 0.00034 & 0.00000 & 0.00006 & 0.00005 \\
 possible $5\%$ CME in $C_{\twp}$ & $C_{\twp}$ & $\pm0.00070$ & $\pm0.00070$ & $\pm0.00053$ & $\pm0.00060$ \\
 HIJING jet-quenching off & $C_{\thp}$ & $\pm0.00064$ & $\pm0.00064$ & $\pm0.00051$ & $\pm0.00140$ \\
 %syst.~in PRC105(2022)014901~\cite{STAR:2021mii} & $Y$ & 0.00070 & 0.00100 & 0.00160 & 0.00029 \\
% \hline total (check Table~\ref{tab}) & -- & 0.0013 & 0.0016 & 0.0011 & 0.0018 \\ 
\hline
\end{tabular}
\end{table*}

%------------------------------------------------------------------------------------------------%

\subsection{Nonflow contamination in $v_2^*$ measurements} %{$\enf$}

\begin{figure*}
	\includegraphics[width=0.325\linewidth]{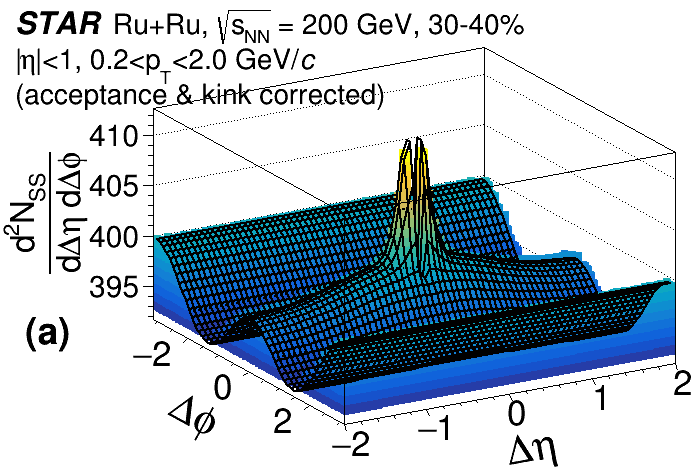}
	\includegraphics[width=0.325\linewidth]{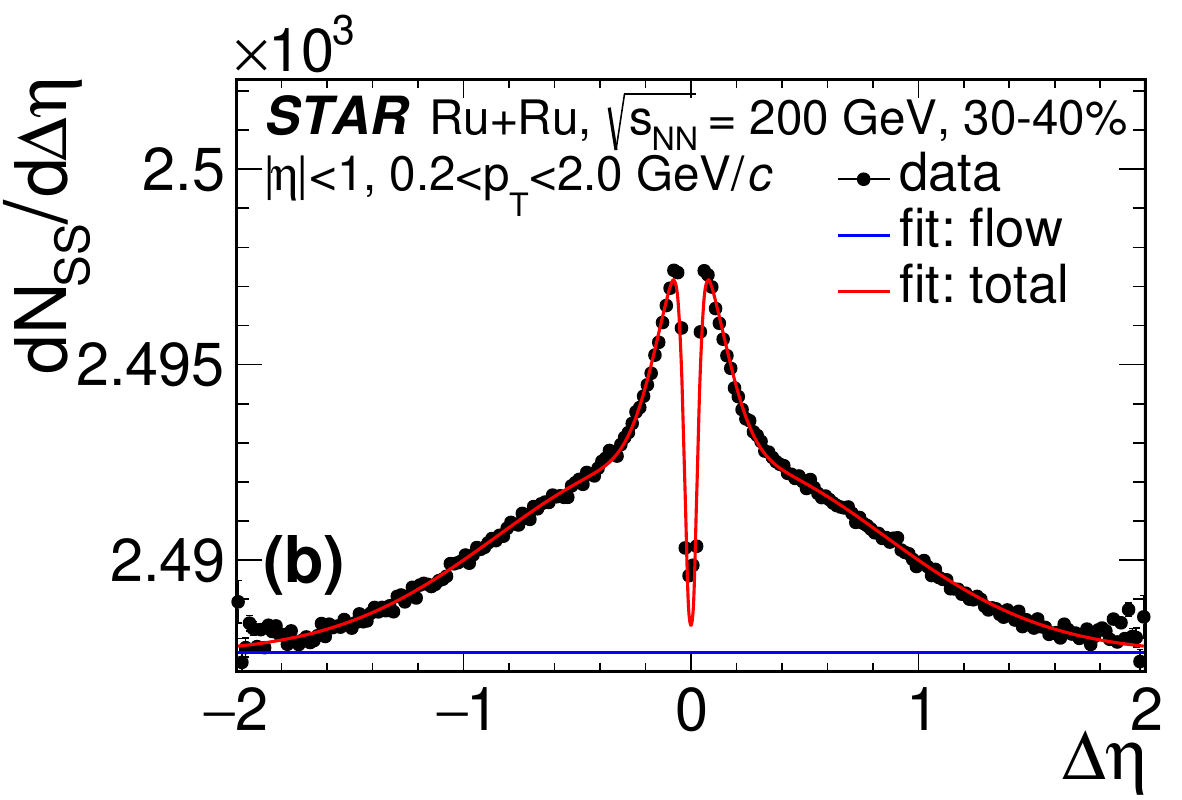}
	\includegraphics[width=0.325\linewidth]{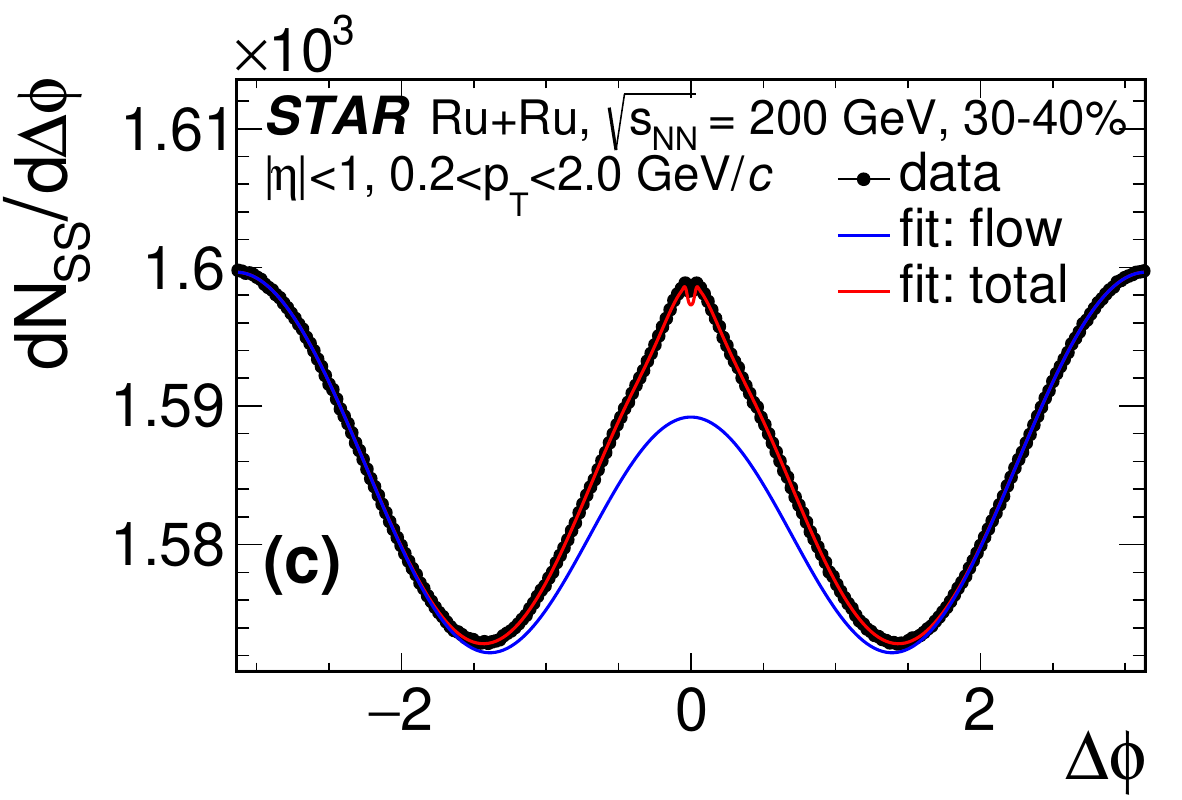}
	\includegraphics[width=0.325\linewidth]{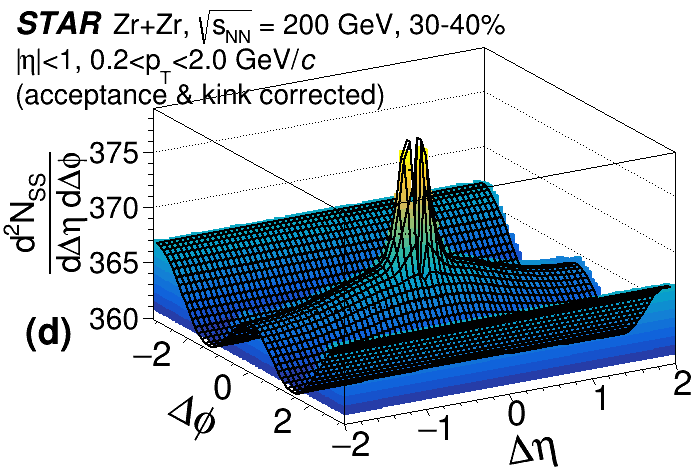}
	\includegraphics[width=0.325\linewidth]{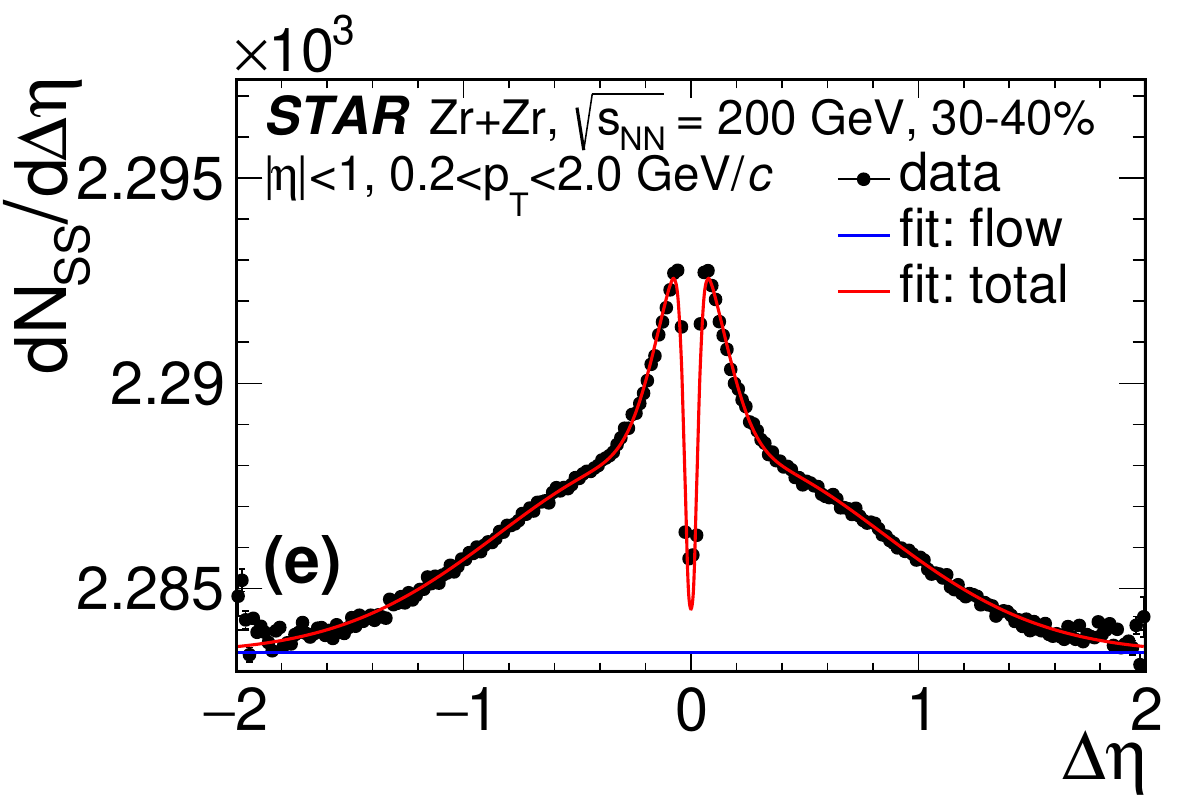}
	\includegraphics[width=0.325\linewidth]{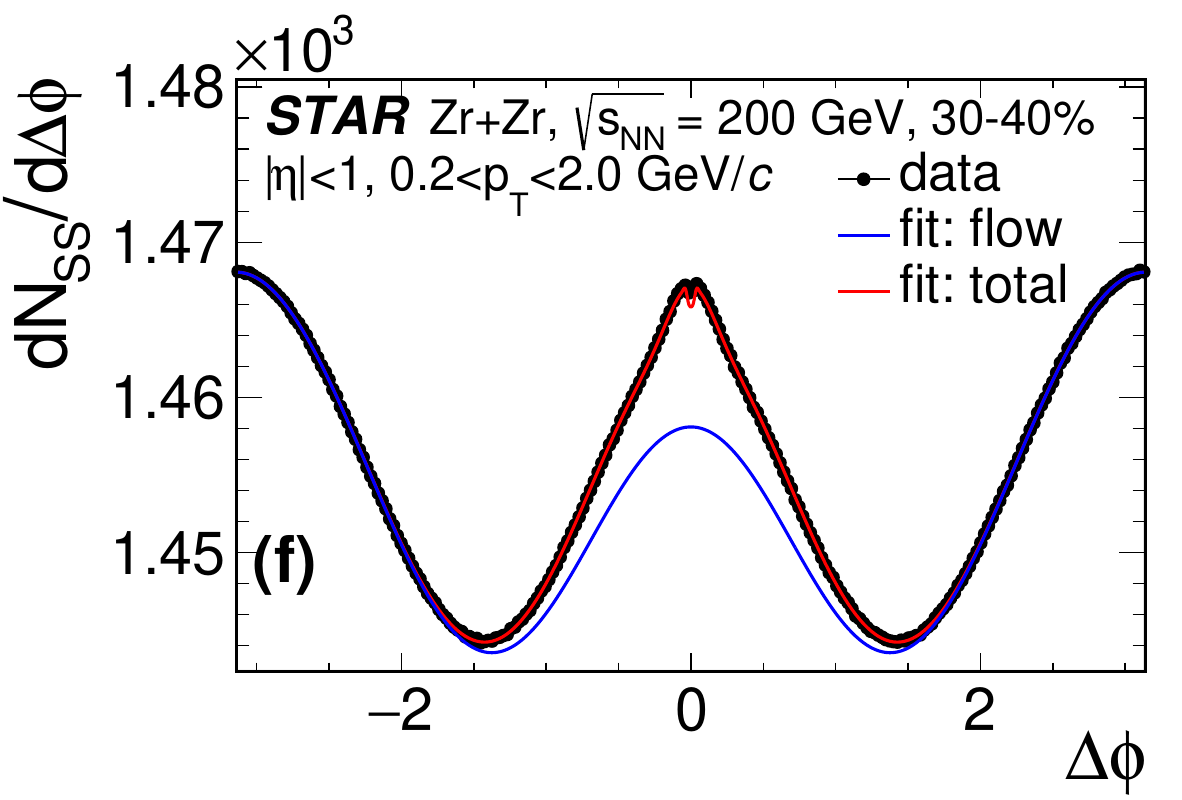}
	\caption{The left plots show the $(\Delta\eta, \Delta\phi)$ 2D distributions (colored) for \pss\ pairs after acceptance and kink corrections with fit result (black mesh). The middle and right plots are the projections of data (black markers) and fit results (red for total fit function, blue for flow component) to $\Delta\eta$, $\Delta\phi$ directions respectively. The upper row is for Ru+Ru and lower row for Zr+Zr. Only the full-event method and \pss\ pairs are shown here. The fitted flow is assumed to be the same between \pos\ and \pss\ pairs, and the same between full-event and subevent methods. The centrality range $30$--$40\%$ is used here, and other centrality ranges are similar.}
	\label{fig:fitproj}
\end{figure*}

\begin{figure*}
	\includegraphics[width=0.45\linewidth]{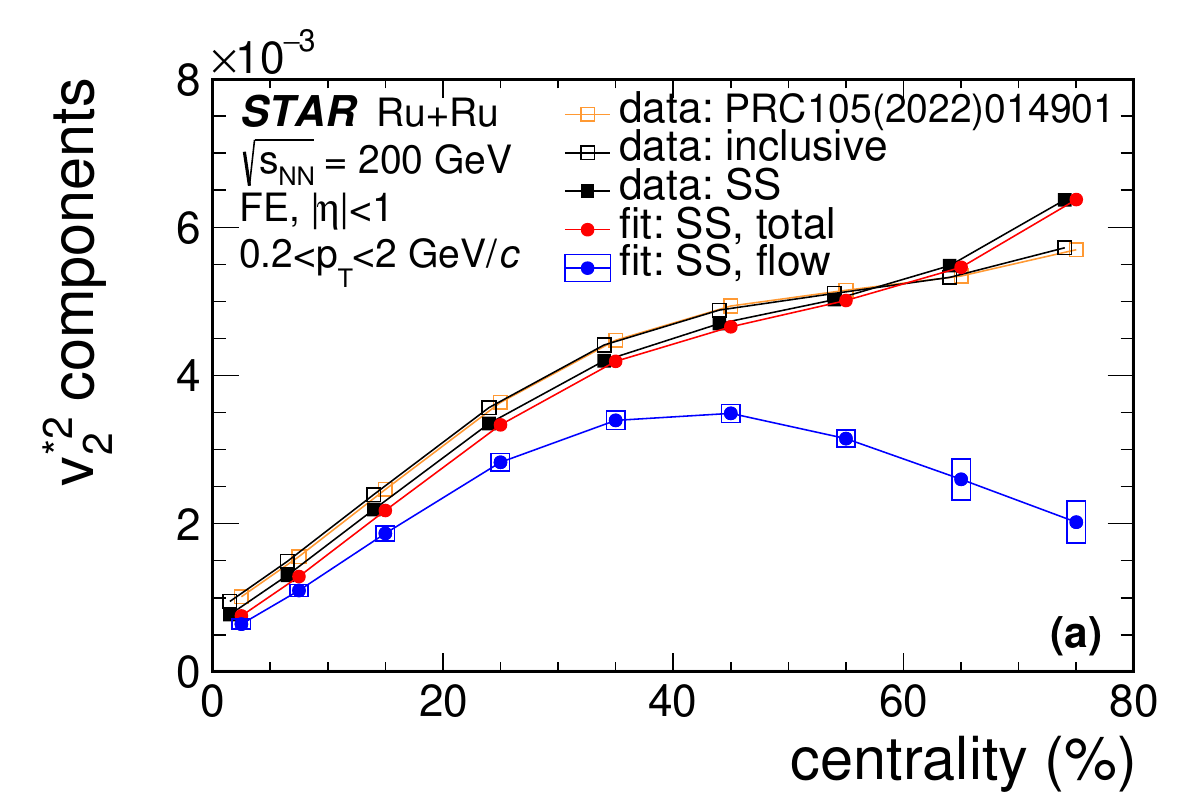}
	\includegraphics[width=0.45\linewidth]{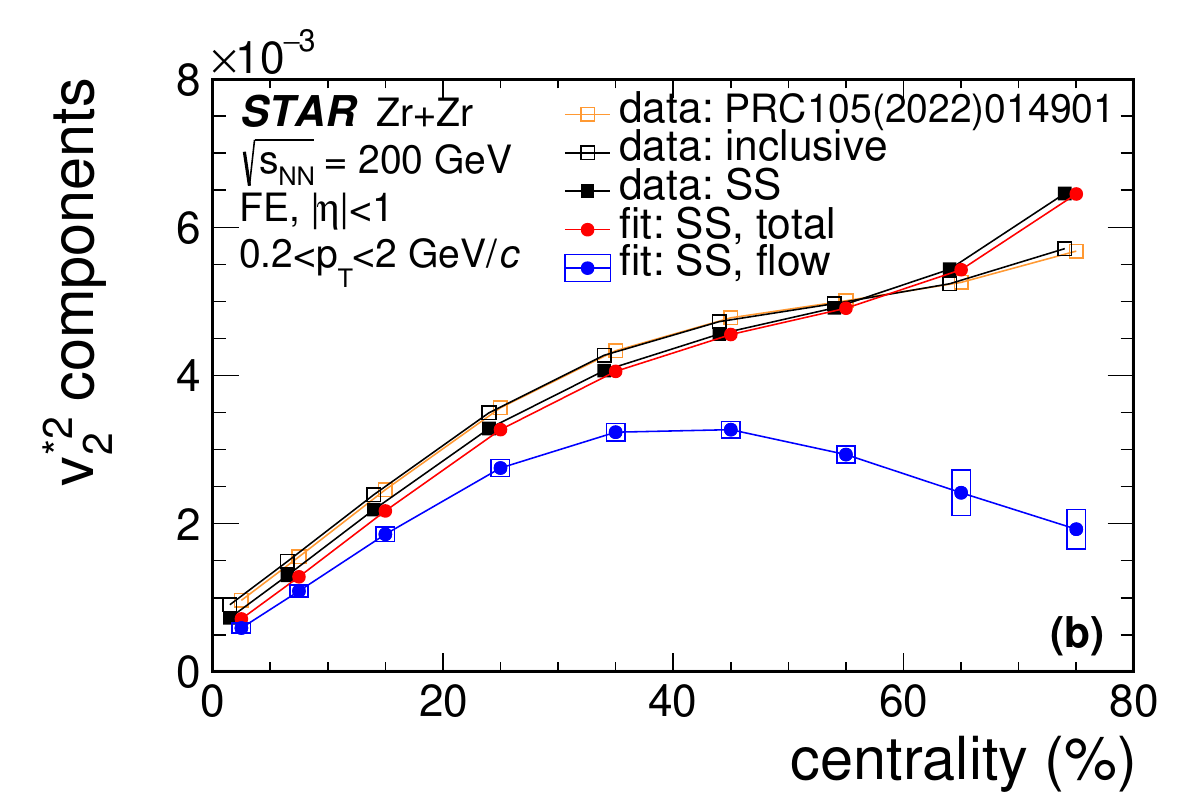}
    %\includegraphics[width=0.325\linewidth]{PduRtFulFlowV2}
	%\caption{${v_{2}^{*}}^{2}$ components, Ru+Ru (left) and Zr+Zr (middle), and the isobar ratio of true flow (right)}
    \caption{
    %${v_{2}^{*}}^{2}$ components, Ru+Ru (left) and Zr+Zr (right)
    The ${v_{2}^{*}}^{2}$ components as functions of centrality for Ru+Ru (left panel) and Zr+Zr (right panel). The blue curves are the fitted $v_{2}^{2}$. The black and red solid markers are the inclusive ${v_{2}^{*}}^{2}$ for \pss\ pairs from data and fit function respectively, and they agree with each other. The black and orange open squares are the inclusive ${v_{2}^{*}}^{2}$ for all pairs from this measurement and the cited Ref.~\cite{STAR:2021mii} Group-3 (see the plot legends), and they also agree with each other. Vertical bars and hollow boxes show the statistical and systematic uncertainties, respectively.
    }
	\label{fig:v2}
\end{figure*}

The two-particle correlation is usually used to calculate the elliptic flow in the TPC. However, this method cannot avoid nonflow backgrounds that also correlate with those two particles. To separate the true flow from this inclusive measurement, a data-driven approach is used in this analysis. We fit the 2D two-particle $(\Delta\eta, \Delta\phi)$ distribution, where $\Delta\eta$ is the pseudorapidity difference between the two particles, and $\Delta\phi$ azimuth difference. Since the global anisotropy, flow, is supposed to be charge-independent, the \pos\ and \pss\ pairs should have the same true flow. We only focus on the \pss\ pairs to avoid the charge-dependent backgrounds that are stronger than charge-independent ones. We assume the true flow is $\eta$-independent at $|\eta|<1$, so we only consider the full event method in this section, and use the same fitted flow value to treat the subevent method. To avoid confusion, we will use ``fitted flow'' or ``fitted $v_{2}$'' to refer to our estimate, which should closely reflect the underlying true flow.

%The leading order of the single particle $\eta$ distribution is uniform in the range $|\eta|<1$.
In the range, $|\eta| < 1$, the single particle $\eta$ distribution is roughly uniform.
Therefore, the $\Delta\eta$ projection of the 2D distribution is mainly a triangle due to the finite $\eta$ range. This acceptance effect also happens in mixed events. Each pair has one particle from the current event and the other from another similar event (in the same centrality bin, same $V_{z}$ bin of width 1 cm).  However, all other correlations do not exist in those mixed events. Thus, we use the 2D distribution from mixed events, with its peak at $\Delta\eta=0$ scaled to one, to correct the acceptance effect in real events by taking the ratio of the real over mixed. We assume that the fitted flow does not have a $\Delta\eta$ dependence, so this operation does not affect the fitted flow measurement. 

After this acceptance correction, the $\Delta\eta$ distribution generally appears flat, revealing the fine structures of nonflow (shown in Figs.~\ref{fig:fitproj}a, \ref{fig:fitproj}d). For each centrality bin, we can get such a corrected 2D distribution, where we see some nonphysical kinks at $|\Delta\eta| = 1 \pm 0.5$. The centrality of this STAR dataset is defined by the charged particle multiplicity within $|\eta|<0.5$~\cite{STAR:2021mii, STAR:2008med}. On the other hand, the \poi\ in this analysis are within $|\eta|<1$.  This distinction means that the centrality bin implicitly imposes an additional constraint on the particle number within $|\eta|<0.5$ but not for $0.5<|\eta|<1$. Consequently, these differences cause those artificial kinks. To eliminate this effect, we project the corrected 2D distribution to $\Delta\eta$ from the away side range of $0.8\pi < |\Delta\phi| < \pi$, where the fine structure of nonflow is small and can be ignored. We then subtract its integral from this projection to remove the pedestal, effectively making it a 2D distribution,  independent of $\Delta\phi$. By taking the difference between the acceptance-corrected 2D distribution and this projection, we can eliminate the kinks and any $\phi$-independent detector effects. It is important to note that this operation does not affect the true flow, as the fitted flow is assumed to be independent of $\Delta\eta$.

With all those operations above, the 2D fit can be conducted, and the fit function from observation and tuning is 
\begin{widetext}
\begin{equation} \label{eq:fit}
\begin{split}
    f(\Delta\eta, \Delta\phi) 
    =& A_{1} G\left(\frac{|\Delta\eta|-\mu}{\sigma_{1}} \right) G\left(\frac{\Delta\phi}{\rho_{1}}\right)
    + A_{2} G\left(\frac{\Delta\eta}{\sigma_{2}}\right) G\left(\frac{\Delta\phi}{\rho_{2}}\right) 
    + \left[A_{3} G\left(\frac{\Delta\eta}{\sigma_{3}}\right) + A_{4} G\left(\frac{\Delta\eta}{\sigma_{4}}\right)\right] G\left(\frac{\Delta\phi}{\rho_{3}}\right) \\
    &+ C\big[ 1 + 2 V_{1} \cos(\Delta\phi) + 2 V_{2} \cos(2\Delta\phi) + 2 V_{3} \cos(3\Delta\phi) \big] \,,
\end{split}
\end{equation}
\end{widetext}
where $G(x) = e^{-x^{2}/2}$ is the Gaussian function. 
The second line represents the flow pedestal, and the parameter $V_{n}$ corresponds to the squared fitted $v_{n}$ ($n=1, 2, 3$) assuming the ``true flow'' does not depend on $\eta$. 
The 2D Gaussians are empirical models for nonflow corrections, guided by the data shape (Fig.~\ref{fig:fitproj}a, \ref{fig:fitproj}d): track merging effect and Coulomb effect for SS pairs could result in the dip at $\Delta\eta=0$; HBT, resonance decays, and intra-jet correlations are short-range shown by the narrow peak; inter-jet correlations are long-range characterized by a wide $\Delta\eta$ Gaussian. 
In Eq.~(\ref{eq:fit}), all the parameters ($A$, $C$, $V$, $\mu$, $\sigma$, $\rho$) are free in the fitting, starting from reasonable initial values. % whose initial values are tuned for optimization. 
Figure~\ref{fig:fitproj} shows the 2D fit results (left) along with their $\Delta\eta$ (middle), $\Delta\phi$ (right) projections in the centrality bin 30-40\% for Ru+Ru (upper) and Zr+Zr (lower) separately. 

%We note that the parameterizations of nonflow by Eq.~(\ref{eq:fit}) or similar functional forms have been performed previously by multiple experiments and authors~\cite{ALICE:2016ccg, PHENIX:2021ubk, PHENIX:2022nht, Nagle:2021rep}. 
We note that correlation studies by 2-dimensional $(\Delta\eta, \Delta\phi)$ distributions have been performed previously~\cite{STAR:2006lbt, STAR:2011ryj, ALICE:2018jco, NA61SHINE:2020tbf, STAR:2015kak, STAR:2023wmd, CMS:2013jlh, ATLAS:2014qaj}. 
The analysis procedure is well established and produces consistent nonflow contributions. Other analyses to quantify nonflow contributions have also been carried out, for example, by utilizing reflection symmetry in $\eta$ in symmetric heavy ion collisions~\cite{Abdelwahab:2014sge}, by varying two-particle or subevent $\eta$ gap~\cite{ALICE:2016ccg, PHENIX:2018lia, PHENIX:2021ubk, Nagle:2021rep}, and by extrapolating from proton-proton, proton-nucleus, and peripheral heavy ion collisions to more central collisions assuming inverse multiplicity scaling of nonflow~\cite{STAR:2004amg}.

Figure~\ref{fig:v2} shows the ${v_{2}^{*}}^{2}$ components as functions of centrality for Ru+Ru and Zr+Zr separately. As listed in the legend, this study closely reproduced the previous STAR measurement for ${v_{2}^{*}}^{2}$ (Ref.~\cite{STAR:2021mii}, Group-3) using all pairs (\pos+\pss), ; the difference is negligible and the numerical difference can be attributed to nonidentical datasets dynamically accessed at run time. 
%Since it only has a minor contribution to the calculation of the total systematics, this difference is not listed in Table~\ref{tab:sys}. 
%It is a small effect compared to other sources of systematics. 
%This difference is included in calculation but not listed in Table~\ref{tab:sys}, since it is a minor effect.
%which could come from accidental computer failures, dynamically dataset revisions, and other trivial technical issues.
The fitted flow is extracted from the \pss\ pair correlations. Thus, the plot also includes the ${v_{2}^{*}}^{2}$ measured only from \pss\ pairs, together with the ${v_{2}^{*}}^{2}$ calculated from the total fit function, both of which are found to be consistent with each other. 
Since multiple corrections are applied before conducting the fits, the fit results have been folded back to be comparable to the data. 
As also shown in the projection plots in Fig.~\ref{fig:fitproj}, the fitted flow is one component of the total fit function, and all the rest are regarded as nonflow components in this study.

\begin{figure*}
	\includegraphics[width=0.45\linewidth]{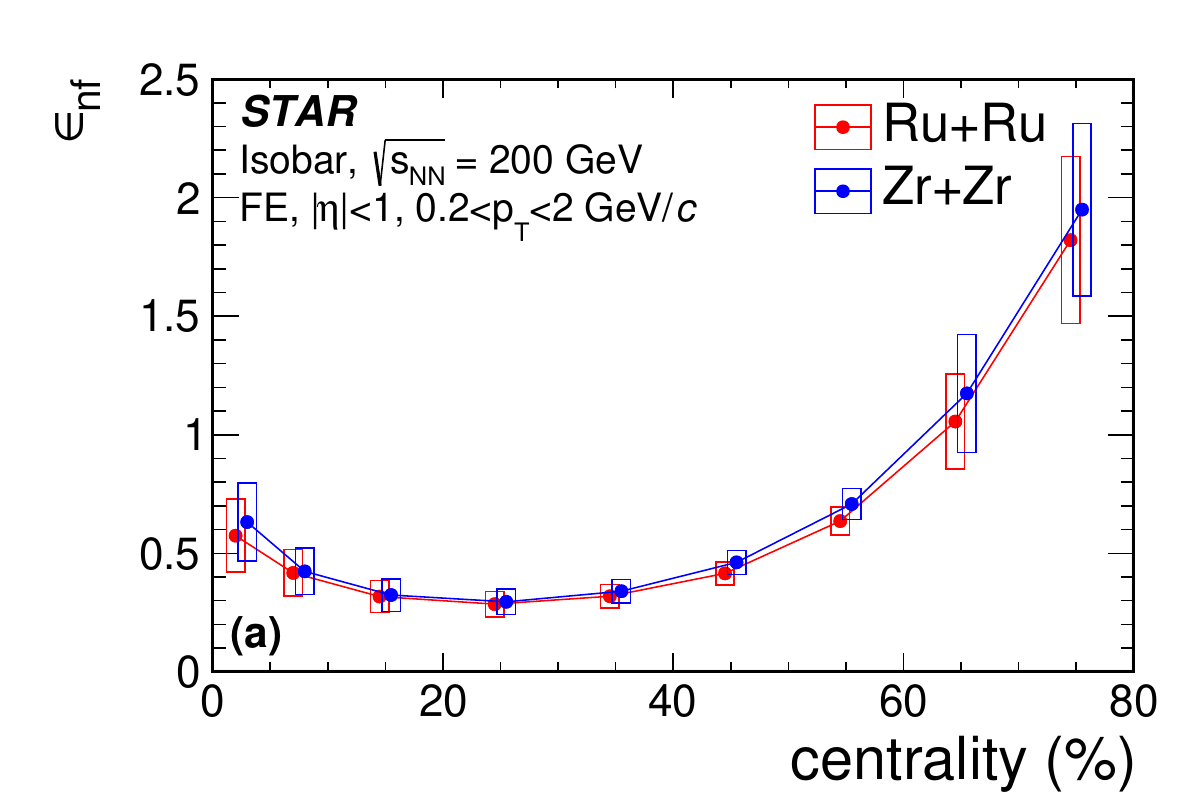}
	\includegraphics[width=0.45\linewidth]{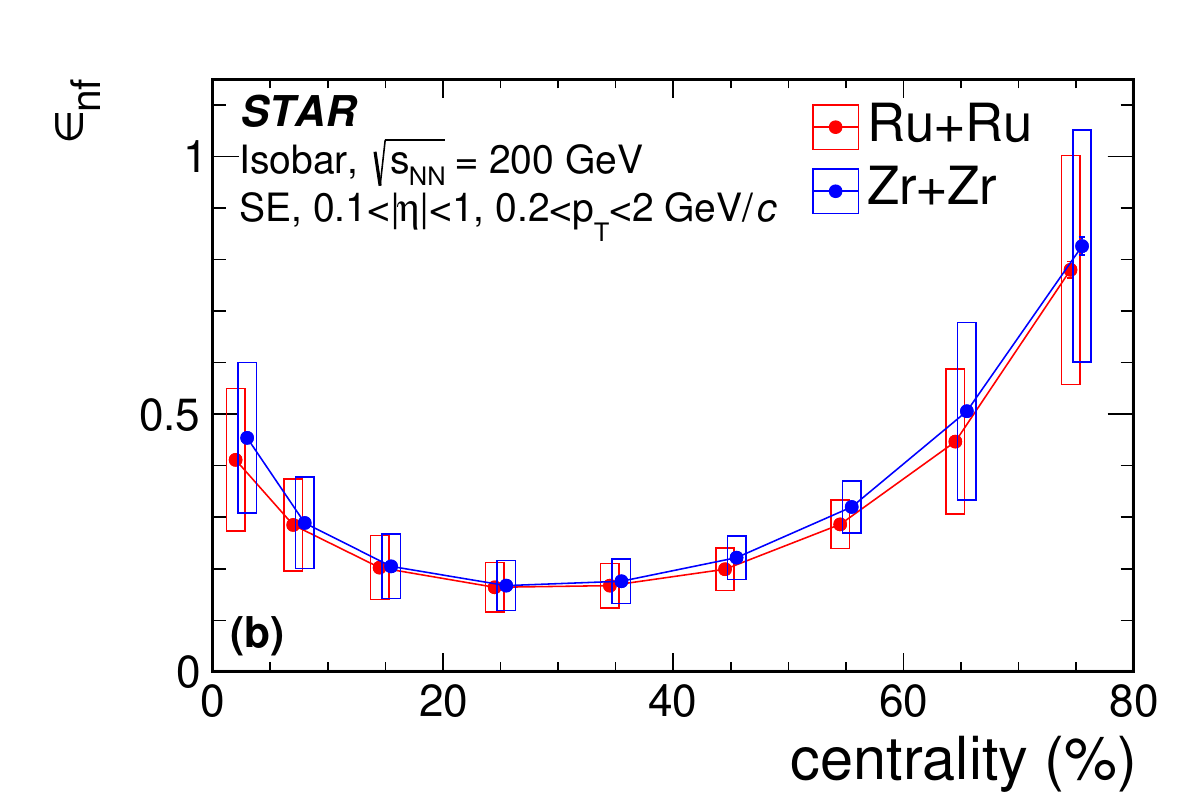}
	\includegraphics[width=0.45\linewidth]{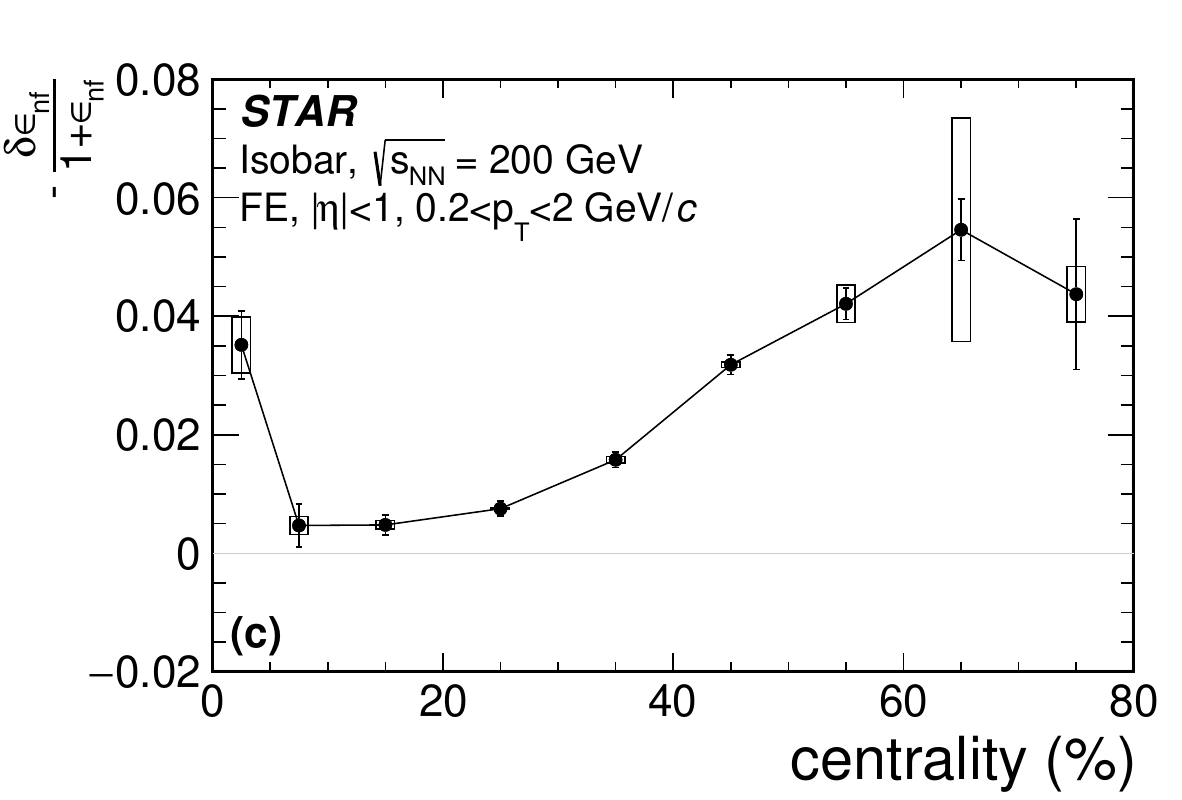}
	\includegraphics[width=0.45\linewidth]{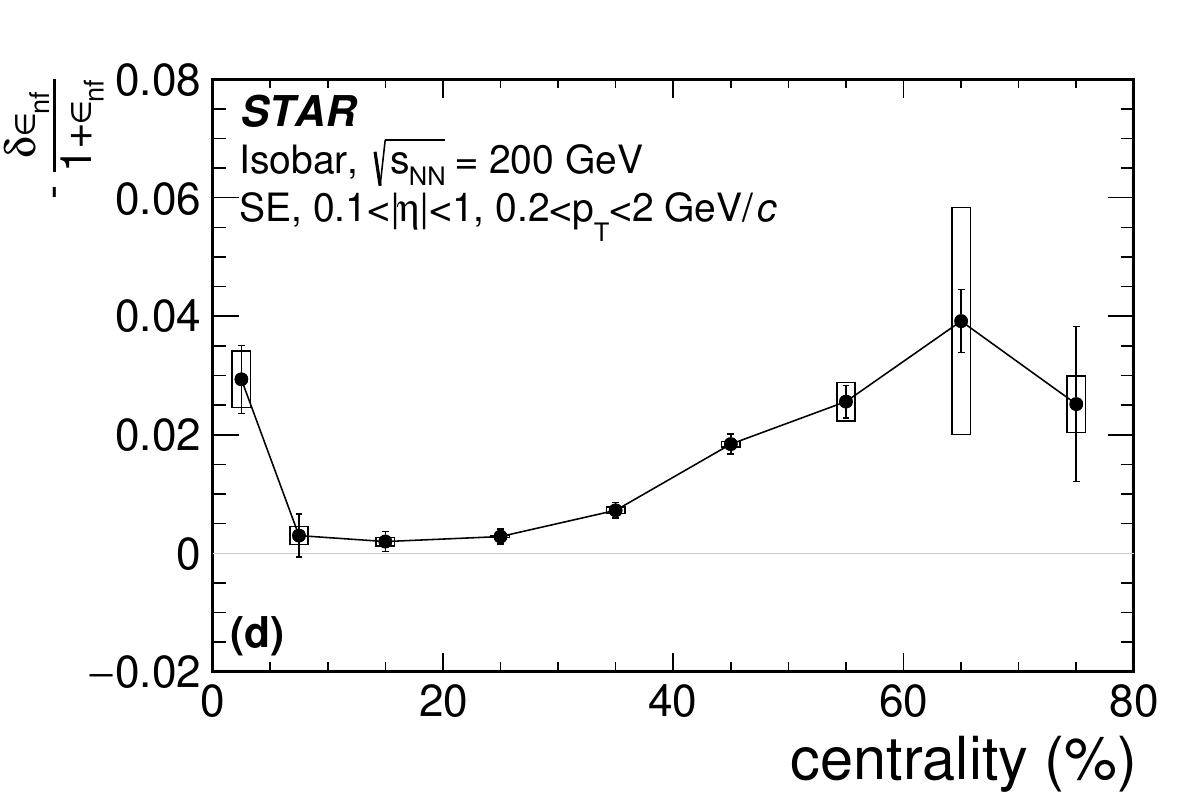}
	\caption{The relative nonflow strength, $\epsilon_{\text{nf}}=v_{2,\nf}^2/v_2^2$, in isobar collisions as functions of centrality for full-event data (upper left panel, with Group-3 cuts) and subevent data (upper right panel, with Group-2 cuts). The lower panels depict the corresponding $-\delta\enf/(1+\enf)$, one of the contributions to the deviation of $Y_{\bkgd}$ from unity; see Eq.~(\ref{eq:Y}). Vertical bars and hollow boxes show the statistical and systematic uncertainties, respectively.}
	\label{fig:enf}
\end{figure*}

The nonflow fraction $\enf$ can therefore be calculated by Eq.~(\ref{eq:enf}) from the fitted flow $v_{2}^{2}$ and the inclusive ${v_{2}^{*}}^{2}$ measurements from the STAR isobar blind analysis~\cite{STAR:2021mii}. Figure~\ref{fig:enf} presents the $\enf$ for the two isobars and the isobar difference $\frac{-\delta\enf}{1+\enf}$ for both full event and subevent cases. The boxes represent the systematic uncertainties. 

Since the STAR data in Ref.~\cite{STAR:2021mii} already include the systematics from selection variations, this study does not duplicate them in the baseline estimation to avoid double counting. Instead, it focuses on considering the systematic uncertainties arising from  sources specific to the background baseline estimation of this work. These sources include fit uncertainties and model dependencies.
The full-event inclusive ${v_{2}^{*}}^{2}$ can be calculated from the 2D distribution (\pos+\pss) in this study, represented by the open black curves in Fig.~\ref{fig:v2}. These results are essentially a replication of the measurement conducted by Group-3 in the STAR isobar blind analysis~\cite{STAR:2021mii}, shown as open orange curves in Fig.~\ref{fig:v2}. 
%The small difference between these two measurements is accounted for in the systematic uncertainty of $v_{2}^{2}$ for the replication study.  
%The inclusive ${v_{2}^{*}}^{2}$ for \pss\ pairs can be calculated from the 2D \pss\ distribution (closed black curves in Fig.~\ref{fig:v2}) and the fit function (closed magenta in Fig.~\ref{fig:v2}), whose small difference is included in the systematic uncertainty of $v_{2}^{2}$ for the fit quality. 
In addition, a $3\%$ flow decorrelation over one unit of pseudorapidity has been observed~\cite{Yan:2022qm}. 
As a result, $\pm 3\%$ variation of $v_{2}^{2}$  is also considered as a systematic uncertainty.  
%Additionally, another method is attempted to correct the kink effect. In this approach, the particle $\eta$ values are randomized, and a more complex 2D function is used for the fit. 
%Additionally, an alternative 2D fit is used with a different function form and corresponding kink correction (by randomizing particle $\eta$ values). Through this method, a fitted flow is obtained, and it is considered as a systematic variation as well. 
Additionally, an alternative 2D fit is used with a different functional form (and corresponding kink correction). The fitted flow deviation from the default is considered as part of the systematic uncertainty.
Those systematic sources and their contributions are listed in Table~\ref{tab:sys}, where the contributions from one-sided variations are expanded to be symmetric in the calculation of systematic uncertainties. 
%\blue{For this alternative method, the $\eta$ is divided into 4 equal ranges $-1 \sim -0.5 \sim 0 \sim 0.5 \sim 1$. If one \poi\ has $\eta$ in one certain range, then its randomized $\eta$ is a random number from this range. Then, we can get the two-particle distribution of this randomized $\eta$ for real event $\fgr(\Delta\eta)$ and for mixed event $\fgm(\Delta\eta)$. Their ratio $\fgc(\Delta\eta) = \fgr(\Delta\eta) / \fgm(\Delta\eta)$ can be used to correct the kinks $h(\Delta\eta, \Delta\phi) = \fhc(\Delta\eta, \Delta\phi) / \fgc(\Delta\eta)$. However, this method cannot remove the kinks completely, and can still have the away side and ridge nonflow correlations, so its fit function is more complicated whereas its fit quality seems worse.}

\begin{figure*}
    \includegraphics[width=0.45\linewidth]{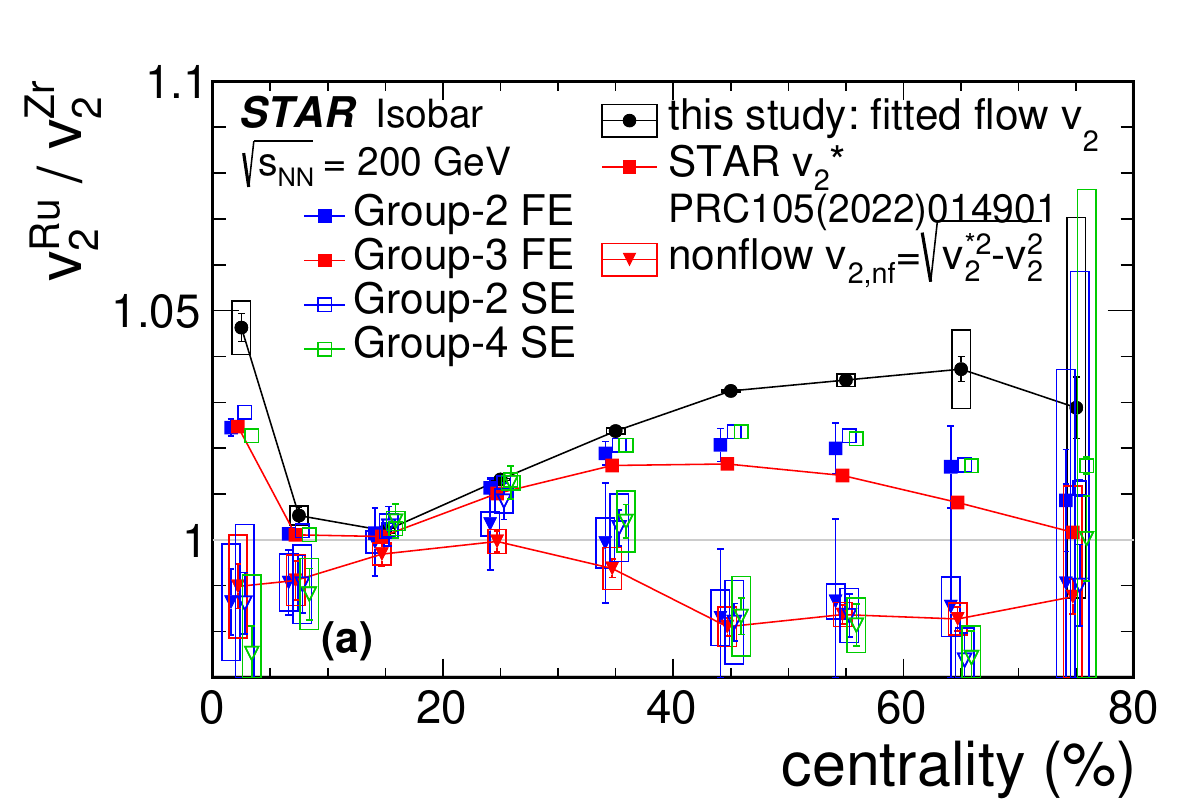}
    \includegraphics[width=0.45\linewidth]{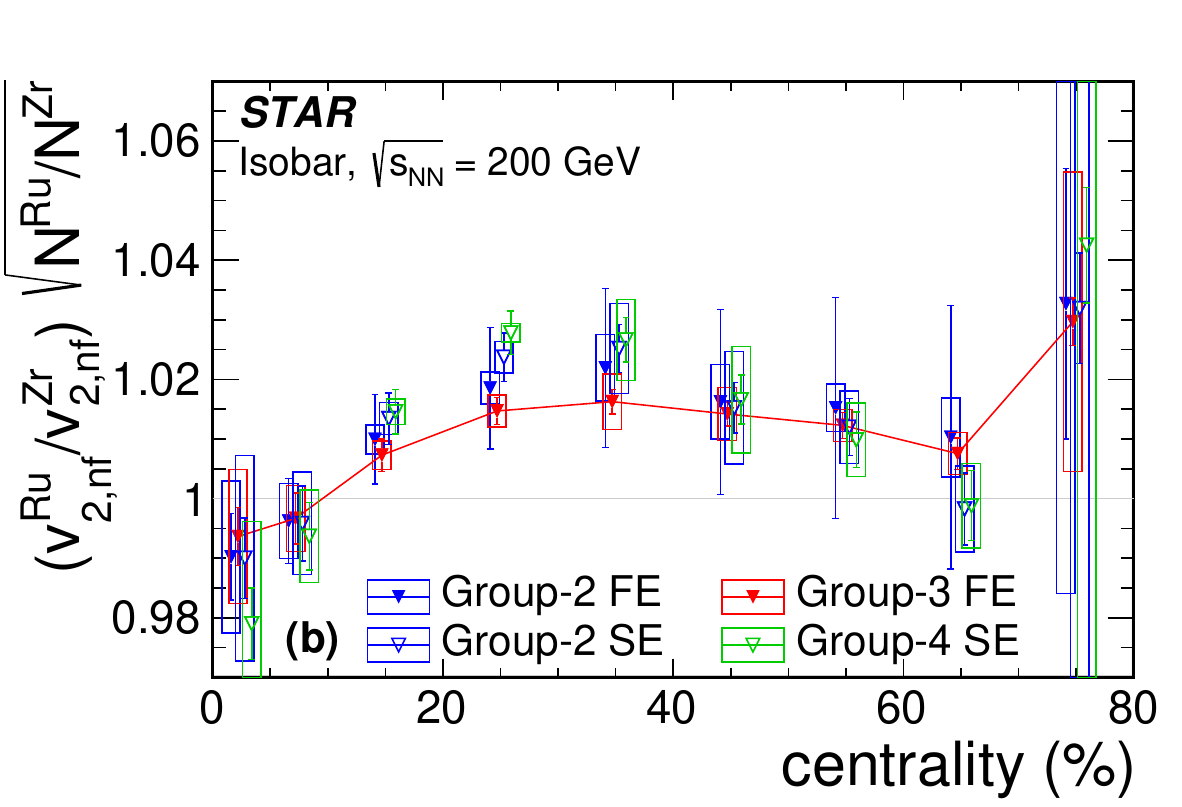}
    \caption{(Left panel) The isobar Ru+Ru/Zr+Zr ratio of the fitted $v_2$ flow parameters as a function of centrality in black solid circles, compared to those of the inclusive $v_{2}^{*}$ measurements from the cited STAR isobar blind analysis~\cite{STAR:2021mii} in colored squares. 
	%(Right panel) The isobar ratio of the extracted nonflow components $v_{2,\nf}$ scaled by the square root of multiplicity $\sqrt{N}$ as a function of centrality. 
	Full-event (FE) measurements are shown in solid squares, and subevent (SE) measurements are shown in open squares where the two particles come from two different subevents. 
	%Vertical bars and hollow boxes show the statistical and systematic uncertainties, respectively.
	Group-2 (FE) exploited Gaussian fits to reduce nonflow contributions in their measurement, resulting in larger statistical uncertainties and isobar ratios closer to the SE measurements. 
	%The points of Group-3 FE is connected by lines to guide the view. 
	(Right panel) The isobar ratio of the extracted nonflow components $v_{2,\nf}$ scaled by the square root of multiplicity as a function of centrality. 
	Vertical bars and hollow boxes show the statistical and systematic uncertainties, respectively. 
	The line connecting one set of the data points is to guide the eye.}
    \label{fig:truev2}
\end{figure*}
It is of interest to examine the relative strength of the fitted flows of the two isobar systems. Figure~\ref{fig:truev2} displays the isobar ratio of the fitted $v_{2}$ parameters, where the systematic uncertainties are represented as boxes. The fitted $v_{2}$ values averaged over the 20-50\% centrality range are 
$0.0561 \pm 0.0010$ and $0.0548 \pm 0.0010$
for Ru+Ru and Zr+Zr collisions, respectively, where the quoted uncertainties are  dominated by systematic uncertainties.
The average ratio within the 20-50\% centrality range is $1.0215 \pm 0.0004 ({\rm stat.}) \pm 0.0009 ({\rm syst.})$. 
The difference in $v_2$ between the isobar systems originates from variations in the initial collision geometries.  These differences can be attributed to distinct nuclear structures,  as predicted by DFT calculations~\cite{Xu:2017zcn,Li:2019kkh,Xu:2021vpn}.
%The ratios deviate from unity in most centrality bins. 
Both hydrodynamic calculations \cite{Xu:2021uar} and transport models~\cite{Li:2018oec} can produce the isobar $v_{2}$ ratios similar to data, including the subtle hump structure of the ratio in the medium centrality region, once the DFT calculated densities are implemented in those models. 
The significant difference in the most central collisions arises predominantly from nuclear deformations.
For comparison, the isobar ratio of the inclusive ${v_{2}^{*}}$ from STAR data~\cite{STAR:2021mii} are also plotted in Fig.~\ref{fig:truev2}.
The measured $v_2^*$ ratios are significantly smaller than those of the fitted, presumably true $v_2$. 
This is because nonflow contamination in Ru+Ru is smaller than that in Zr+Zr due to the higher charged particle multiplicity dilution.
This is shown in Fig.~\ref{fig:truev2} by the triangles where most data points are below unity.
Factoring out the multiplicity dilutions, Fig.~\ref{fig:truev2}(b) shows the ratio of $\sqrt{N}v_{\rm 2,nf}$, which reflects the genuine difference in nonflow correlations between the isobar systems.  
The ratio is in fact larger than unity for most centralities, which is consistent with the presumably larger energy density achieved in Ru+Ru than Zr+Zr collisions.

Azimuthal anisotropies in central heavy-ion collisions are particularly sensitive to nuclear deformations~\cite{Giacalone:2019pca, Giacalone:2021uhj}. 
It is noteworthy that the difference between the measured $v_2^*$ ratio and the fitted one is significant also in the most central collisions, as shown in Fig.~\ref{fig:truev2}(a). 
This suggests that using comparisons of the measured $v_2^*$ ratio to hydrodynamic or other model calculations, which often fail to describe nonflow contributions, to infer nuclear deformations should be taken with caution~\cite{Zhang:2021kxj, Bhatta:2023cqf}. 
It is interesting to observe that there is no significant difference between full-event and subevent $v_2^*$ results in the most central collisions. This similarity arises because the relative nonflow contribution to $v_2^*$ is similar between full-event and subevent for those central collisions, as shown in Fig.~\ref{fig:enf}.

%------------------------------------------------------------------------------------------------%
\subsection{Three-particle correlation background} %{$C_{\thp}$}
\label{sec:3p}

\begin{figure*}
	\includegraphics[width=0.45\linewidth]{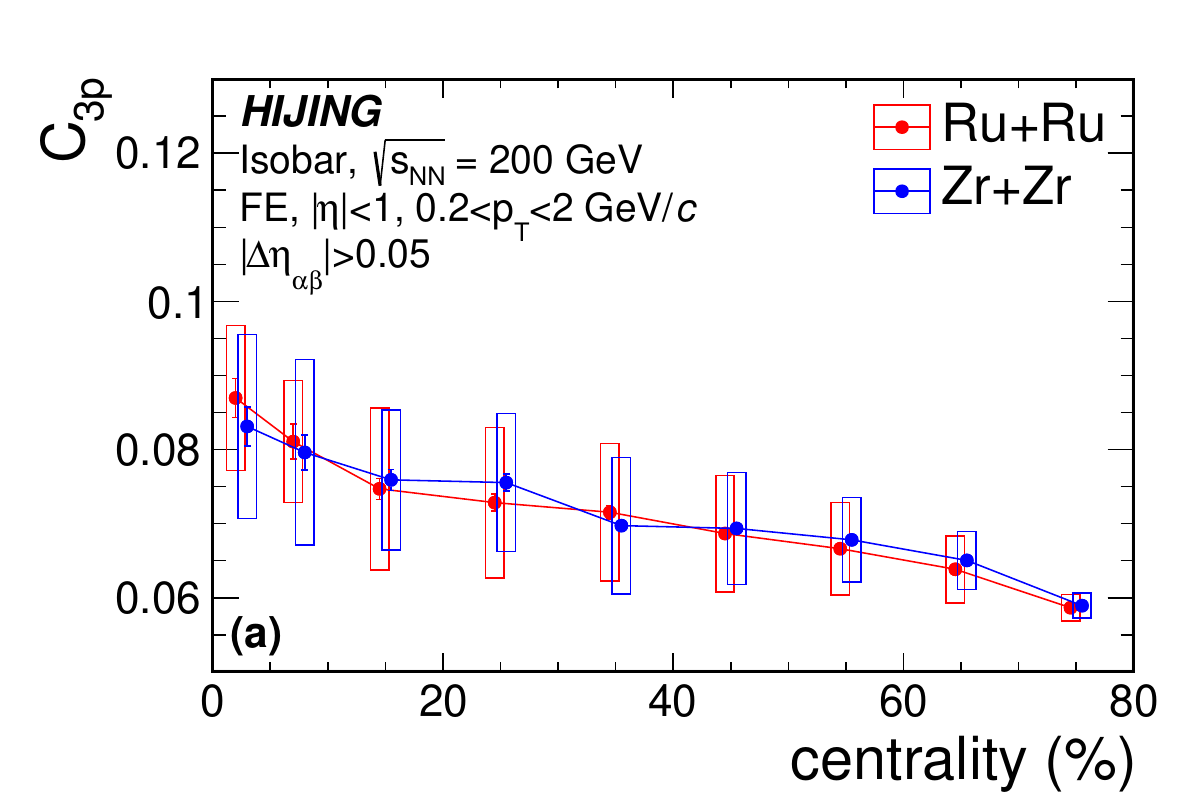}
 	\includegraphics[width=0.45\linewidth]{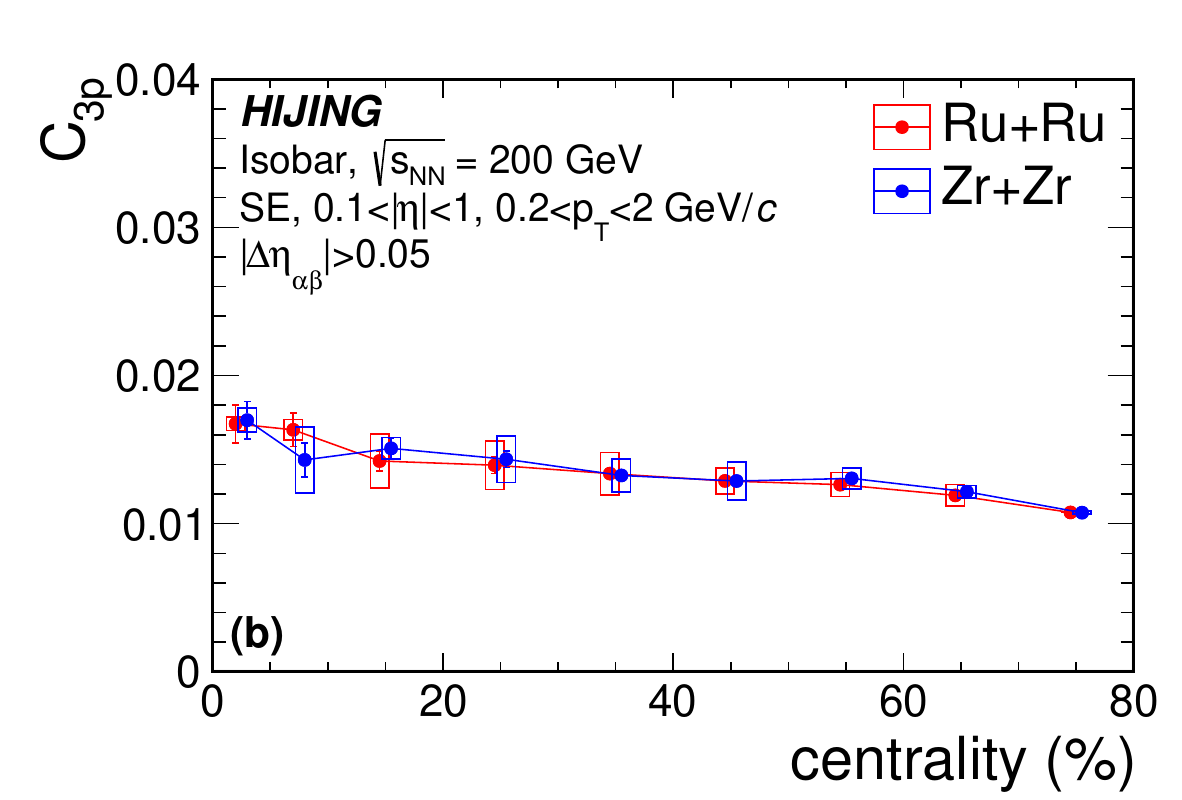}
    \includegraphics[width=0.45\linewidth]{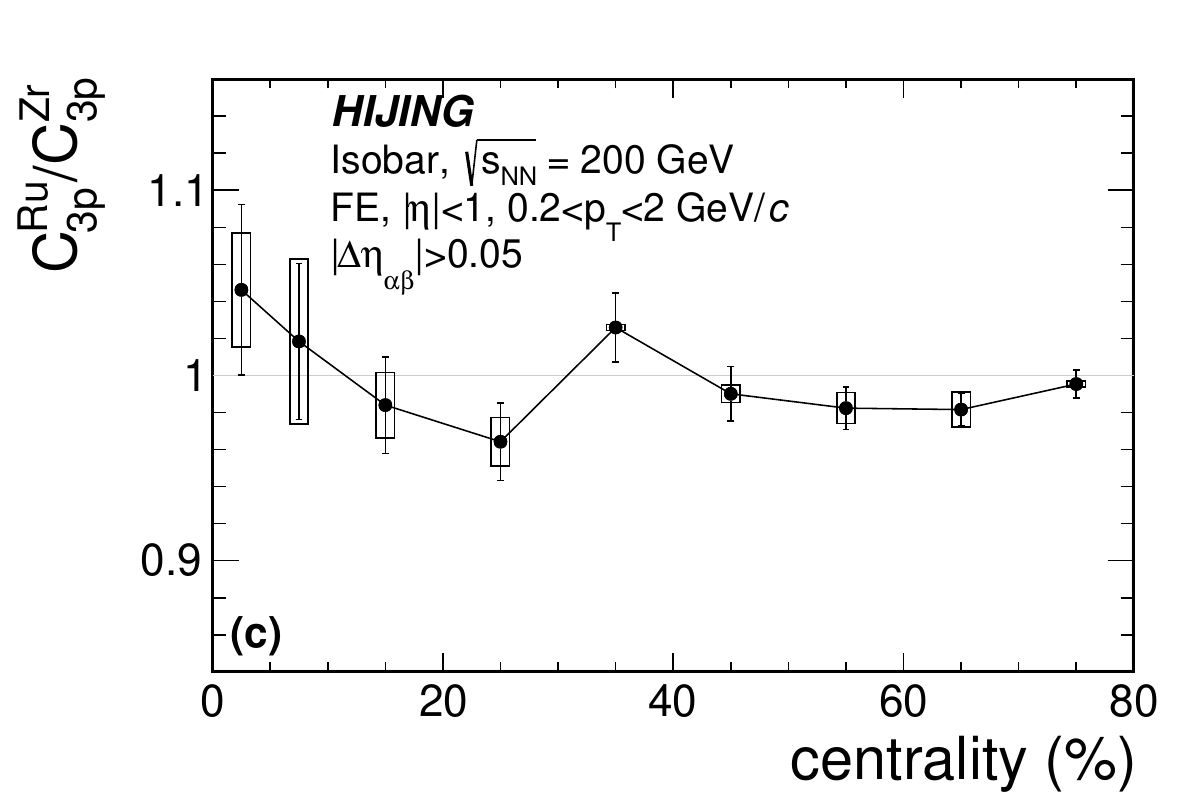}
 	\includegraphics[width=0.45\linewidth]{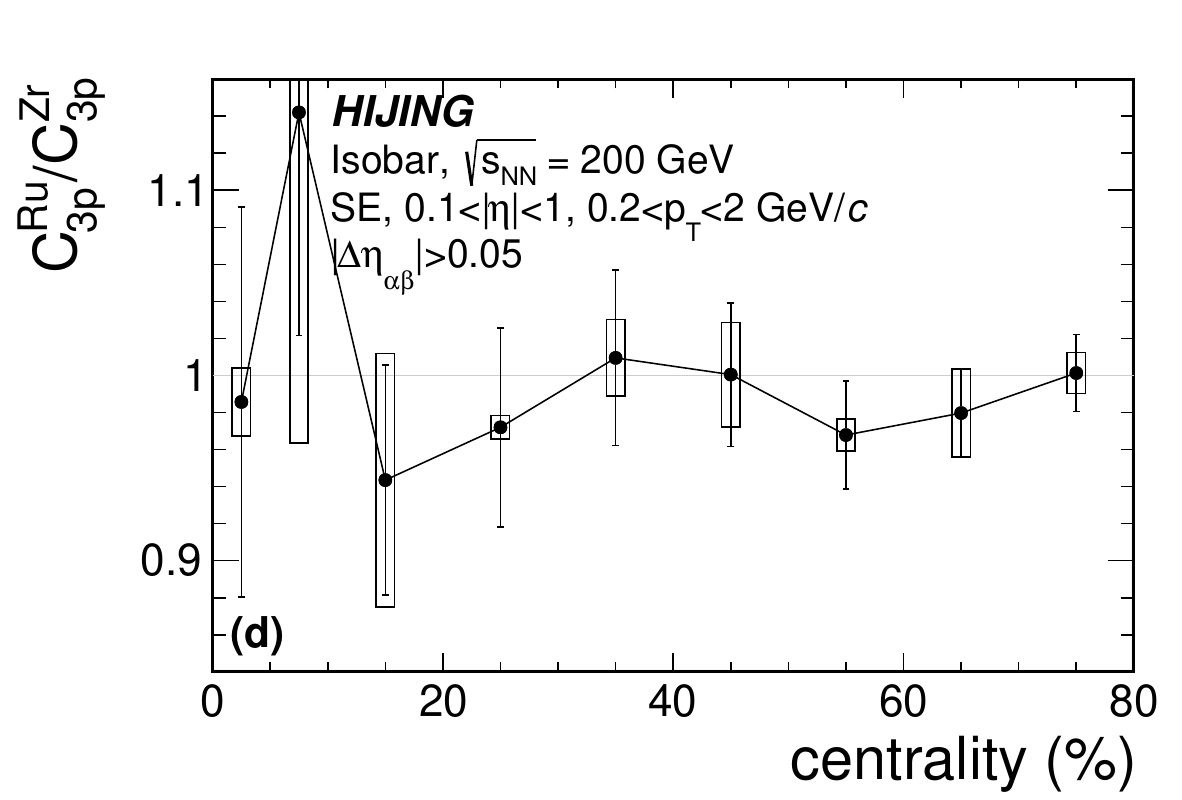}
	\caption{The $C_{\thp}$ in Ru+Ru and Zr+Zr collisions as functions of centrality for full-event (left panel, with Group-3 cuts) and subevent (right panel, with Group-2 cuts) analysis, obtained from \hijing\ simulations of 7.0 billion minimum-bias events for each system. The lower panels show the Ru+Ru over Zr+Zr ratio of $C_{\thp}$. Vertical bars indicate statistical uncertainties. The hollow boxes show systematic uncertainties, estimated from \hijing\ quenching-off simulations. The two \poi's have an $\eta$ gap, $\Delta\eta_{\alpha\beta}>0.05$. For the subevent method, the two \poi's come from the same subevent, while the reference particle comes from the other subevent.}
	\label{fig:c3p}
\end{figure*}

The three-particle (3p) background correlation is the last piece needed to form the background baseline estimate, but it is challenging to measure due to the significant combinatorial background. We resort to the \hijing\ (Heavy Ion Jet INteraction Generator) model~\cite{Wang:1991hta,Gyulassy:1994ew}, which simulates parton-parton hard scatterings based on perturbative QCD and gives a reasonable description of partonic energy loss in the QGP medium (jet quenching). Since \hijing\ does not have collective flow, the correlator $C_{3}\equiv C_{3,\pos}-C_{3,\pss}$ in \hijing\ is entirely composed of genuine 3-particle correlations, $C_{3,\pos}=\frac{N_{\thp,\pos}}{N_\pos N}C_{\thp,\pos}, C_{3,\pss}=\frac{N_{\thp,\pss}}{N_\pss N}C_{\thp,\pss}$ (c.f.~Eq.~(\ref{eq:c3bkgd})). The pathlength-dependent jet quenching does produce some degree of anisotropy in the final-state particle azimuthal distribution, sensitive to the initial geometry and indistinguishable from collective anisotropy. This anisotropy, however, is negligible compared to the effect of the 3p correlations~\cite{Feng:2021pgf}. The 3p correlation strength, $C_{\thp}$, can be readily obtained from the 3p correlator in \hijing,
\begin{equation} \label{eq:c3nn}
    C_\thp=N^2 C_3\,.
\end{equation}

We use \hijing\ version v1.411 and simulate 7.0 billion events each for Ru+Ru and Zr+Zr collisions at $\snn = 200$~GeV. The nuclear structure density distributions are given by energy density functional theory calculations~\cite{Xu:2017zcn,Li:2018oec,Xu:2021vpn}, and they are implemented in the initial geometry setup in \hijing.
Only the final-state charged pions, kaons, protons and their antiparticles are used for centrality definition and \poi.
The centrality is defined by the multiplicity distribution of particles with $|\eta|<0.5$.
The same analysis cuts as the STAR isobar data analysis~\cite{STAR:2021mii} (see Table~\ref{tab}) are used to process \hijing\ simulation data.
Figure~\ref{fig:c3p} shows the $C_\thp$ in isobar collisions as functions of centrality, as obtained from \hijing\ simulations, and the Ru+Ru over Zr+Zr ratios of $C_\thp$. The centrality dependence is weak, as one would expect for $C_\thp$, which is defined for the correlated triplets with the dilution effect factored out as in Eq.~(\ref{eq:c3p}).
The $C_\thp$ is larger in full-event than subevent because the number of triplets drops with acceptance more rapidly than single multiplicity.

%\red{For all the \hijing\ results, the STAR detector efficiency ($\sim 85\%$) is always taken into consideration.}

\begin{figure*}
	\includegraphics[width=0.45\linewidth]{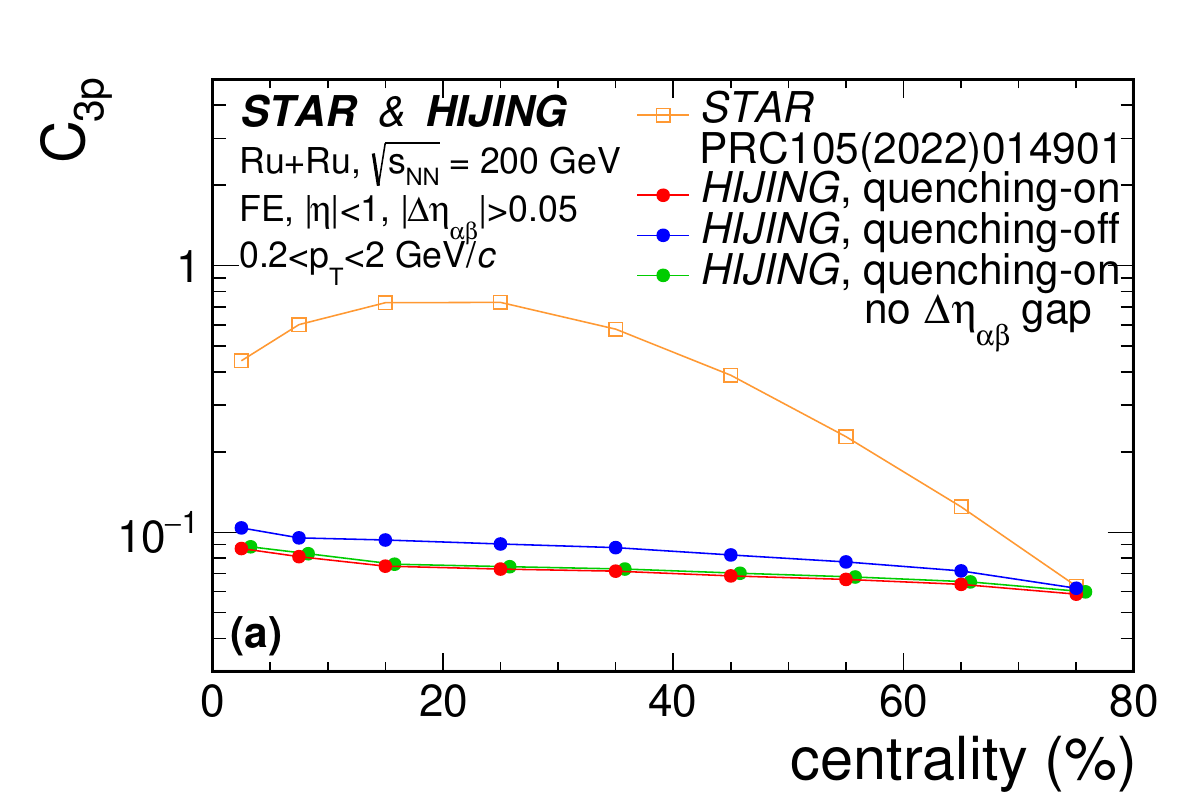}
	\includegraphics[width=0.45\linewidth]{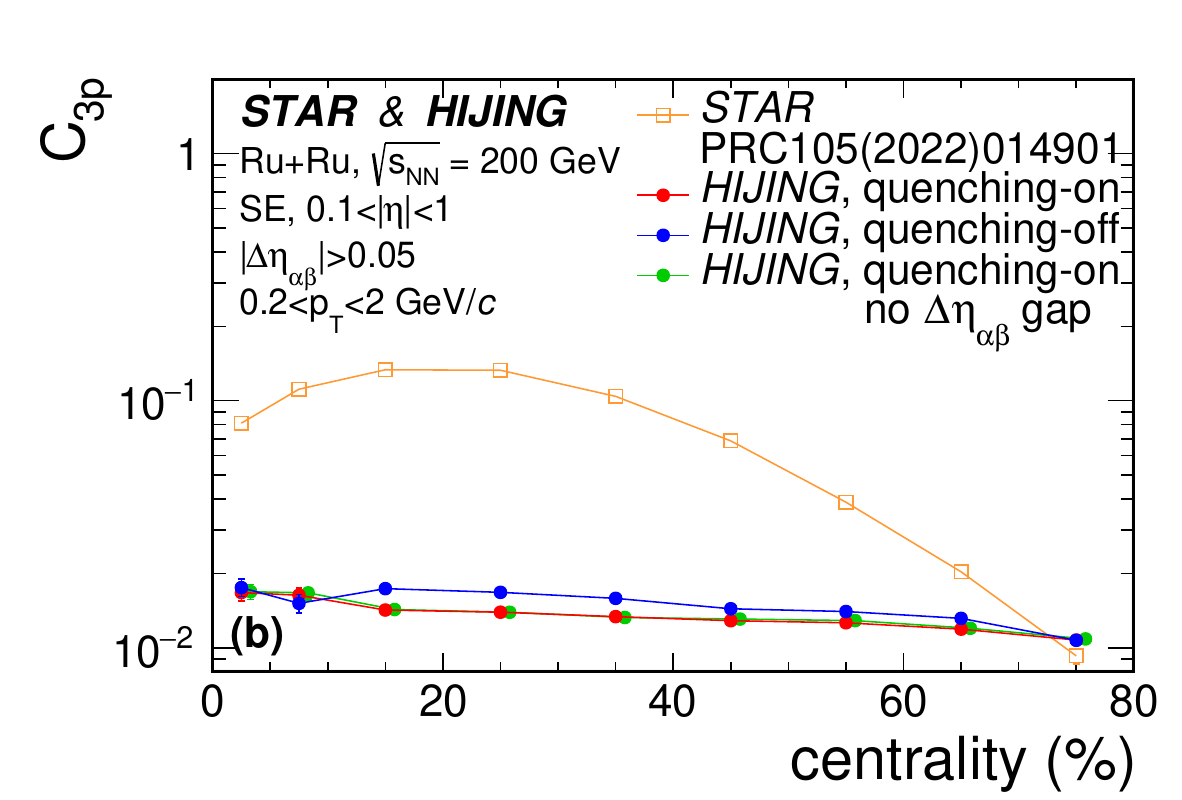}
	\caption{Comparison of $C_{\thp}$ between \hijing\ (jet quenching on) and the cited STAR measurements~\cite{STAR:2021mii} in Ru+Ru collisions for full-event (left panel, with Group-3 cuts) and subevent (right panel, with Group-2 cuts) analyses (Zr+Zr is similar). The data measurements on the plots use Eq.~(\ref{eq:c3nn}), which are inclusive. The peripheral data are similar to the \hijing\ result, suggesting that the peripheral data are dominated by 3p correlation background. The non-peripheral data are dominated by flow-induced background, absent from \hijing. The $C_\thp$ results from \hijing\ simulation with quenching turned off are also shown, which is taken as the maximum systematic uncertainty. 
    %\blue{The current quenching-off is old. The current $C_{\thp}$ used for baseline is also old. I will update that when \hijing\ is ready.} \red{let's use logy.}}
	The two \poi's have an $\eta$ gap, $\Delta\eta_{\alpha\beta}>0.05$. For the subevent method, the two \poi's come from the same subevent, while the reference particle comes from the other subevent.
    To give a magnitude assessment, the \hijing\ (quenching on) results without the $\deta_{\alpha\beta}>0.05$ cut (as in Group-4 data analysis~\cite{STAR:2021mii}) are also displayed.
	}
	\label{fig:comparec3p}
\end{figure*}
An important question is how well \hijing\ describes data in terms of 3p correlations? The question cannot be directly answered because of the difficulties to access the 3p correlations in real data as aforementioned. However, there are a few checks one can make. We first examine how well \hijing\ describes the $C_3$ in peripheral collisions, where the nonflow effects dominate due to smaller multiplicity dilution. We show the measured $C_3$ in data and in \hijing\ in Fig.~\ref{fig:comparec3p} for full-event and subevent analysis. 
%The full-event magnitude is larger than the subevent magnitude because of the trivial acceptance effect. \textbf{[not finalized]}. 
As shown in Fig.~\ref{fig:comparec3p}, the 70-80\% peripheral data are reasonably well described by \hijing. This suggests that the peripheral $C_3$ data are dominated by 3p correlations, the flow-induced background is small in peripheral collisions. 
%\hijing\ is a good model for jets and jet quenching, and also a good parameterization for soft physics~\cite{Wang:1991hta,Gyulassy:1994ew}, so it is a suitable description here for 3p correlations.
\hijing\ is a reasonable model for (mini-)jet production as well as soft physics via string fragmentation, so it is considered as suitable description for 3p correlations.
%\red{\sout{We have also examined Ks/Lambda-h correlations... (I'll write more on this)}}

The default setup of our \hijing\ simulations include jet quenching. We also simulate \hijing\ with jet quenching turned off. The quenching-off $C_{\thp}$ is about 20\% higher than the quenching-on result, as shown in Fig.~\ref{fig:comparec3p}. We take this difference as the maximum systematic uncertainty on 3p correlation estimate. 
As another check, we also show in Fig.~\ref{fig:comparec3p} the \hijing\ $C_3$ without cutting on $\deta_{\alpha\beta}$ (as in the Group-4 data analysis~\cite{STAR:2021mii}). The difference is small, and we conclude that \hijing\ with quenching-on and -off %provide a safe estimate of the maximum systematic uncertainty. 
provide a safe estimate of the one-side maximum systematic uncertainty. We thus assign their difference divided by $\sqrt{3}$, assuming a uniform probability for the systematic uncertainty, to be the one standard deviation systematic uncertainty on our 3p correlation background estimate, and expand it to be symmetric, as listed in Table~\ref{tab:sys}. 
%Since quanching-on and -off are two extreme cases, the systematic uncertainty is their difference divided by $\sqrt{3}$. 
%We thus assign their difference divided by $\sqrt{3}$ to be the systematic uncertainty on our 3p correlation background estimate, where uniform distribution is assumed and $\sqrt{3}$ is the ratio of its boundary over standard deviation. 

\begin{figure*}
	\includegraphics[width=0.45\linewidth]{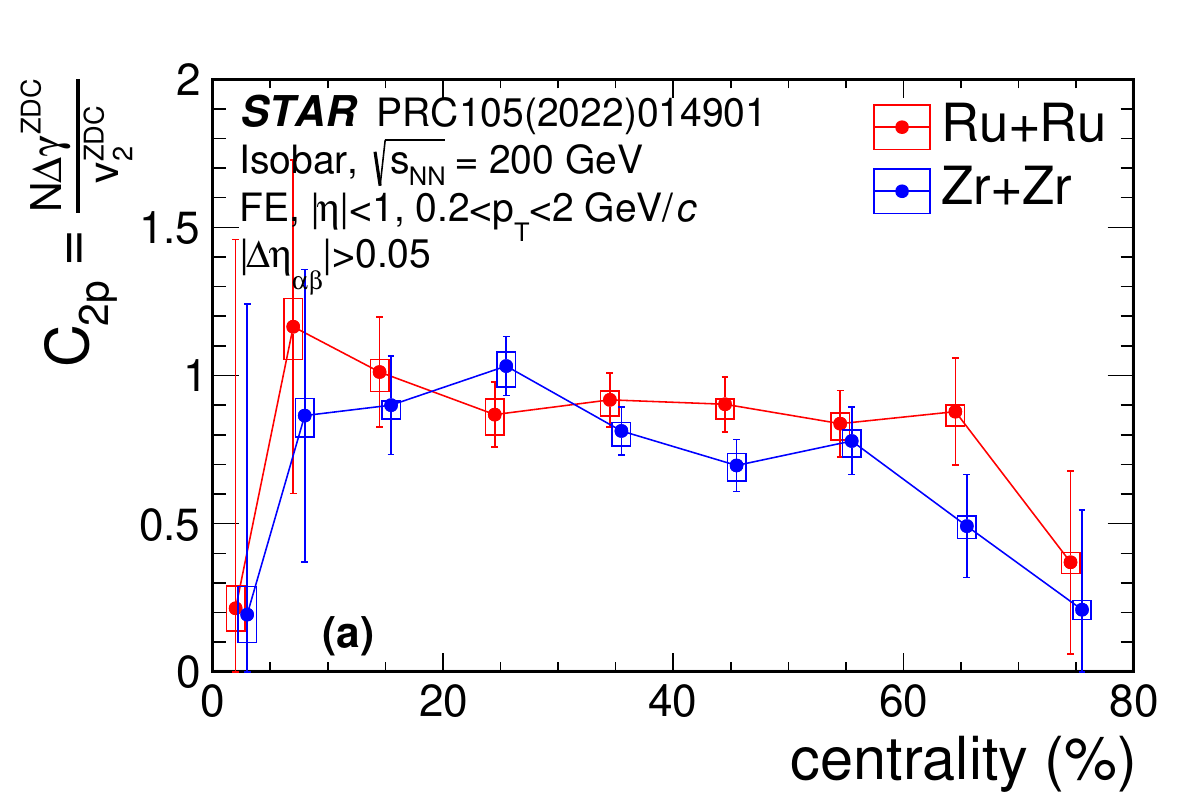}
 	\includegraphics[width=0.45\linewidth]{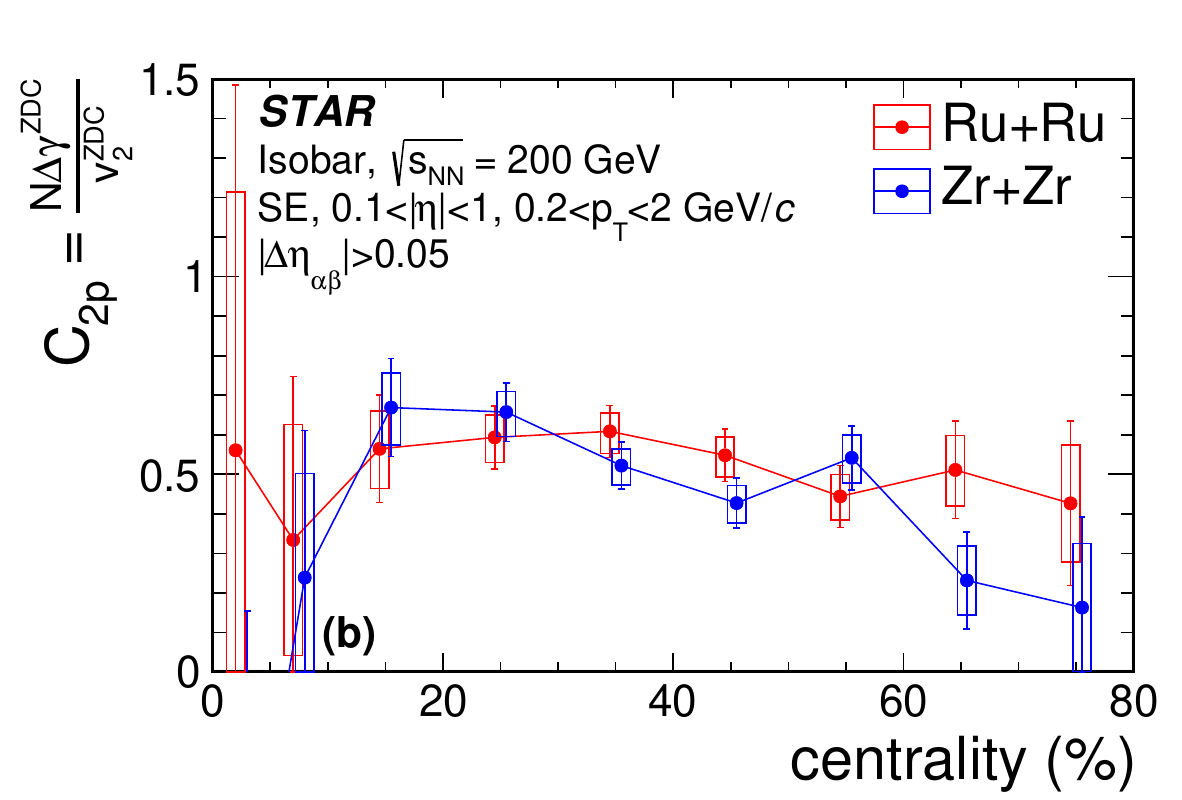}
	\caption{The $C_{\twp}$ in isobar collisions for the full-event (left panel, from the Group-3 isobar result) and subevent (right panel, with Group-2 cuts) analyses, obtained from the ZDC measurements of $\dg/v_2$ multiplied by the efficiency-corrected multiplicity. Vertical bars indicate statistical uncertainties. The asymmetric systematic uncertainty, shown by boxes, is composed of the data systematic uncertainty from the cited Ref.~\cite{STAR:2021mii} and a one-sided systematic uncertainty of $-5\%$ to account for possible CME contributions in the ZDC measurements. %The horizontal lines show the averages over the 20-50\% centrality range.
    %\red{I think we need to include the data syst. errors because the zdc data are not used in Y measurement so no double counting is involved.}
	The two \poi's in $\Delta\gamma$ have an $\eta$ gap, $|\Delta\eta_{\alpha\beta}|>0.05$. For subevent method, the \poi\ pair comes from the same subevent. %\sout{STAR data~\cite{STAR:2021mii} for subevent with $|\Delta\eta_{\alpha\beta}|>0.05$ gap (right panel) uses a wider $\eta$ range $0.05<|\eta|<1$, which is consistent with results from the default setup with $0.1<|\eta|<1$. For the baseline calculations, we do not repeat the systematics in Ref.~\cite{STAR:2021mii} to avoid double counting, and use the default $\eta$ range for subevent.}}
    }
	\label{fig:c2p}
\end{figure*}
\begin{figure*}
    \includegraphics[width=0.45\linewidth]{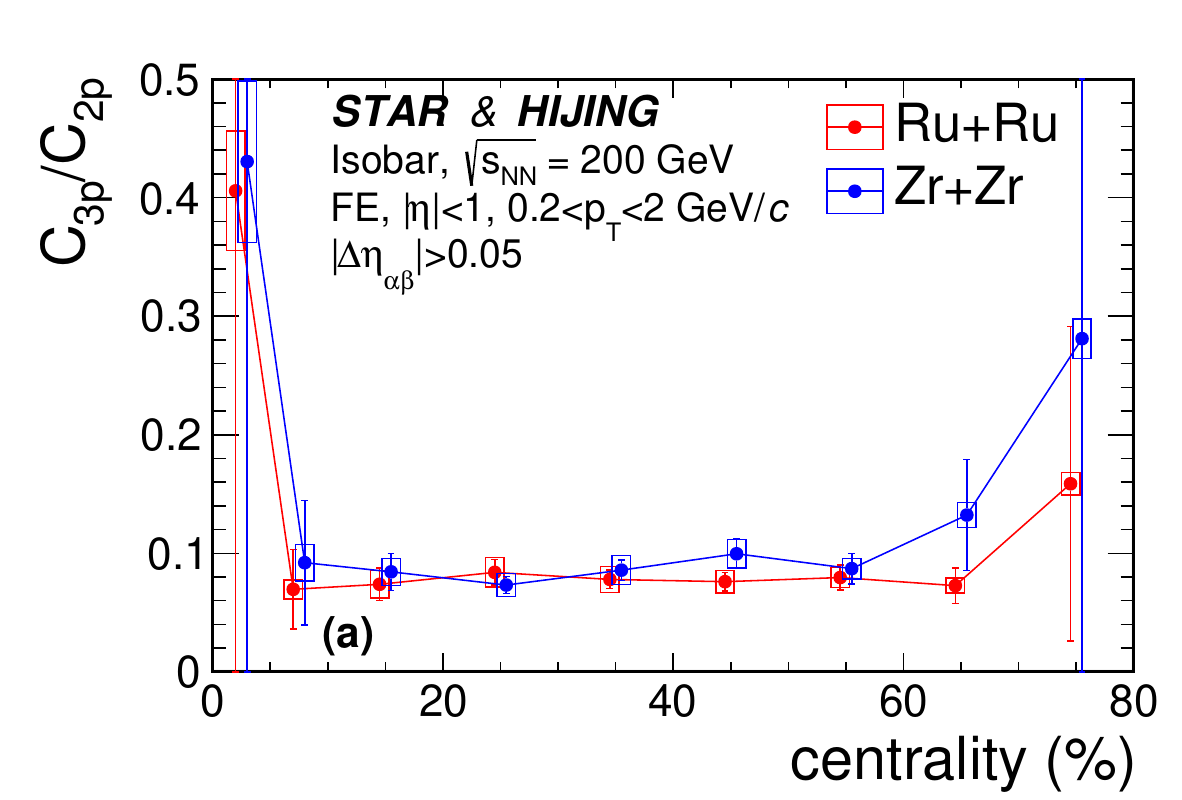}
    \includegraphics[width=0.45\linewidth]{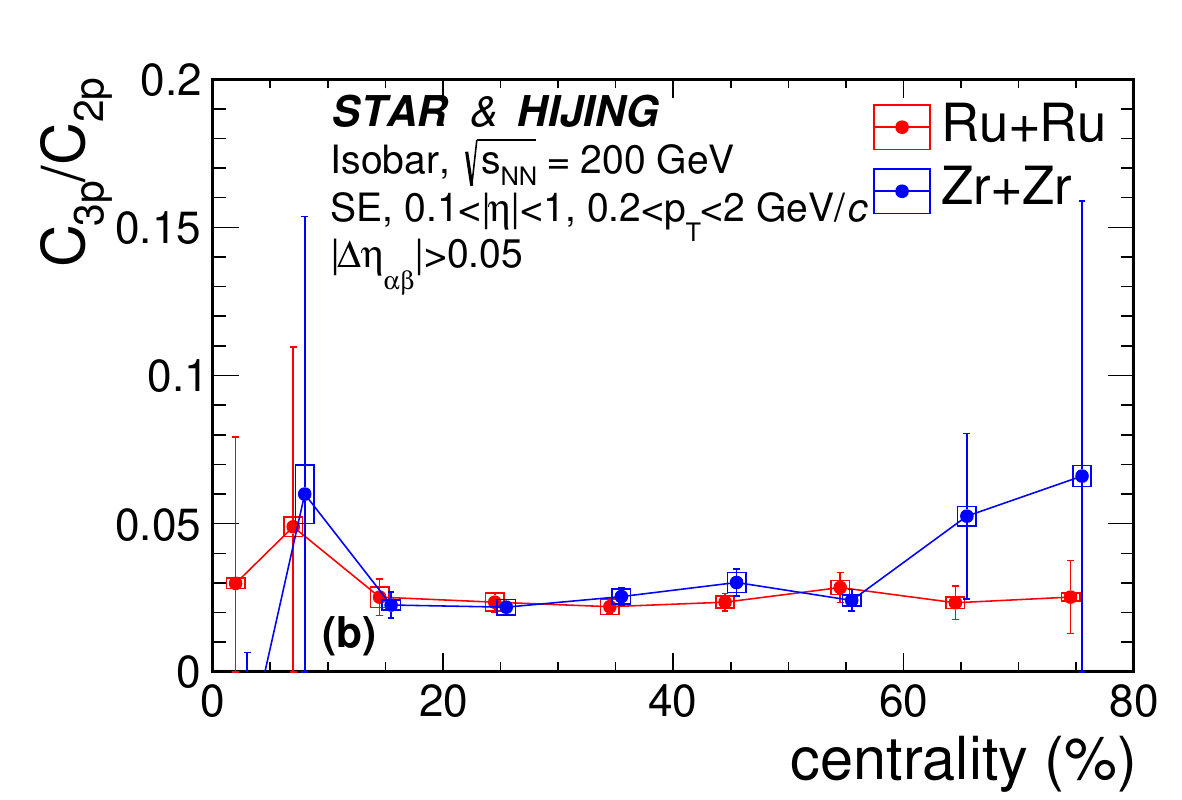}
    \caption{The ratio $C_{\thp} / C_{\twp}$ as functions of centrality for full-event (left panel, with Group-3 cuts) and subevent methods (right panel, with Group-2 cuts). Some of the central and peripheral data points are off the plots with large error bars. Vertical bars and hollow boxes show the statistical and systematic uncertainties, respectively.
	%\red{let's zoom into 0-0.2}
	}
    \label{fig:c3pc2p}
\end{figure*}

In order to estimate the baseline contribution from 3p correlations, we also need the quantity $C_\twp$ (Eq.~(\ref{eq:Y})). %, the denominator of the last term).
As shown in Eq.~(\ref{eq:c2p}), $C_{\twp}$ represents the 2p nonflow background correlations in $C_{3}$ measurements. If the azimuth of \srp\ is known, then Eq.~(\ref{eq:v2:rp}) gives the true elliptic flow and Eq.~(\ref{eq:dg:rp}) gives $\Delta\gamma$ without 3p nonflow by definition. If so, Eq.~(\ref{eq:dgbkgd}) simply becomes
\begin{equation} \label{eq:dgbkgd:rp}
    %N \frac{\Delta\gamma_{\bkgd}\{\srp\}}{v_{2}\{\srp\}} = C_{\twp} \frac{v_{2}^{2}}{v_{2}^{2}\{\srp\}} = C_{\twp}\,.
    N \frac{\Delta\gamma_{\bkgd}\{\srp\}}{v_{2}\{\srp\}} = C_{\twp}\,.
\end{equation}
In STAR, the ZDC is separated from TPC by a large $\eta$ gap and at high collision energies measures only spectator neutrons. Thus, the ZDC measurements of the event plane are not correlated with \poi, and the $N\Delta\gamma / v_{2}$ w.r.t.~ZDC can be used to estimate $C_{\twp}$. 
Figure~\ref{fig:c2p}(a) shows the full-event ZDC measurement from the STAR isobar blind analysis (Group-3)~\cite{STAR:2021mii}. 
Figure~\ref{fig:c2p}(b) shows the subevent ZDC measurement from this analysis because the subevent ZDC measurement in the STAR blind analysis~\cite{STAR:2021mii} used $0.05<|\eta|<1$ instead of $0.1<|\eta|<1$.

The \poi\ multiplicities have been corrected for the centrality- and $\pt$-dependent track reconstruction efficiencies of the STAR TPC. The average efficiency is on the order of $85\%$. The efficiency is obtained from Monte Carlo tracks simulated by \geant\ in the STAR detector and embedded into the isobar data on the pixel level with proper detector response simulations. 
%The efficiency is taken to be the same for Ru+Ru and Zr+Zr collisions as a function of $\pt$. 
%This efficiency is folded with the measured $\pt$ distributions in each individual isobar system to obtained the efficiency-corrected multiplicity. Since the $\pt$ spectra of the isobar collisions are similar, the corrected multiplicity ratio is similar to the measured one.
In this embedding sample, the input tracks and reconstructed tracks are distributed as functions of centrality, particle species, charge, and $\pt$, $\eta$, which are slightly different between the two isobars.
For each centrality bin, we obtain the integral (total number of $p\bar{p}$, $K^{\pm}$, $\pi^{\pm}$) inside our cuts (\poi) for the reconstructed and input tracks, and their ratio gives us the efficiency. 
Due to the $\eta$ gap, the subevent has slightly different efficiency compared to the full event. 
%Since the isobar blind analysis~\cite{STAR:2021mii} did not have efficiency correction, this study also does not apply it to the track level. 
%Instead, we only use efficiency to correct the \poi\ number, so that it can be compared with HIJING model. 

The systematic uncertainties shown in Fig.~\ref{fig:c2p} include those on the ZDC measurements from the blind analysis~\cite{STAR:2021mii} for both full-event and subevent analyses. 
The ZDC measurements could contain some CME signal, possibly on the order of a few percent~\cite{STAR:2021pwb,Feng:2021oub}. 
Because $C_\twp$ refers to 2p background correlation, we additionally assign a one-sided systematic uncertainty of $-5\%$ on $C_{\twp}$. 
%Only this one-sided systematic uncertainty is included in the estimate of those on $Y_{\bkgd}$ (Table~\ref{tab:sys}). The systematic uncertainties from data are not included on $Y_{\bkgd}$ but only on $Y$, to avoid double counting. 
This one-sided uncertainty is expanded to be symmetric in the calculation of systematic uncertainties on $Y_{\bkgd}$, as listed in Table~\ref{tab:sys}. 
The uncertainties on the ZDC measurement of $C_{\twp}$ are not included in those on $Y_{\bkgd}$ but only on $Y$~\cite{STAR:2021mii}, to avoid double counting. 
% The systematic uncertainties from the STAR ZDC measurements are not included in the $C_\twp$ for background estimate, but in the $Y$ because the $\Delta\gamma / v_{2}^{*}$ measurements already have those and it is not necessary to double count that in background baseline estimate. 
%\blue{Since the systematics for possible CME signal dominates over other sources, the subevent result in Fig.~\ref{fig:c2p}(b) only takes this as systematic uncertainty.}

Figure~\ref{fig:c3pc2p} shows the ratios of $C_\thp/C_\twp$ as a function of centrality in both full-event and subevent analyses. The ratios are on the order of a few percent. The ratio in the full event is larger than in the subevent due to a simple acceptance effect--the smaller the acceptance, the smaller the high-order correlations.

\begin{figure*}
	\includegraphics[width=0.325\linewidth]{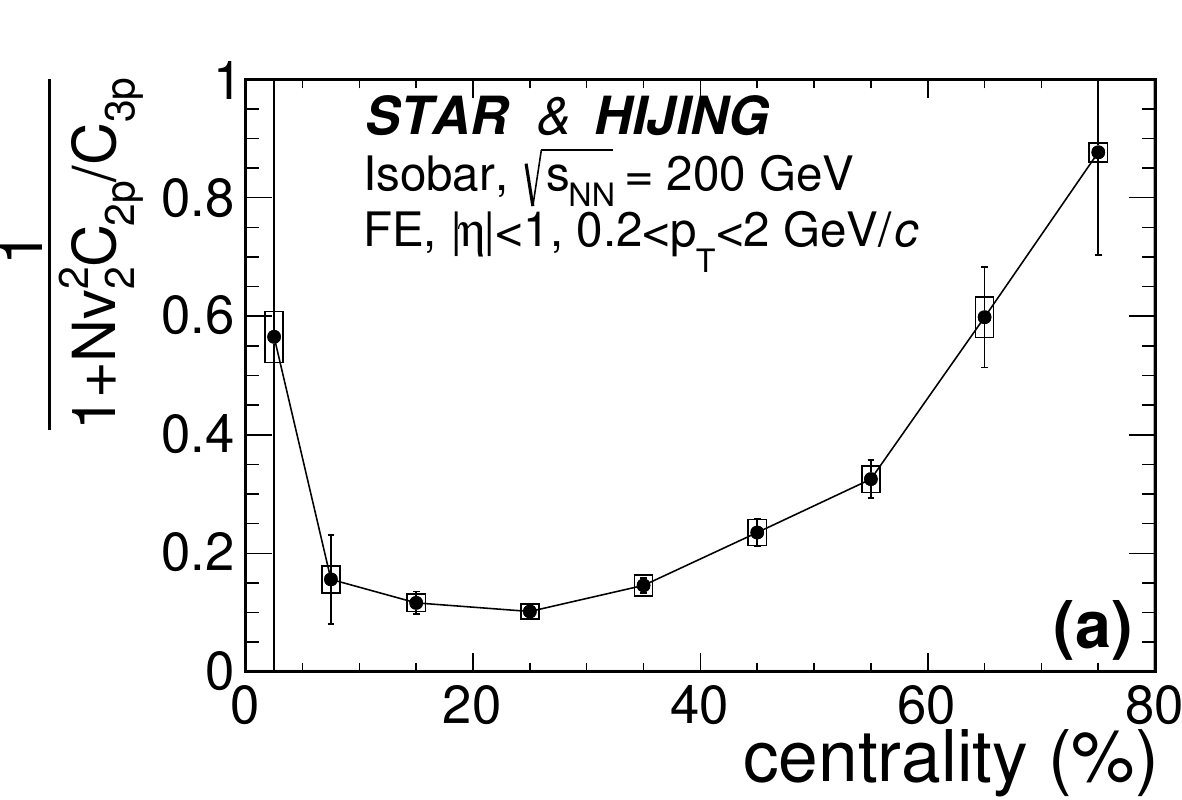}
	\includegraphics[width=0.325\linewidth]{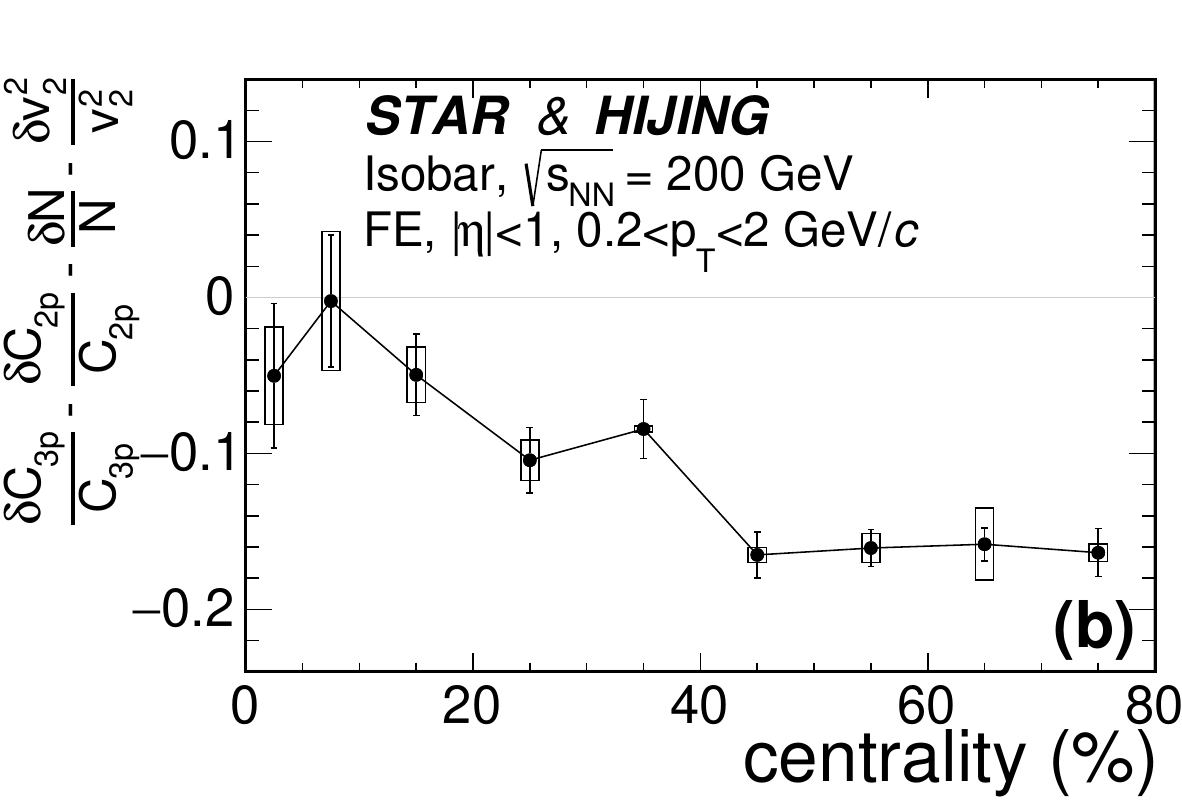}
	\includegraphics[width=0.325\linewidth]{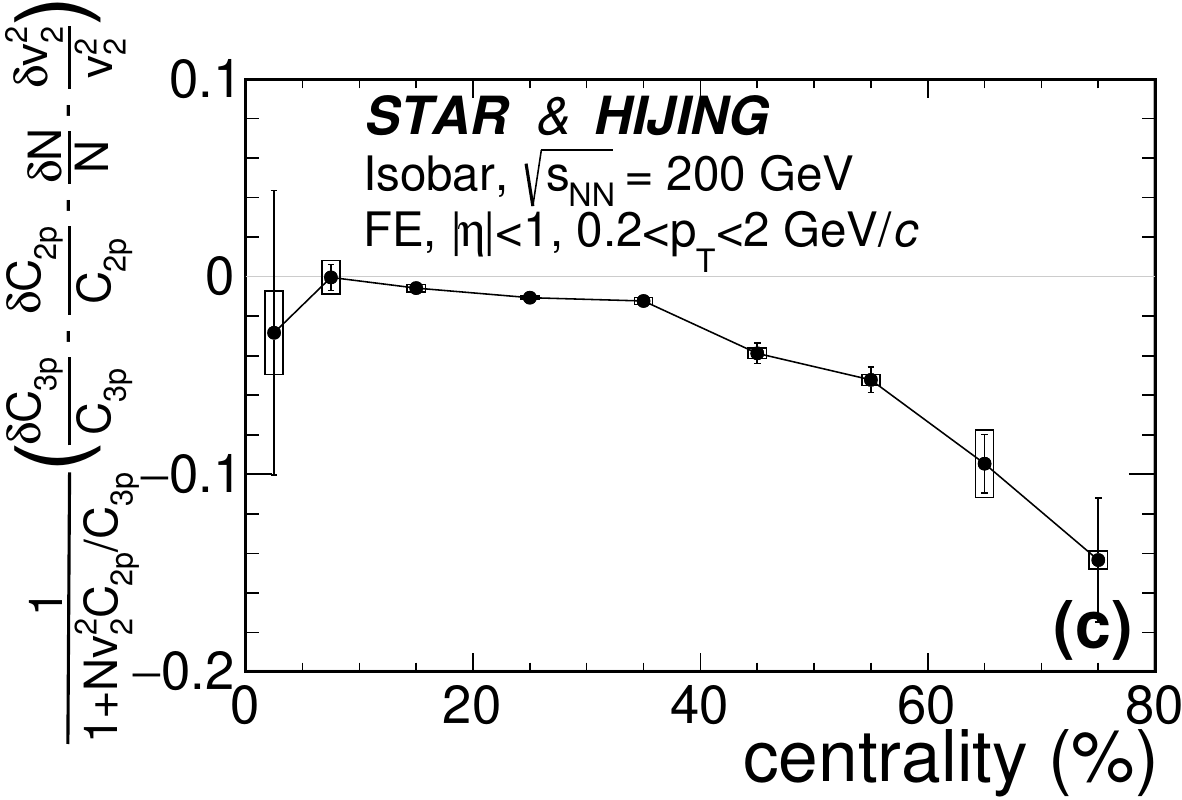}
	\includegraphics[width=0.325\linewidth]{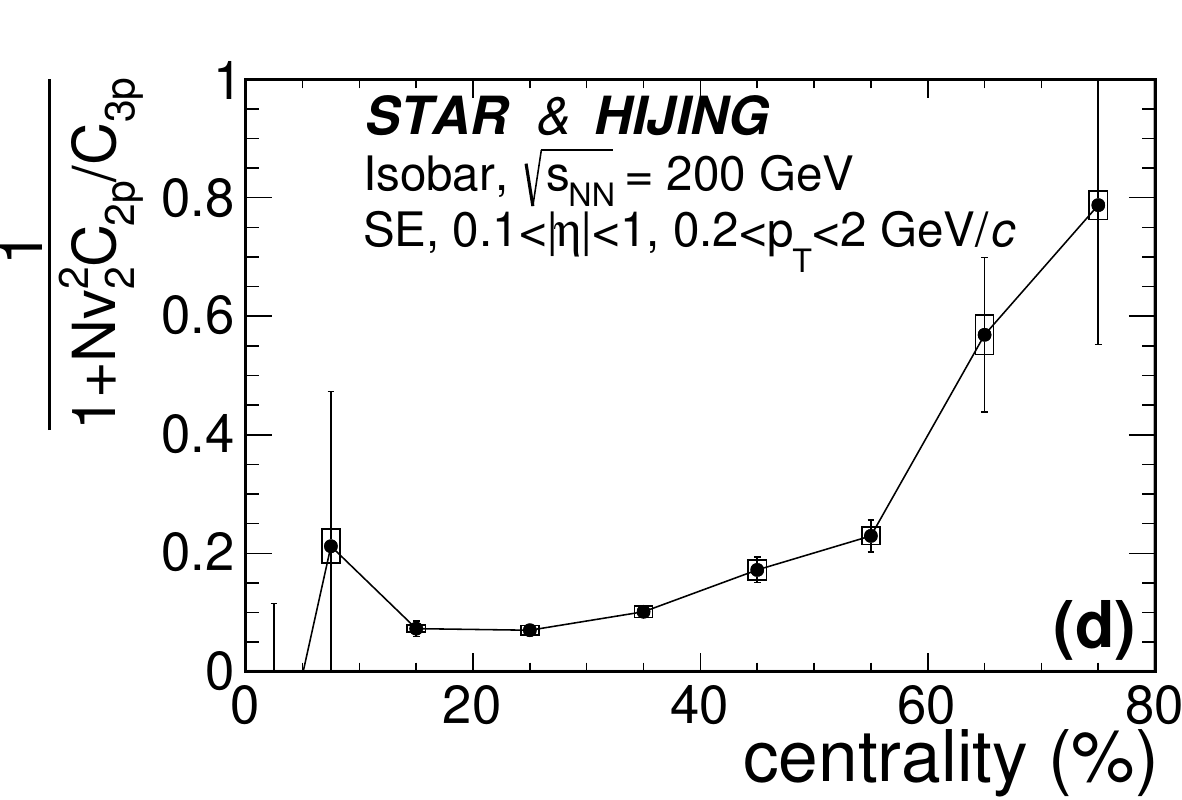}
	\includegraphics[width=0.325\linewidth]{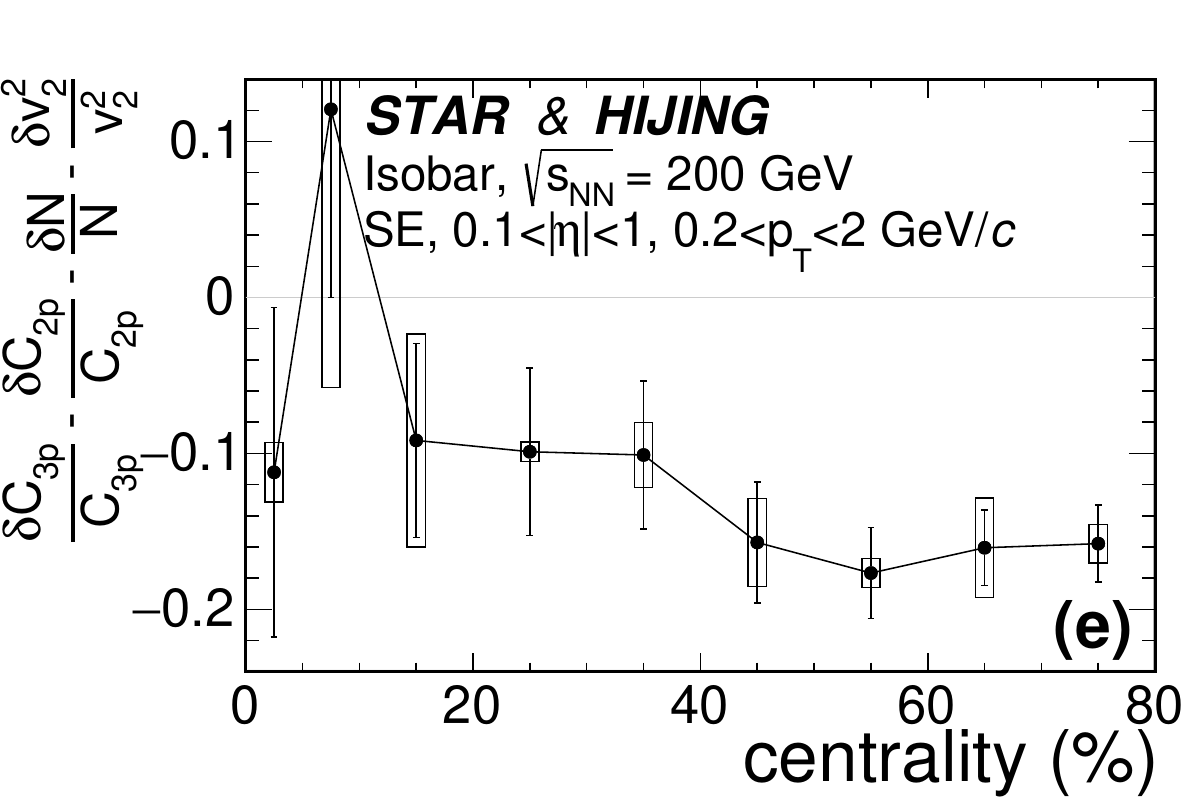}
	\includegraphics[width=0.325\linewidth]{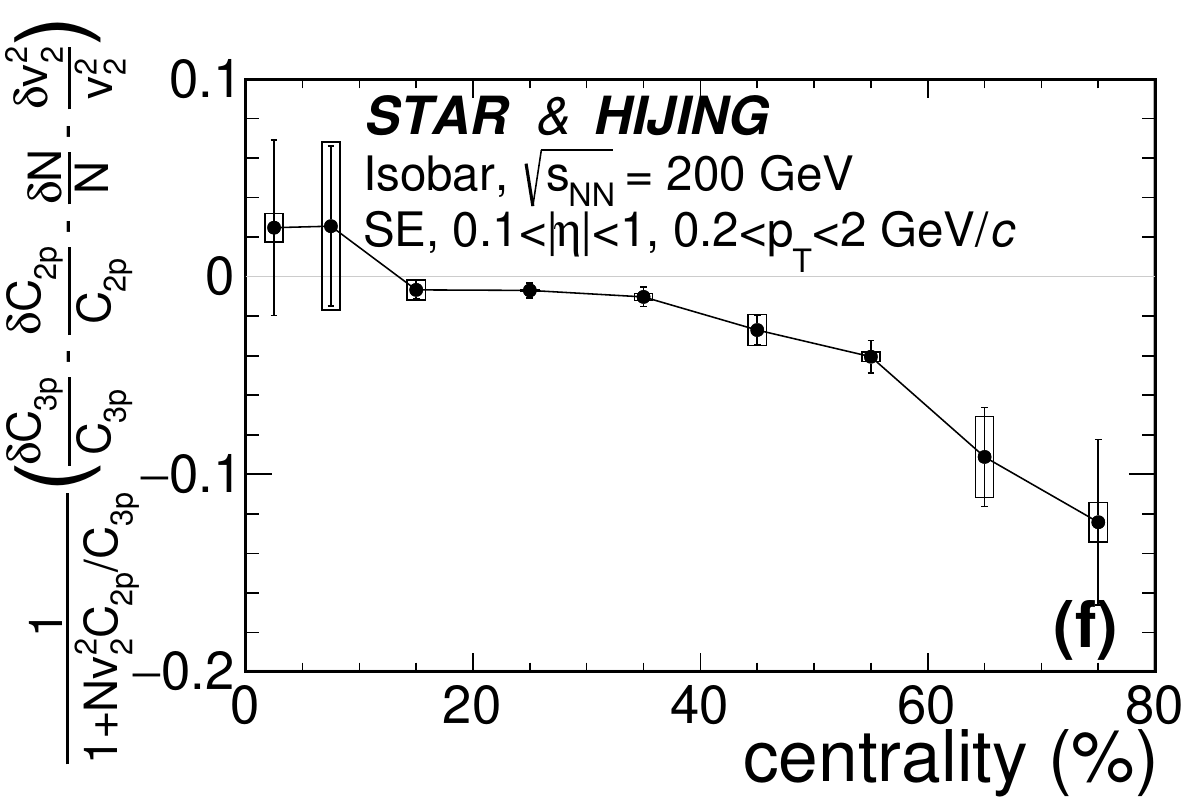}
	\caption{The nonflow components related to 3p nonflow as functions of centrality for full-event (upper row, with Group-3 cuts) and subevent (lower row, with Group-2 cuts) methods. The multiplicity $N$ has been corrected for tracking efficiency. Vertical bars and hollow boxes show the statistical and systematic uncertainties, respectively.}
	\label{fig:nf3p}
\end{figure*}

With all the ingredients ready, as in Figs.~\ref{fig:rn}, \ref{fig:c2p}, \ref{fig:truev2}, and \ref{fig:c3p}, one can easily calculate the background contribution from 3p correlations. The results are shown in Fig.~\ref{fig:nf3p}. The prefactor is depicted in the left column, the sum of the various isobar differences is depicted in the middle column, and the final 3p background difference is depicted in the right column.

%------------------------------------------------------------------------------------------------%

\section{Result} \label{sec:result}

\begin{figure*}
	\includegraphics[width=0.45\linewidth]{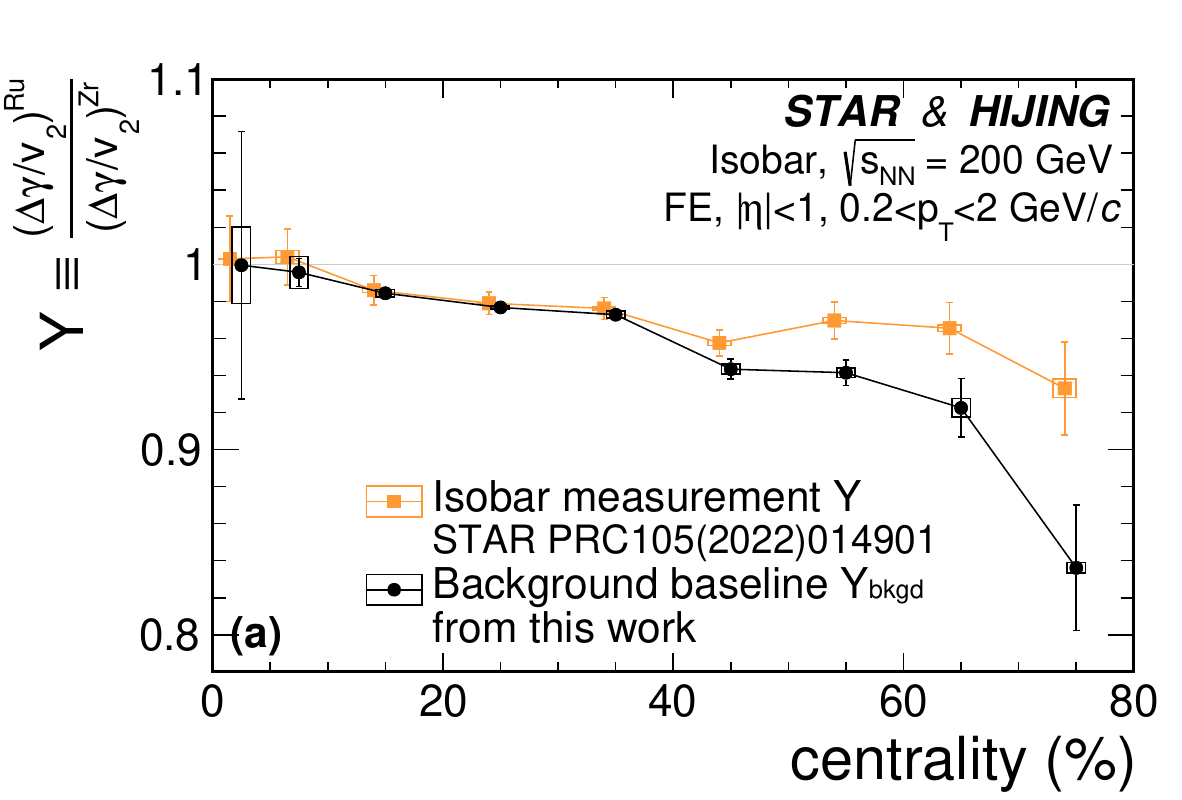}
	\includegraphics[width=0.45\linewidth]{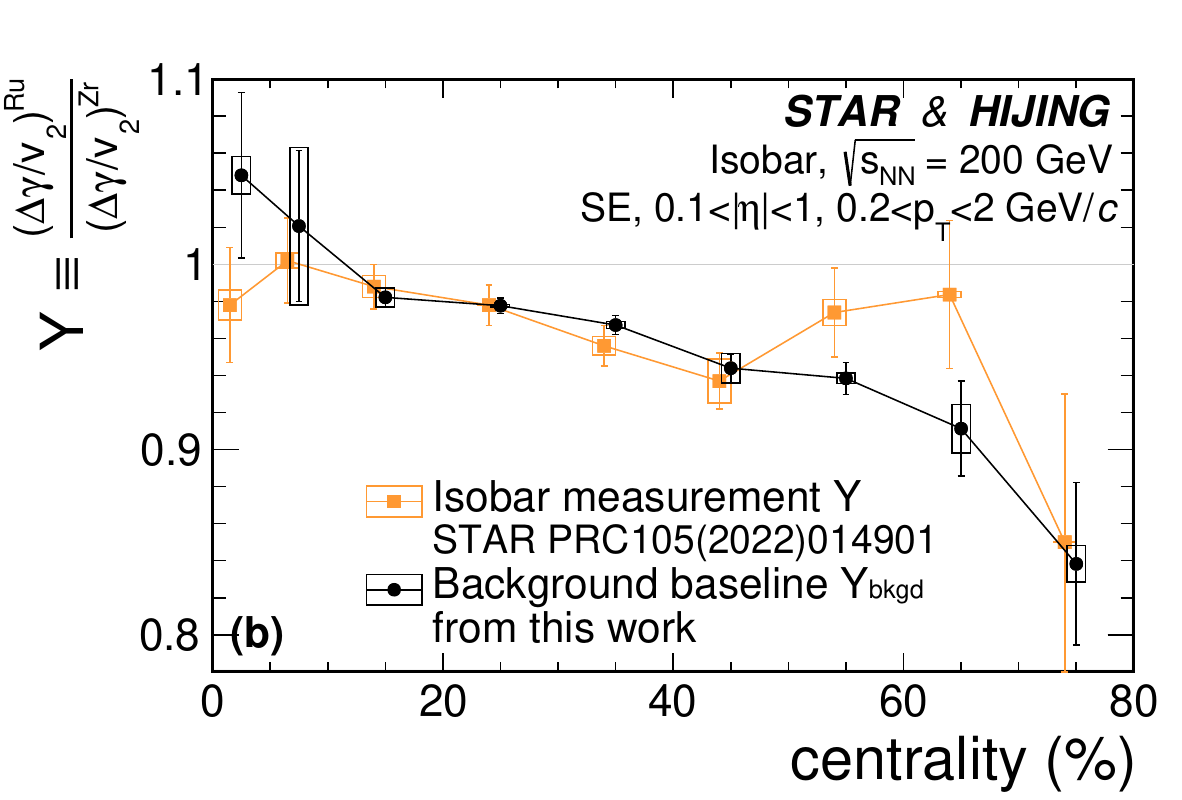}
	\caption{
        The $Y\equiv\frac{(\Delta\gamma/v_{2}^{*})^{\Ru}}{(\Delta\gamma/v_{2}^{*})^{\Zr}}$ measurements from the cited Ref.~\cite{STAR:2021mii} (orange curves) and background baseline $Y_{\bkgd}$ from this study (black curves) as functions of centrality for full-event (left panel, with Group-3 cuts) and subevent (right panel, with Group-2 cuts) methods. Vertical bars and hollow boxes show the statistical and systematic uncertainties, respectively.
	}
	\label{fig:dgv2}
\end{figure*}

The previous section discusses all the ingredients needed for background estimation. %This section plugs them into Eq.~(\ref{eq:Y}) and calculates some intermediate steps and the final results. 
Besides the unity, there are three terms in Eq.~(\ref{eq:Y}). 
%The first two terms have been presented in Fig.~\ref{fig:rn} and Fig.~\ref{fig:enf} respectively. 
These terms are presented in Fig.~\ref{fig:rn}, Fig.~\ref{fig:enf}, and Fig.~\ref{fig:nf3p}, separately. 
The sum of all those terms by Eq.~(\ref{eq:Y}) is the background baseline estimate.
It is shown in Fig.~\ref{fig:dgv2} as a function of centrality, together with the STAR data from Group-3 full-event and Group-2 subevent measurements. 
We also apply the same procedure for the other two measurements in the STAR isobar blind analysis~\cite{STAR:2021mii}, namely Group-2 full-event and Group-4 subevent measurements. 
%In Fig.~\ref{fig:dgv2}, the difference between data and baseline at 50-80\% centrality is $0.0305\pm0.0100\pm0.0041$ ($2.8\sigma$) for full events, and $0.0355\pm0.0215\pm0.0046$ ($1.6\sigma$) for subevents. This discrepancy at peripheral collisions is still probably due to fluctuations. 

\begin{figure*}
        \includegraphics[width=1.0\linewidth]{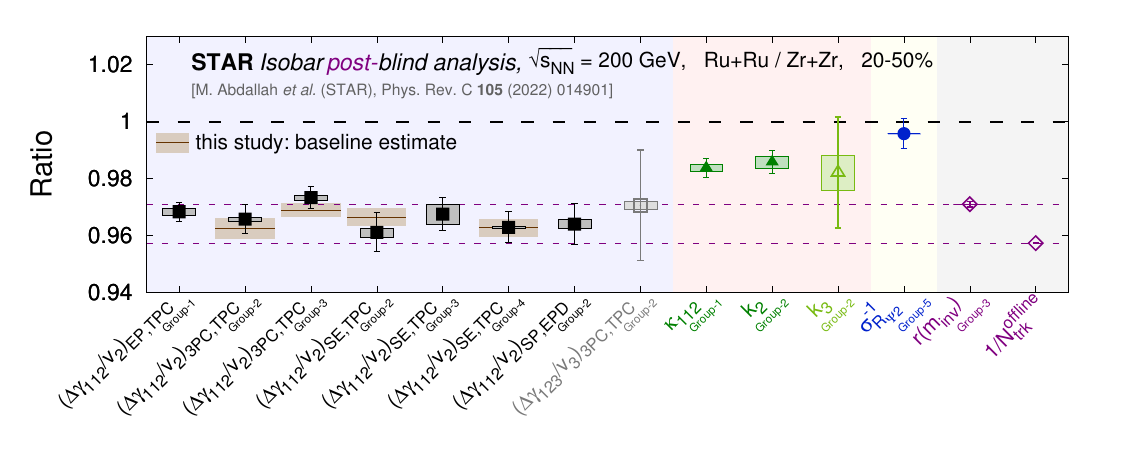}
	\caption{Compilation of the Ru+Ru to Zr+Zr isobar ratios of the $\dg/v_2$ measurements, $Y$ (black squares, with statistical uncertainties indicated by the vertical bars and systematic uncertainties by hollow boxes), and the $r$ and $1/N$ measurements (purple diamonds), with the left coordinate labeling, from the cited STAR blind analysis~\cite{STAR:2021mii}. %\red{\sout{The isobar ratios of multiplicity scaled quantity $N\dg/v_2$, i.e., the $Y$ values of the black squares are read according to the right coordinate.} [do we still need right axis?]} 
    %\red{\sout{The vertical lines indicate statistical uncertainties whereas boxes indicate systematic uncertainties.}}  
    %The colors in the background are intended to separate two types of measures.
    The estimated background baselines from this analysis, $Y_{\bkgd}$, for the four cumulant measurements of the isobar $\dg/v_2$ ratios 
    %\red{\sout{(or equivelently $N\dg/v_2$)}} 
    are shown by the horizontal bars (central values) and the shaded areas (total uncertainties). The total uncertainties are the quadratic sum of the statistical and systematic uncertainties on the background baseline estimates~\cite{STAR:2021mii}. %\sout{points assessed by analysis cut variations~\cite{STAR:2021mii}; they are uncorrelated.
    %The various $\dg/v_2$ measurements were from the same data sample~\cite{STAR:2021mii}; the systematic uncertainties on the data and on the background baseline estimates are correlated across the data points.}
    %\red{the left label should be $\mean{\frac{(\dg/v_2)^\Ru}{(\dg/v_2)^\Zr}}$ and the right label should simply be $\mean{\frac{\Ntrk^\Ru}{\Ntrk^\Zr}}\cdot\mean{\frac{(\dg/v_2)^\Ru}{(\dg/v_2)^\Zr}}$}
   	}
	\label{fig:moneyplot}
\end{figure*}

We compute an average background baseline over the centrality range of 20-50\%. 
Each of the three background terms in Eq.~\ref{eq:Y} is averaged first, weighted by the corresponding inverse squared statistical uncertainty.  Then, the three terms are added to yield the average $Y_{\bkgd}$ baseline.
The average $Y_{\bkgd}$ baselines are tabulated in Table~\ref{tab} for the four measurements and plotted on the summary plot in Fig.~\ref{fig:moneyplot}.
% \begin{equation}
% \begin{split}
% 	\langle Y_{\bkgd} \rangle &= 1 + \left\langle\frac{\delta (C_\twp / N)}{C_\twp / N}\right\rangle
% 	%+\left\langle\frac{\delta F_{2}}{F_{2}}\right\rangle 
% 	-\left\langle\frac{\delta \enf}{1+\enf}\right\rangle \\
%     &+\left\langle\frac{1}{1+\frac{Nv_2^2}{C_\thp/C_\twp}}\left(\frac{\delta C_\thp}{C_\thp}-\frac{\delta C_\twp}{C_\twp}-\frac{\delta N}{N}-\frac{\delta v_2^2}{v_2^2}\right)\right\rangle\,,
% \end{split}
% \end{equation}
% where $\langle \cdots \rangle$ here means 20-50\% average.
\begin{table*}[]
    \caption{Isobar measurements in the 20-50\% centrality range from the STAR blind analysis~\cite{STAR:2021mii} and the corresponding background baseline estimates from this work, for the four measurements using cumulant analysis techniques (two-particle cumulant for $v_2^*$ and three-particle cumulant for $\dg$). Kinematic cuts~\cite{STAR:2021mii} are $|\eta|<1$ for full event (FE) and $0.1<|\eta|<1$ for subevent (SE), both with $0.2 <\pt<2$~\gevc. The first block lists the important analysis cuts~\cite{STAR:2021mii}, slightly differing among the four measurements, along with the isobar ratios of the $\mean{\dg/v_2}$ measurements~\cite{STAR:2021mii} (denoted as $Y$). The second block tabulates the ingredients used for the background baseline estimates, either from data or \hijing\ simulations with the corresponding analysis cuts, along with final background baseline ($Y_{\rm bkgd}$) as well as the magnitudes of its three components in Eq.~(\ref{eq:Y}). %\red{\sout{Note the numbers do not necessarily add up exactly because of difference between average of ratios and ratio of averages.}} 
    The last block lists the background subtracted signal ($Y_{\signal}$) and the CME upper limit at the 95\% confidence level assuming the CME signal difference between Ru+Ru and Zr+Zr is 15\%. 
	The first quoted uncertainty is statistical and the second systematic.
	%The numbers of the format $vv\pm xx \pm yy$ show the value, statistical, and systematic uncertainties in sequence.
    %\\
    %\blue{$\delta C_{\thp} / C_{\thp}$ comes from our old \hijing, where there is no $\Delta\eta_{\alpha\beta}$ gap and range $0.05<|\eta|<1$ for subevent. The systematics of $C_{\thp}$ come from the difference between old \hijing\ with quenching on and off.}
	%\blue{We are still accumulating \hijing\ events, so the numbers here will change in further.}
	}
    \label{tab}
	\resizebox{1.0\linewidth}{!}{
	\begin{tabular}{l|cc|cc}\hline
         &  Group-2 FE & Group-3 FE & Group-2 SE & Group-4 SE \\ \hline
         %kinematics & \multicolumn{2}{c|}{$|\eta|<1$, $0.2 <\pt<2$~GeV} & \multicolumn{2}{c}{$0.1<|\eta|<1$, $0.2 <\pt<2$~GeV} \\ %\hline
         $\Delta\gamma$ cuts & \multicolumn{2}{c|}{$|\Delta\eta_{\alpha\beta}|>0.05$} & $|\Delta\eta_{\alpha\beta}|>0.05$ & {\bf --} \\ %\hline
         $v_{2}^*$ cuts & $|\Delta\eta_{c}|>0.05$ \& Gaus.~fit & {\bf --} & {\bf --} & Same-sign only \\ 
         $Y \equiv \mean{\frac{(\dg/v_2^*)^\Ru}{(\dg/v_2^*)^\Zr}}$ & $0.9658\pm0.0050\pm0.0007$ & $0.9733\pm0.0040\pm0.0010$ & $0.9611\pm0.0070\pm0.0016$ & $0.9629\pm0.0050\pm0.0003$ \\ %this line will be removed in the final paper
         %$Y\equiv\mean{\frac{N^\Ru}{N^\Zr}}\mean{\frac{(\dg/v_2^*)^\Ru}{(\dg/v_2^*)^\Zr}}$ & $1.0091\pm0.0050\pm0.0007$ & $1.0166\pm0.0040\pm0.0010$ & $1.0037\pm0.0070\pm0.0016$ & $1.0055\pm0.0050\pm0.0003$ \\
         \hline
         %$v_2$ & \multicolumn{2}{c|}{$xx\pm xx\pm xx$} & \multicolumn{2}{c}{$xx\pm xx\pm xx$} \\
         $\mean{\enf}$ & $0.2528\pm0.0027\pm0.0489$ & $0.3419\pm0.0008\pm0.0555$ & $0.1812\pm0.0007\pm0.0464$ & $0.1948\pm0.0007\pm0.0465$ \\
         $\mean{C_\twp}$ (ZDC) & \multicolumn{2}{c|}{$0.8302\pm0.0511\pm0.0415$} & $0.5236\pm0.0375\pm0.0262$ & $0.6097\pm0.0540\pm0.0305$ \\ 
         $\mean{C_\thp}$ (\hijing) & \multicolumn{2}{c|}{$0.0707\pm0.0005\pm0.0084$} & $0.0133\pm0.0002\pm0.0013$ & $0.0135\pm0.0002\pm0.0013$ \\ 
         $\mean{\delta C_\thp/C_\thp}$ (\hijing) & \multicolumn{2}{c|}{$-0.0054\pm0.0102\pm0.0049$} & $-0.0035\pm0.0262\pm0.0065$ & $-0.0180\pm0.0241\pm0.0073$ \\
         %\hline
         %$\mean{\delta r/r+\delta N/N}$ & \multicolumn{2}{c|}{$0.0106\pm0.0003\pm0.0007$} & $0.0119\pm0.0004\pm0.0006$ & $0.0104\pm0.0004\pm0.0006$ \\
         $\mean{\delta r/r}$ & \multicolumn{2}{c|}{$-0.0329\pm0.0003\pm0.0007$} & $-0.0308\pm0.0004\pm0.0006$ & $-0.0323\pm0.0003\pm0.0006$ \\
         $\mean{-\delta\enf/(1+\enf)}$ & $0.0097\pm0.0028\pm0.0001$ & $0.0162\pm0.0008\pm0.0013$ & $0.0080\pm0.0008\pm0.0007$ & $0.0075\pm0.0008\pm0.0008$ \\
         the third term & \multicolumn{2}{c|}{$-0.0144\pm0.0017\pm0.0011$} & $-0.0108\pm0.0028\pm0.0008$ & $-0.0125\pm0.0025\pm0.0016$ \\ 
         $Y_{\bkgd}$ & $0.9625\pm0.0033\pm0.0013$ & $0.9689\pm0.0019\pm0.0016$ & $0.9664\pm0.0029\pm0.0011$ & $0.9628\pm0.0027\pm0.0018$ \\ 
         \hline
         $Y_{\signal}$ & $0.0033\pm0.0060\pm0.0014$ & $0.0044\pm0.0043\pm0.0019$ & $-0.0052\pm0.0075\pm0.0020$ & $0.0001\pm0.0061\pm0.0018$\\ 
         $\fcme^{\Ru}$ upper limit & $11.5\%$ & $10.3\%$ & $8.3\%$ & $9.8\%$ \\ 
         \hline
    \end{tabular}
	}
\end{table*}

The difference between the STAR data from Ref.~\cite{STAR:2021mii} and the baseline from this study, $Y_{\signal}=Y-Y_{\bkgd}$, reflects the relative isobar difference of the possible CME signals in the inclusive $\dg$ measurements. %The main systematic sources and their uncertainties to $Y_{\signal}$ are listed in Table~\ref{tab:sys}. 
Simple algebra indicates
%\begin{equation}
%    Y_{\signal} / Y_{\bkgd} = \frac{\delta \fcme}{1 - \fcme^{\Ru}}\,.
%\end{equation}
\begin{equation}
\begin{split}
	Y_{\rm signal} % \equiv& Y - Y_{\bkgd} = \frac{(\Delta\gamma/v_{2}^{*})^{\Ru}}{(\Delta\gamma/v_{2}^{*})^{\Zr}} - Y_{\bkgd} \\
	%=& \frac{[\Delta\gamma_{\bkgd}/v_{2}^{*} / (1-\fcme)]^{\Ru}}{[\Delta\gamma_{\bkgd}/v_{2}^{*} / (1-\fcme)]^{\Zr}} - Y_{\bkgd} \\
	%=& Y_{\bkgd} \left( \frac{[1/(1-\fcme)]^{\Ru}}{[1/(1-\fcme)]^{\Zr}} - 1 \right) \\
	=& Y_{\bkgd} \frac{\delta \fcme}{1 - \fcme^{\Ru}} %\\
	%=& Y_{\bkgd} \left( \frac{b}{uY_{\bkgd}} - 1 \right) \fcme = \fcme \left(\frac{b}{u} - Y_{\bkgd}\right) \\
	%=& \fcme \left( \frac{(B^{2}/v_{2}^{*})^{\Ru}}{(B^{2}/v_{2}^{*})^{\Zr}} - Y_{\bkgd} \right)
\end{split}
\end{equation}
where $\fcme \equiv \Delta\gamma_{\cme}/\Delta\gamma = 1- \Delta\gamma_{\bkgd}/\Delta\gamma$ is the fraction of CME signal in the $\Delta\gamma$ measurement, and $\delta \fcme \equiv \fcme^{\Ru} - \fcme^{\Zr}$ is its difference between Ru+Ru and Zr+Zr collisions.
%where $\delta\fcme\equiv\fcme^{\Ru}-\fcme^{\Zr}$, and $\fcme^{\Ru}$ and $\fcme^{\Zr}$ are the fractions of CME signals in the respective $\dg$ measurements in Ru+Ru and Zr+Zr collisions. 
Assuming the CME signal is proportional to the squared magnetic field, $\Delta\gamma_{\cme} \propto B^2$, 
%\begin{equation} \label{eq:f-ratio}
%\begin{split}
%	\frac{\fcme^{\Zr}}{\fcme^{\Ru}} 
%	=& \frac{\Delta\gamma_{\cme}^{\Zr}}{\Delta\gamma_{\cme}^{\Ru}}
%	\cdot \frac{\Delta\gamma^{\Ru} / v_{2}^{*\Ru}}{\Delta\gamma^{\Zr} / v_{2}^{*\Zr}}
%	\cdot \frac{v_{2}^{*\Ru}}{v_{2}^{*\Zr}} 
%	= \frac{(B^{2})^{\Zr}}{(B^{2})^{\Ru}} Y \frac{v_{2}^{*\Ru}}{v_{2}^{*\Zr}} 
%	\,,
%\end{split}
%\end{equation}
then after a few steps of algebra we obtain
%\begin{equation}
%    \fcme^{\rm Ru} = \frac{Y_{\signal}}{Y} \left[ 1 - Y_{\rm bkgd} \left(1-\frac{\delta v_2^*}{v_2^*}\right) \left(1-\frac{\delta B^2}{B^2}\right) \right]^{-1}\,,
%\end{equation}
\begin{equation} \label{eq:fcme-Y}
\begin{split}
    %\fcme^{\Zr} =& Y_{\signal} \left/ \left( \frac{(B^{2}/v_{2}^{*})^{\Ru}}{(B^{2}/v_{2}^{*})^{\Zr}} - Y_{\bkgd} \right) \right. \\
    %\approx& Y_{\signal} \left/ \left[\frac{\delta B^{2}}{B^{2}} - \frac{\delta v_{2}^{*}}{v_{2}^{*}} + 1 - Y_{\bkgd} \right] \right. \,, \\
    %\fcme^{\Ru} =& \frac{(B^{2}/v_{2}^{*})^{\Ru}}{(B^{2}/v_{2}^{*})^{\Zr}} \frac{1}{Y} \fcme^{\Zr} \,, \\
    \fcme^{\Ru} =& \frac{Y_{\signal}}{Y} \left/ \left[ 1 - Y_{\bkgd} \left/ \frac{(B^{2}/v_{2}^{*})^{\Ru}}{(B^{2}/v_{2}^{*})^{\Zr}} \right. \right] \right. \\
    \approx& \frac{Y_{\signal}}{Y} \left/ \left[ 1 - Y_{\bkgd} \left( 1 + \frac{\delta v_{2}^{*}}{v_{2}^{*}} - \frac{\delta B^{2}}{B^{2}} \right) \right] \right.
\end{split}
\end{equation}
where $\delta B^2/B^2\equiv(B^2_{\Ru}-B^2_{\Zr})/B^2_{\Zr}$ is the relative squared magnetic field difference between the isobar collisions.
It reduces to the simple relationship 
%$\fcme=\frac{Y_{\signal}}{Y}/\frac{\delta B^2}{B^2}$
$\fcme^{\Ru} = Y_{\signal} \left/ \frac{\delta B^{2}}{B^{2}} \right.$ if neglecting small quantities.

\begin{figure*}
    \centering
    \includegraphics[width=0.45\linewidth]{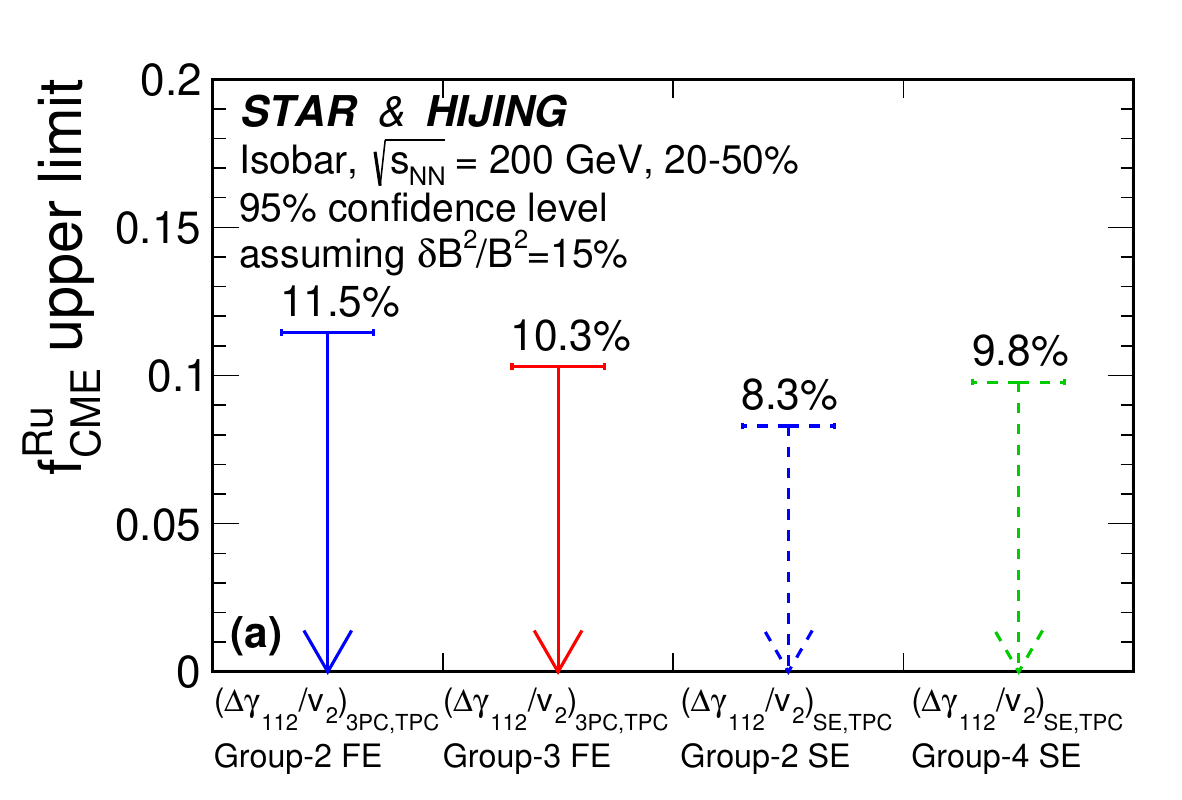}
    \includegraphics[width=0.45\linewidth]{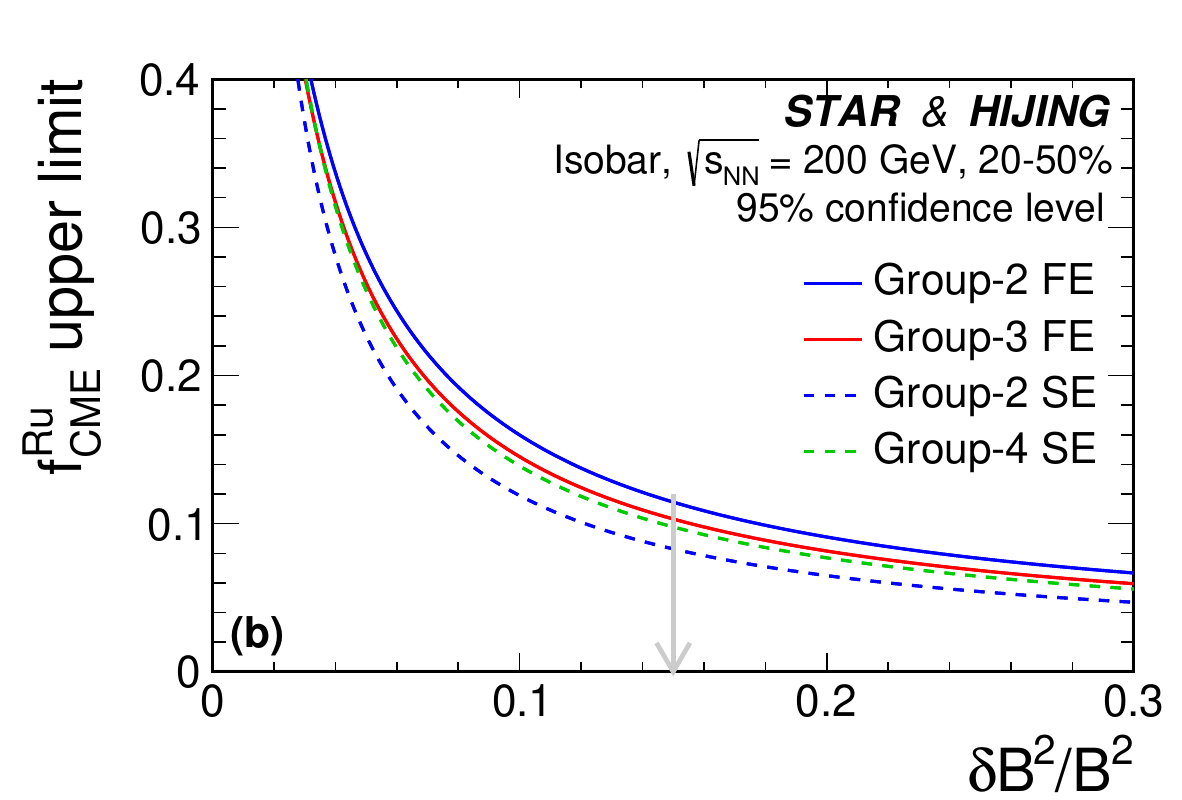}
    \caption{(Left panel) Upper limits at 95\% confidence level on the CME fraction $\fcme^{\rm Ru}$ in the inclusive $\dg$ measurement in 20--50\% centrality Ru+Ru collisions extracted from the isobar blind analyses~\cite{STAR:2021mii} with the background baseline estimates for four measurements from this work, assuming a squared magnetic field difference of $\delta B^{2}/B^{2} = 15\%$ between the two isobar systems~\cite{Deng:2016knn, Voloshin:2010ut}.
    (Right panel) Extracted $\fcme^{\rm Ru}$ upper limits as a function of $\delta B^{2}/B^{2}$. 
    %The $\fcme^{\Ru}$ is calculated from Eq.~(\ref{eq:fcme-Y}) before any approximations.
    The statistical and systematic uncertainties are added in quadrature in extracting the upper limits.
    }
    \label{fig:limit}
\end{figure*}

Our results for $Y_{\signal}$ and $\fcme$ are consistent with zero. 
Assuming $\delta B^2/B^2=15\%$~\cite{Voloshin:2010ut, Deng:2016knn} and that $\fcme$ has a Gaussian probability distribution with a lower bound at the origin~\cite{Feldman:1997qc} ($\fcme \ge 0$), we extract an upper limit on $\fcme^{\Ru}$ of roughly 10\% for all four measurements at 95\% confidence level. 
(The upper limit on $\fcme^{\Zr}$ is accordingly smaller.) 
%(The upper limit of $\fcme^{\Zr}$ is a little bit smaller than $\fcme^{\Ru}$, which is not shown in the paper to avoid confusion.) 
The upper limits are listed in Table~\ref{tab} and illustrated in Fig.~\ref{fig:limit}(a). 
% \begin{equation}
% 	\Delta\gamma_{\cme}^{\Ru} - \Delta\gamma_{\cme}^{\Zr} = 15\% \Delta\gamma_{\cme}^{\Zr}\,,
% \end{equation}
% \red{Then, $f_{\cme}$ can be estimated from $Y - Y_{\bkgd}$
% \begin{equation}
% 	Y - Y_{\bkgd} =? \frac{\Delta\gamma_{\cme}^{\Ru} - \Delta\gamma_{\cme}^{\Zr}}{\Delta\gamma^{\Zr}} = 15\% \frac{\Delta\gamma_{\cme}^{\Zr}}{\Delta\gamma^{\Zr}} = 15\% f_{\cme}\,
% \end{equation}
% as the CME signal fraction in $\Delta\gamma$ measurement from Zr+Zr.
Figure~\ref{fig:limit}(b) shows the $\fcme^{\Ru}$ upper limits extracted for a range of values of $\delta B^2/B^2$. 

We note that in Fig.~\ref{fig:dgv2} the difference between data and baseline in the 50-80\% centrality range is $0.0305 \pm 0.0100 \pm 0.0041$ ($2.8\sigma$) for full events and $0.0355 \pm 0.0215 \pm 0.0046$ ($1.6\sigma$) for subevents. In the blind analysis~\cite{STAR:2021mii} and in this work, we have concentrated on the mid-central 20-50\% centrality range where the CME is predicted to be more probable than peripheral or central collisions~\cite{Kharzeev:2004ey,Kharzeev:2007jp,Fukushima:2008xe}. We speculate that the peripheral collision results are likely due to fluctuations.

%------------------------------------------------------------------------------------------------%

\section{Summary}

%------------------------------------------------------------------------------------------------%

In this study, we have estimated nonflow contributions in $v_{2}$ by fitting two-particle $(\Delta\eta,\Delta\phi)$ distributions and analyzing the deviations from simple multiplicity scaling of the three-particle correlator using the STAR isobar data. 
The 3-particle nonflow correlation contributions to $C_{3}$ are evaluated using \hijing\ simulations. 
With these inputs, we have obtained an improved background estimate of the Ru+Ru to Zr+Zr ratio of the $\Delta\gamma / v_{2}$ variable. The estimated background baselines are found to be consistent with the STAR measurements for both the full-event and subevent methods.
We have also extracted an upper limit of the CME fraction of approximately $10\%$ with a $95\%$ confidence level in isobar collisions at 200 GeV. 

%This paper focuses on the STAR isobar experiments. On the other hand, we need to point out that STAR had CME studies in the previous Au+Au experiments~\cite{STAR:2020gky, STAR:2021pwb}, where indications of possible CME signal were seen~\cite{STAR:2021pwb}. As an outlook, STAR will continue CME search in Au+Au collisions in beam energy scan and high-statistics top energy (200 GeV) dataset taken in year 2023 and 2025~\cite{STAR:2022bur}. 
%On the other hand, there’s indication of possible CME signal in Au+Au collisions (Ref.[16] Phys.Rev.Lett.~128 (2022) 092301). STAR will continue CME search in Au+Au collisions in beam energy scan and high-statistics top energy (200 GeV) dataset taken in 2023+2025. 
This paper focuses on the STAR isobar experiments. On the other hand, the Au+Au collision data from STAR indicate a possible finite CME signal~\cite{STAR:2021pwb}. 
This is consistent with the expectation that the signal to background ratio is approximately a factor of three larger in Au+Au collisions than in isobar collisions~\cite{Feng:2021oub}. 
To outlook, an order of magnitude increase in statistics is expected from future data taking of Au+Au collisions at 200 GeV~\cite{STAR:2022bur}. STAR will continue the CME search with those data as well as data from the beam energy scan~\cite{STAR:2023bur}. 
%As elaborated in the recent Beam-User-Request for the years 2024-25, STAR will continue CME search in Au+Au collisions in beam energy scan and high-statistics top energy (200 GeV) dataset taken in 2023+2025 which shows the prospects of achieving $5\sigma$ measurements. STAR will also continue to analyze the Beam Energy Scan-II data. The novel analysis technique developed here will be highly useful for future analyses.

%------------------------------------------------------------------------------------------------%

\section*{acknowledgments}

We thank the RHIC Operations Group and RCF at BNL, the NERSC Center at LBNL, and the Open Science Grid consortium for providing resources and support.  This work was supported in part by the Office of Nuclear Physics within the U.S. DOE Office of Science, the U.S. National Science Foundation, National Natural Science Foundation of China, Chinese Academy of Science, the Ministry of Science and Technology of China and the Chinese Ministry of Education, the Higher Education Sprout Project by Ministry of Education at NCKU, the National Research Foundation of Korea, Czech Science Foundation and Ministry of Education, Youth and Sports of the Czech Republic, Hungarian National Research, Development and Innovation Office, New National Excellency Programme of the Hungarian Ministry of Human Capacities, Department of Atomic Energy and Department of Science and Technology of the Government of India, the National Science Centre and WUT ID-UB of Poland, the Ministry of Science, Education and Sports of the Republic of Croatia, German Bundesministerium f\"ur Bildung, Wissenschaft, Forschung and Technologie (BMBF), Helmholtz Association, Ministry of Education, Culture, Sports, Science, and Technology (MEXT), Japan Society for the Promotion of Science (JSPS) and Agencia Nacional de Investigaci\'on y Desarrollo (ANID) of Chile.

%------------------------------------------------------------------------------------------------%

\bibliography{./ref}

%------------------------------------------------------------------------------------------------%

\end{document}